\documentclass[fleqn,usenatbib,hyperref]{aa}
\PassOptionsToPackage{pdfpagelabels=true}{hyperref}
\usepackage[dvipsnames,svgnames,x11names,hyperref]{xcolor}
\pdfoutput=1
\usepackage[colorlinks]{hyperref}
\hypersetup{colorlinks=true,linkcolor=[rgb]{0.796,0.263,0.208},citecolor=[rgb]{0.18,0.525,0.757},filecolor=[rgb]{0.796,0.263,0.208},urlcolor=[rgb]{0.796,0.263,0.208}}
\usepackage{graphicx,color}
\usepackage{epstopdf}
\usepackage{rotating}
\usepackage{graphicx}
\usepackage{txfonts}
\usepackage[graphicx]{realboxes}
\usepackage[T1]{fontenc}
\usepackage{ae,aecompl}
\usepackage{graphics}
\usepackage{amsfonts}
\usepackage{amsmath}
\usepackage{multicol}
\usepackage{layout}
\usepackage{amssymb}
\usepackage[english]{babel}
\usepackage{scalerel}
\usepackage{times}
\usepackage{xcolor}
\usepackage[export]{adjustbox}
\usepackage{soul}

\begin{document}

\title{AMICO galaxy clusters in KiDS-DR3: cosmological constraints
        from large-scale stacked weak lensing profiles}

\author{
  Carlo Giocoli\inst{\ref{uno},\ref{due},\ref{tre}},\thanks{email: \href{mailto:carlo.giocoli@inaf.it}{carlo.giocoli@inaf.it}}
  Federico Marulli\inst{\ref{due},\ref{uno},\ref{tre}},
  Lauro Moscardini\inst{\ref{due},\ref{uno},\ref{tre}},
  Mauro Sereno\inst{\ref{uno},\ref{tre}},
  Alfonso Veropalumbo\inst{\ref{quattro},\ref{cinque}},  
  Lorenzo Gigante\inst{\ref{due}},
  Matteo Maturi\inst{\ref{sei},\ref{sette}},
  Mario Radovich\inst{\ref{otto}},
  Fabio Bellagamba\inst{\ref{due}},
  Mauro Roncarelli\inst{\ref{due}},
  Sandro Bardelli\inst{\ref{uno}},
  Sofia Contarini\inst{\ref{due},\ref{uno}},  
  Giovanni Covone\inst{\ref{nove},\ref{dieci},\ref{undici}},
  Joachim Harnois-D\'{e}raps\inst{\ref{dodici},\ref{tredici}},  
  Lorenzo Ingoglia\inst{\ref{nove}}, 
  Giorgio F. Lesci\inst{\ref{due},\ref{uno}}, 
  Lorenza Nanni\inst{\ref{due},\ref{quattordici}},
  Emanuella Puddu\inst{\ref{dieci}}}

\institute{
\label{uno} INAF - Osservatorio di Astrofisica e Scienza dello Spazio di Bologna, via Gobetti 93/3, I-40129 Bologna, Italy \and
\label{due} Dipartimento di Fisica e Astronomia "Augusto Righi", Alma Mater Studiorum Universit\`{a} di Bologna, via Gobetti 93/2, I-40129 Bologna, Italy \and
\label{tre} INFN - Sezione di Bologna, viale Berti Pichat 6/2, I-40127 Bologna, Italy \and
\label{quattro} Dipartimento di Fisica, Universit\`a degli Studi Roma Tre, via della Vasca Navale 84, I-00146 Rome, Italy \and
\label{cinque} INFN - Sezione di Roma Tre, via della Vasca Navale 84, I-00146 Rome, Italy \and
\label{sei} Zentrum f\"ur Astronomie, Universit\"at Heidelberg, Philosophenweg 12, D-69120 Heidelberg, Germany \and
\label{sette} ITP, Universit\"at Heidelberg, Philosophenweg 16, D-69120 Heidelberg, Germany \and
\label{otto} INAF - Osservatorio Astronomico di Padova, vicolo dell'Osservatorio 5, I-35122 Padova, Italy \and
\label{nove} Dipartimento di Fisica "E. Pancini", Universit\'a di Napoli Federico II, C.U. di Monte Sant'Angelo, via Cintia, I-80126 Napoli, Italy \and
\label{dieci} INAF - Osservatorio Astronomico di Capodimonte, Salita Moiariello 16, I-80131, Napoli, Italy  \and
\label{undici} INFN - Sezione di Napoli, via Cintia, I-80126, Napoli, Italy  \and
\label{dodici} Astrophysics Research Institute, Liverpool John Moores University, 146 Brownlow Hill, Liverpool, L3 5RF, UK \and
\label{tredici} School of Mathematics, Statistics and Physics, Newcastle University, Herschel Building, NE1 7RU Newcastle-upon-Tyne, United Kingdom \and
\label{quattordici} Institute of Cosmology \& Gravitation, University of Portsmouth, Dennis Sciama Building, Portsmouth, PO1 3FX, UK 
}

\date{2021}

\abstract {The large-scale mass distribution around dark matter haloes
  hosting galaxy clusters provides sensitive cosmological information.}
          {In this work, we make use of a large photometric galaxy
            cluster sample, constructed from the public Third Data
            Release of the Kilo-Degree Survey, and the corresponding
            shear signal, to assess cluster masses and test the
            concordance $\Lambda$-cold dark matter ($\Lambda$CDM)
            model.  In particular, we study the weak gravitational
            lensing effects on scales beyond the cluster virial
            radius, where the signal is dominated by correlated and
            uncorrelated matter density distributions along the
            line-of-sight. The analysed catalogue consists of $6962$
            galaxy clusters, in the redshift range $0.1\le z \le0.6$
            and with signal-to-noise ratio larger than $3.5$.}
          {We perform a full Bayesian analysis to model the stacked
            shear profiles of these clusters. The adopted likelihood
            function considers both the small-scale 1-halo term, used
            primarily to constrain the cluster structural properties,
            and the 2-halo term, that can be used to constrain
            cosmological parameters.}
          {We find that the adopted modelling is successful to assess
            both the cluster masses and the total matter density
            parameter, $\Omega_M$, when fitting shear profiles up to
            the largest available scales of 35 Mpc/$h$. Moreover, our
            results provide a strong observational evidence of the
            2-halo signal in the stacked gravitational lensing of
            galaxy clusters, further demonstrating the reliability of
            this probe for cosmological studies. The main result of
            this work is a robust constraint on $\Omega_M$, assuming a
            flat $\Lambda$CDM cosmology. We get $\Omega_M = 0.29 \pm
            0.02$, estimated from the full posterior probability
            distribution, consistent with the estimates from cosmic
            microwave background experiments.}
          {}
    
    \titlerunning{AMICO-KiDS-DR3: $\Omega_M$ from large-scale stacked profiles}
    \authorrunning{Giocoli et al. 2021}    
    
    \maketitle

\section{Introduction}
\label{intro_label}

The standard $\mathrm{\Lambda}$-cold dark matter
($\mathrm{\Lambda}$CDM) cosmological model provides the most
convenient and simplest parametrisation of the so-called Concordance
Model.  Recent wide field observational surveys have provided
stringent constraints about the two most contributing energy-density
constituents of our Universe: the dark energy and the dark matter.
While the first contributes to about $\sim 68\%$ of the total
mass-energy, the latter constitutes about $\sim 27\%$
\citep{wmap9,planck1_14,planck16a,planck18}.  The current consensus
view is that this dark matter component is mainly cold, being composed
of particles moving at non-relativistic speeds.  Only $\sim 5\%$ of
the Universe is made up of ordinary matter (baryons), and even less of
radiation ($\sim 10^{-5}$).

The $\mathrm{\Lambda}$CDM model is not based on any particular
assumption about the nature and physical origin of the dark
components.  Nevertheless, it is currently the best cosmological model
available to describe the large-scale structure of the Universe and
its evolution.  Moreover, different observations
\citep{hikage19,hamana20,bautista21} show that the geometry of our
Universe is flat on large scales, so that the Concordance Model is
often also called the flat $\mathrm{\Lambda}$CDM model.

In this scenario, clusters of galaxies play an important role both for
constraining cosmological parameters
\citep{sheth99b,tormen98a,borgani11}, and for the study of the
formation and evolution of the galaxies in their environment
\citep{springel01b,borgani04,andreon10,angulo12,zhang16,zenteno16}.
In fact, galaxy clusters are hosted by the most massive virialised CDM
haloes, and their redshift evolution -- number density and spatial
distribution -- is highly sensitive to different cosmological
parameters \citep[see e.g.][]{planckxxiv, vikhlinin09, veropalumbo14,
  veropalumbo16, sereno15b, pacaud18, marulli18, marulli20,
  costanzi19, moresco20, lesci20, nanni21}.

By exploiting the optical and near infrared bands, it is possible to
identify clusters by searching for the cluster-scale over-densities in
the three-dimensional galaxy distribution of wide field observations.
This technique allows to cover large and deep areas of the sky through
photometric surveys, giving the possibility to collect a sufficiently
large number of clusters and use them as cosmological tool.  Ongoing
and planned photometric surveys will allow us to increase by orders of
magnitude the present census of galaxy clusters, expanding their
detection toward lower masses and higher redshifts \citep{lsst,
  euclidredbook, wfirst, sartoris16}.

However, in order to use galaxy clusters as probe to infer
cosmological parameters it is crucial to derive precise and reliable
cluster mass estimates \citep{meneghetti10b,giocoli14}.  In
particular, the measured number of galaxy cluster members, accessible
in photometric or spectroscopic observations, are typically used as
mass-proxy, to be linked to the total mass of the cluster
\citep{simet17,murata19}.  Calibrating the relationship between
mass-proxy and the true three-dimensional mass is often challenging
and sometime hidden in complex, and not yet well understood,
astrophysical processes \citep{rasia12,rasia13,angelinelli20}.

The most reliable approach to estimate cluster masses is currently
provided by the gravitational lensing signal on background galaxies
\citep{bartelmann01,umetsu20}.  The robustness of this method relies
on the fact that it does not require any assumption on the dynamical
state of the luminous and dark matter, being sensitive only to the
total projected matter density distribution between the sources and
the observer.  In particular, the weak lensing effect consists of
slight distortions of the source shapes (called shear) and the
intensification of their apparent brightness (magnification), caused
by the deflection and the convergence of light rays travelling from
the sources to the observer.  The weak lensing comes into play when
the source is at large angular distances from the geometric centre of
the lens.  Hence it is also very useful to probe the lensing cluster
profiles up to large radii.  All of these properties made the weak
lensing extremely powerful to probe the mass density distributions in
clusters on different scales \citep{sereno18}.

A small number of background galaxies limits the study of individual
cluster masses down to few times $10^{14}M_{\odot}$ for typical
current optical surveys.  Indeed, in order to reach a large
signal-to-noise in the weak lensing mass estimate, it is a common
practice to stack the weak lensing signal -- averaging the number of
slightly distorted galaxy images in radial bins to increase the shear
estimate and reduce the noise -- of systems in the same redshift and
observable bins: the resultant signal is that of the mass-weighted
ensemble.  Precisely, the stacked gravitational lensing represents the
cross-correlation between foreground deflector positions and
background galaxy shears.  Given a shallow but broad survey, stacking
the signal around a large number of lenses is an efficient method to
compensate for the low number density of lensed sources. For example,
this approach was applied with great success to the Sloan Digital Sky
Survey (SDSS) data to measure the ensemble averaged lensing signal
around groups and clusters \citep{mandelbaum06c,sheldon09}.  By
splitting the sample of clusters into subsets based on an observable
property, such as amplitude or optical richness, it is possible to
determine the scaling relations between the observable and the mean
mass of the sample \citep[see, e.g.,][]{bellagamba19}.

In this work we will exploit the data-set from the Kilo Degree Survey
(KiDS hereafter) \citep{dejong13}.  The principal scientific goal of
KiDS is to exploit the weak lensing and photometric redshift
measurements to map the matter large-scale distribution in the
Universe.

Appropriate algorithms have to be used to identify galaxy clusters in
photometric surveys.  In this work we exploit the Adaptive Matched
Identifier of Clustered Objects (\texttt{AMICO}) algorithm, presented and
tested in \citet{bellagamba18}, and used to extract the cluster sample
from KiDS-DR3 by \citet{maturi19}.  \texttt{AMICO} is an algorithm for the
detection of galaxy clusters in photometric surveys where the data-set is
affected by a noisy background.  It is built upon a linear optimal
matched filter method \citep{maturi05,maturi07,viola10,bellagamba11},
which maximises the signal-to-noise ratio ($S/N$) taking advantage of
the statistical properties of the field and cluster galaxies.

We build the shear profile of our lens samples following the
conservative approach detailed in \citep{bellagamba19}, then stack
the profile signal to increase the $S/N$.  We determine the weighted
ellipticity of the background sources in radial bins, that is the
tangential and cross components of the shear.  In our subsequent
analyses we use only the tangential component.  Hence are effectively
measuring the shear-cluster correlation function.

For most of the clusters in the sample, the $S/N$ in the shear data is
too low to constrain the density profile and, consequently, to measure
a reliable mass. So we are tied to measure the mean mass for ensembles
of objects chosen according to their observables and redshift.  In
short words, we exploit the shear-cluster correlation (or stacked
lensing).

The main goal of this work is to assess both the cluster masses and
the total matter density parameter, $\Omega_M$, modelling the
shear-cluster correlation up to the largest available scales.

The statistical numerical code has been built with the
\texttt{CosmoBolognaLib} ($V 5.3$), a set of {\em free software}
C++/Python libraries (Marulli et al. 2016) that provide all the
functions used for the estimation of the various cosmological
components of our fitting model.

This work is organised as follows.  In Sec.~\ref{sec:Data} we present
the observational data-set: shear and cluster catalogues.  We
introduce our theoretical model in Sec.~\ref{models_label}, that we
adopt to describe the stacked tangential shear profile of the cluster
sample.  In Sec.~\ref{results_label} we display our results, then
summarise and conclude in Sec.~\ref{summary_label}.

In all the analyses presented in this work, we assume a flat
$\Lambda$CDM model with baryon density $\Omega_b=0.0486$ and scaled
Hubble parameter $h=0.7$. The matter density parameter is left free
except in the first analysis stage where, to infer reference cluster
masses while modelling the 1-halo term, it is fixed to
$\Omega_M=0.3$. When inferring $\Omega_M$ from the 2-halo modelling,
we assume anyway a flat $\Lambda$CDM, hence
$\Omega_{\Lambda}=1-\Omega_M$ is a derived parameter.  The symbol
$\log$, if not specified otherwise, refers to the logarithm base $10$
of the quantity.

\begin{figure*}
  \includegraphics[width=0.45\hsize]{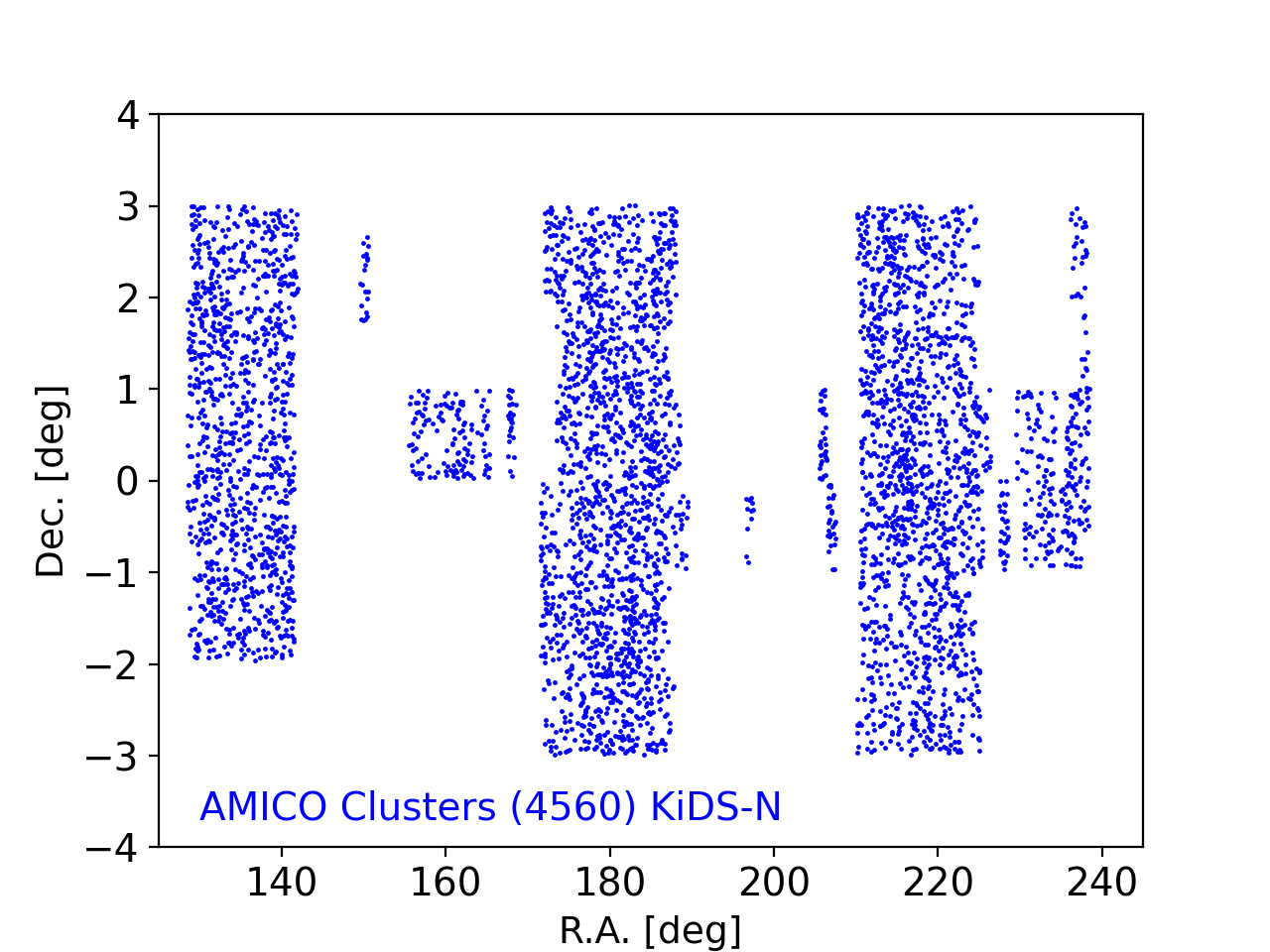}
  \includegraphics[width=0.45\hsize]{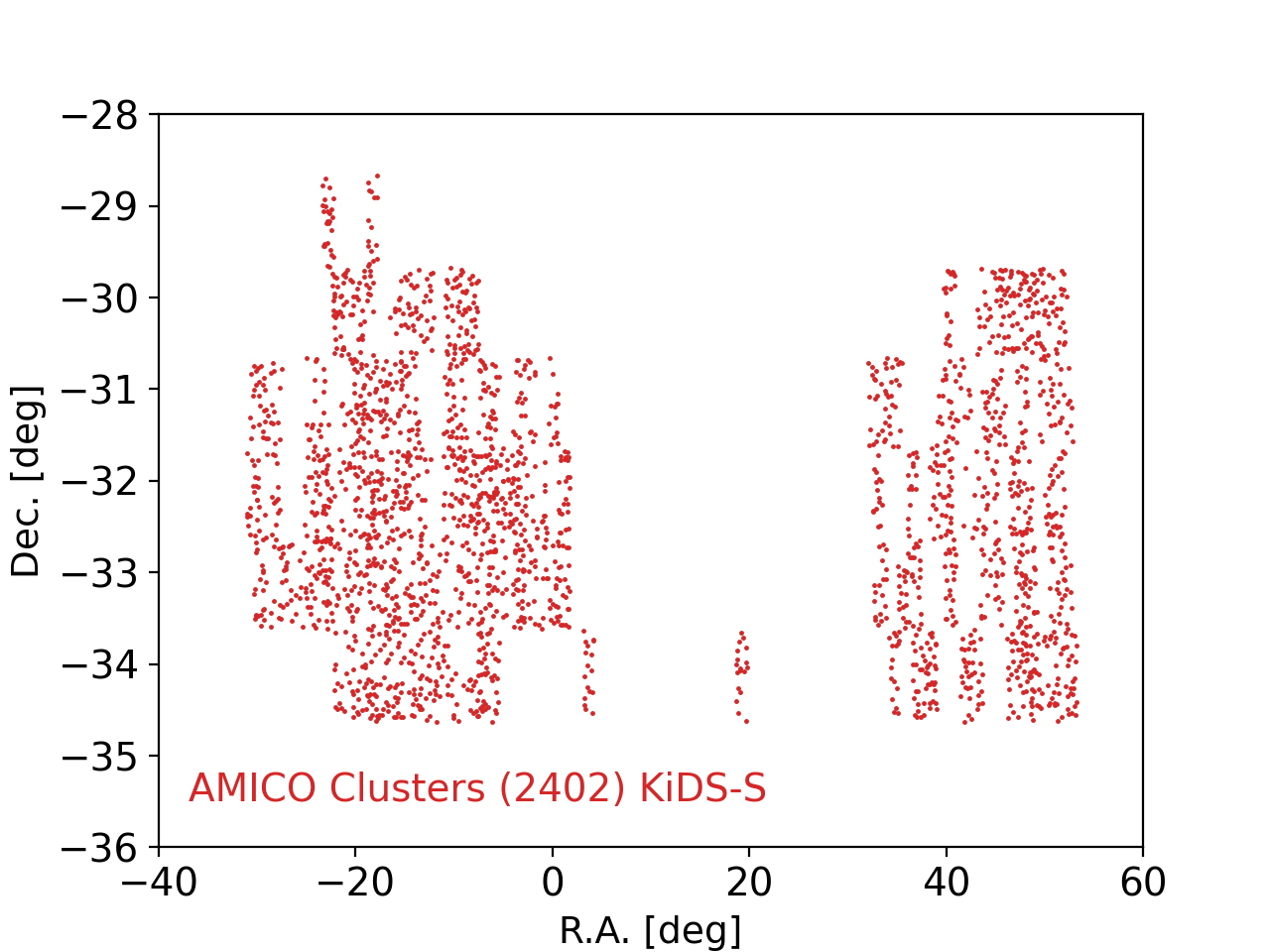}
  \caption{Distribution of cluster centres in angular coordinates of
    the 6962 detections by \texttt{AMICO} divided in the KiDS-N (left)
    and KiDS-S (right) areas.  Systems with $0.1\le z \le 0.6$ and
    with $S/N>3.5$ are displayed. \label{figAMICOpositions}}
\end{figure*}

\section{Data} \label{sec:Data}

KiDS \citep{dejong13} is an optical wide-field imaging survey aiming
at mapping 1350 square degrees of extra-galactic sky in two stripes (an
equatorial one, KiDS-N, and one centred around the South Galactic
Pole, KiDS-S) in four broad-band filters ({\em u, g, r, i}). The
observations are performed with the 268 Megapixels OmegaCAM wide-field
imager \citep[][]{omegacameso11}, owning a mosaic of 32 science CCDs
and positioned on the Very Large Telescope (VLT) Survey Telescope
(VST). This is an ESO telescope of 2.6 meters in diameter, located at
the Paranal Observatory \citep[for further technical information about
  VST see][]{capaccioli11}. The principal scientific goal of KiDS is
to exploit weak lensing and photometric redshift measurements to map
the large-scale matter distribution in the Universe.
The VST-OmegaCAM is an optimal choice for such a survey, as it was
expressly created to give excellent and uniform image quality over a
large ($1 \deg^2$) field of view with a resolution of 0.21
arcsec/pixel. Note that a field of view $1\text{ deg}$ wide
corresponds to $\sim12.5 \text{ Mpc}/h$ at the median redshift of our
lens sample, $z\sim0.35$.

The DR3-KiDS catalogue is composed of about 100,000 sources per square
degree, for a total of almost 50 million sources over the full survey
area. These data are attentively and uniformly calibrated and made
usable in an easily accessible dedicated archive.

We exploit the Third Data Release survey
\citep[KiDS-DR3,][]{dejong17}, which has been obtained with the same
instrumental settings of the other two previous releases
\citep{dejong15} that were made public in 2013 (DR1) and 2015 (DR2).
KiDS-DR3 extends the total released data-set coverage to approximately
$447 \deg^2$ with 440 survey tiles. Therefore, with respect to the
other previous releases, it provided a considerable survey area
extension. This release also includes photometric redshifts with the
corresponding probability distribution functions, a global improved
photometric calibration, weak lensing shear catalogues and
lensing-optimised image data. Source detection, positions and shape
parameters used for weak lensing measurements are all derived from the
stacked $r$-band images, while magnitudes are measured in all filters
using forced photometry.

One of the main improvements of KiDS-DR3, compared to DR1 and DR2, is
the enhanced photometric calibration and the inclusion of photometric
redshift distribution probabilities \citep[e.g., see][]{kuijken15,
  dejong17}.  The combined set of the $440$ survey tiles of DR3 mostly
comprises a small number of large contiguous areas, see
Fig.~\ref{figAMICOpositions}.  This enables a refinement of the
photometric calibration that exploits both the overlap between
observations within a filter as well as the stellar colours across
filters \citep{dejong17}.\footnote{The data products that constitute
the main DR3 release (stacked images, weight and flag maps, and
single-band source lists for 292 survey tiles, as well as a multi-band
catalogue for the combined DR1, DR2 and DR3 survey area of 440 survey
tiles), are released via the ESO Science Archive, and they are also
accessible via the Astro-WISE system
(\href{http://kids.strw.leidenuniv.nl/DR1/access\_aw.php}{http://kids.strw.leidenuniv.nl/DR1/access\_aw.php})
and the KiDS website
\href{http://kids.strw.leidenuniv.nl/DR3}{http://kids.strw.leidenuniv.nl/DR3}.}

\subsection{Galaxy catalogue}

The KiDS galaxy catalogue provides the spatial coordinates, the 2
arcsec aperture photometry in four bands (u, g, r, i) and photometric
redshifts for all galaxies down to the 5$\sigma$ limiting magnitudes
of 24.3, 25.1, 24.9 and 23.8 in the four bands, respectively.  We
chose to avoid the use of galaxy colours to detect clusters in order
to minimise the dependence of the selection function on the presence
(or absence) of the red-sequence of cluster galaxies
\citep{maturi19}. For more information about the shear and redshift
properties of the entire galaxy catalogue see \cite{dejong17} and
\cite{hildebrandt17}.

\subsubsection{Galaxy shape measurements}
The original shear analysis of the KiDS-DR3 data has been described in
\cite{kuijken15} and \cite{hildebrandt17}.
The shape measurements has been performed with {\it lens}fit
\citep{miller07,miller13} and successfully calibrated for the KiDS-DR3
data in \citet{fenechconti17}.

In this work, the data used for shape measurements are those in the
$r$-band, being the ones with the best seeing properties and with
highest source density. The error on the multiplicative shear
calibration, estimated from simulations with {\it lens}fit and
benefiting from a self-calibration, is of the order of 1\%
\citep{hildebrandt17}. The final catalogue KiDS-DR3 gives shear
measurements for $\sim15$ million of galaxies, with an effective
number density of $n_{\rm eff}=8.53$ galaxies arcmin${^{-2}}$
\citep[as defined in][]{heymans12b}, over a total effective area of
$360 \deg^2$.

\subsubsection{Galaxies photometric redshifts}
The original properties of photometric redshifts (photo-$z$) of KiDS
galaxies are described in \cite{kuijken15} and \cite{dejong17}. The
photo-$z$s were extracted with BPZ \citep{benitez00, hildebrandt12}, a
Bayesian photo-$z$ estimator based upon template fitting, from the
4-band ({\em u, g, r, i}). Furthermore, BPZ returns a photo-$z$
posterior probability distribution function which \texttt{AMICO} fully
exploits. When compared with spectroscopic redshifts from the Galaxy
And Mass Assembly \citep[GAMA,][]{liske15} spectroscopic survey of low
redshift galaxies, the resultant accuracy is $\sigma_z\sim0.04(1+z)$,
as shown in \cite{dejong17}.

\subsection{Cluster catalogue}
As already mentioned, the galaxy cluster catalogue we use in this work
has been extracted from KiDS-DR3 using the \texttt{AMICO}
algorithm. In particular, we analyse the same cluster catalogue
adopted by \cite{bellagamba19}. Its full description and validation
can be found in \cite{maturi19}. Here, we present the main properties
of this cluster catalogue. Let us premise that the cluster detection
through \texttt{AMICO} on KiDS-DR3 is an enhancement of the work by
\cite{radovich17} on DR2, both in terms of total covered area (438
deg$^2$ against 114 deg$^2$) and of detection algorithm with respect
to the previous matched filter method \citep{bellagamba11}. From the
initial covered 438 deg$^2$, all clusters belonging to those regions
heavily affected by satellite tracks, haloes created by bright stars,
and image artefacts are rejected \citep{maturi19}. Furthermore, we
select only detections with $S/N > 3.5$\footnote{The method used to
  assess the quality of the detections exploits realistic mock
  catalogues constructed from the real data themselves, as described
  in Section 6.1 of \citet{maturi19}. These mock catalogues are used
  to estimate the uncertainties on the quantities characterising the
  detections, as well as the purity and completeness of the entire
  sample.}. Finally, we narrowed the redshift range to $0.1\le z\le
0.6$, for a final sample of 6962 galaxy clusters. Objects at $z<0.1$
are discarded because of their low lensing power, while those at
$z>0.6$ are excluded because the background galaxies density in KiDS
data is too small to allow a robust weak lensing
analysis. Furthermore, collecting clusters with $0.1\le z \le 0.6$ we
can robustly separate the clustered objects from the population of
background galaxies. For each cluster we select galaxy population,
excluding the galaxies whose most likely redshift $z_s$ is not
significantly higher than the lens one $z_l$:
\begin{equation}
  z_{s,min} > z_l + \Delta z,
\end{equation}
where $z_{s,min}$ is the lower bound of the region including 
the $2\sigma$ of the probability density distribution and 
$\Delta z$ is set to 0.05, similar to the typical uncertainty on
photometric redshifts in the galaxy catalogue and well larger than the
uncertainty on the cluster redshifts.

The criterion based on the photometric redshift distribution aims at
keeping only galaxies which have a well-behaved redshift probability
distribution and have a negligible probability of being at redshift
equal to or lower than the cluster.  In addition we consider $ \leq
z_s \leq 1$ and the flag ODDS in BPZ to be larger than $0.8$: the ODDS
parameter indicates the quality of the BPZ photometric redshifts ( see
\citet{kuijken15} for details in the case of KiDS DR3 data).

Left and right panel of Fig.~\ref{figAMICOpositions} display the
angular distribution of galaxy clusters detected by \texttt{AMICO} in
the (R.A., Dec) coordinate system in the KiDS-N and KiDS-S sky area
covered by the DR3 data, respectively. We use the masks from the two
sky areas to generate random samples of positions, which are needed to
construct tangential shear profiles around random points (see the
discussion on systematics in Section \ref{results_label}).

\texttt{AMICO} searches for cluster candidates by convolving the 3D
galaxy distribution with a redshift-dependent filter, defined as the
ratio between a cluster signal and a noise model. Such a convolution
is able to create a 3D amplitude map, where every peak constitutes a
possible detection. Then, for each cluster candidate, \texttt{AMICO}
returns angular positions, redshift, $S/N$ and the signal amplitude
$A$, which is a measure of the cluster galaxy abundance. In
\citet{bellagamba18} it has been demonstrated, using simulations, that
the amplitude is a well-behaved mass-proxy, provided that the model
calibration is accurate enough. The signal amplitude is defined as:
\begin{equation}
A(\theta_c,z_c)\equiv\alpha^{-1}(z_c)\sum_{i=1}^{N_{gal}}\frac{M_c(\theta_i-\theta_c,m_i)p_i(z_c)}{N(m_i,z_c)}-B(z_c), 
\label{AMICOAmpl}
\end{equation}
where $\theta_c$ are the cluster sky coordinates, $M_c$ is the cluster
model (i.e the expected density of galaxies per unit magnitude and
solid angle) at the cluster redshift $z_c$, $N$ the noise
distribution, $p_i(z)$, $\theta_i$ and $m_i$ indicate the photometric
redshift distribution, the sky coordinates and the magnitude of the
$i$-th galaxy, respectively; the parameters $\alpha$ and $B$ are
redshift dependent functions providing the normalisation and the
background subtraction, respectively.

The cluster model $M_c$ is described by a luminosity function and a
radial density profile, obeying to the formalism presented in
\cite{bellagamba18}; derived from the observed galaxy population of
clusters detected through the SZ-effect \citep{henning17}, as
described in detail in \cite{maturi19}. Furthermore, \texttt{AMICO}
assigns to each galaxy in a sky region a probability to be part of a
given detection, defined as:
\begin{equation}
P(i\in j)\equiv P_{f,i}
\frac{A_jM_{c,j}(\theta_i-\theta_g,m_i)p_i(z_j)}{A_jM_{c,j}(\theta_i-\theta_j,m_i)p_i(z_j)+N(m_i,z_j)}
, \label{Member prob.}
\end{equation}
where $A_j$, $\theta_j$ and $z_j$ are the amplitude, the sky
coordinates and the redshift of the $j$-th detection,
respectively. $P_{f,i}$ is the field probability of the $i$-th galaxy
before the $j$-th detection is defined. Indeed, in this application to
KiDS-DR3 data, the adopted filter includes galaxy coordinates,
$r$-band magnitudes and the full photometric redshift distribution
$p(z)$.

The distributions in redshift and amplitude of this sample are shown
in \cite{bellagamba19}. In the following analysis we will divide the
sample in three redshift bins:
\begin{itemize}
\item $0.1\le z<0.3$, 
\item $0.3\le z<0.45$,
\item  $0.45\le z\le0.6$.
\end{itemize}
Those bins have been chosen with aim of having almost the same number
of clusters and numerous enough also to study the amplitude trends, in
each of them.

\subsection{Measuring the tangential shear profile}
Using the tangential component of the shear signal $\gamma_t$ we can write the
excess surface mass density as:
\begin{equation}
\Delta \Sigma (r) = \bar{\Sigma}(<r) - \Sigma(r) \equiv \Sigma_{\rm
  crit} \gamma_t,
\end{equation}
where $\Sigma(r)$ represents the mass surface density of the lens at
distance $r$ and $\bar{\Sigma}(<r)$ its mean within $r$;
$\Sigma_{\rm crit}$ indicates the critical surface density
\citep{bartelmann01} that can be read as:
\begin{equation}
  \Sigma_{\rm crit} = \dfrac{c^2}{4 \pi G} \dfrac{D_l}{D_s D_{ls}}\,,
  \label{eq_scrit}
\end{equation}
with $D_l$ $D_s$ and $D_{ls}$ are the angular diameter distances
observer-lens, observer-source and source-lens, respectively;
$c$ represents the speed of light and $G$ the
universal gravitational constant.

Selecting all background sources of each cluster-lens, as underlined
before in the text, we can compute the tangential component of the
shear with respect to the cluster centre $\gamma_t$. This allow us to
construct the excess surface mass density profile at distance $r_j$
from the following relation:
\begin{equation}
  \Delta \Sigma (r_j) = \left( \dfrac{ \sum_{i \in j} \left( w_i \Sigma^{-2}_{\mathrm{crit},i} \right)  \gamma_{t,i}\Sigma_{\mathrm{crit}, i}}{\sum_{i \in j} \left( w_i \Sigma^{-2}_{\mathrm{crit},i}\right) } \right)  \dfrac{1}{1 + K(r_j) }
\end{equation}
where $w_i$ indicates the weight assigned to the measurement of the
source ellipticity and $K(r_j)$ the average correction due to the
multiplicative noise bias in the shear estimate as in Eq.~(7) by
\citet{bellagamba19}. To compute the critical surface density for the
$i$-th galaxy we use the most probable source redshift as given by
BPZ.

When stacking the excess surface mass density profile of clusters in equal
amplitude and redshift bin we refer to the following relation:
\begin{equation}
  \Delta \Sigma_N(r_j) = \dfrac{\sum_{n\in N} W_{n,j} \Delta \Sigma_{n}(r_j)
  }{\sum_{n \in N} W_{n,j}}\,,
\end{equation}
where $N$ represents the bin in which we perform the stacking and
$W_{n,j}$ indicates the weight for the $j$-th radial bin of the $n$-th
cluster:
\begin{equation}
 W_{n,j} = \sum_{i \in j} w_i \Sigma^2_{\mathrm{crit},i}\,.
\end{equation}

\section{Models}
\label{models_label}

We model the stacked lensing signal with the projected halo model
formalism. The total density profile is constructed considering the
signal coming from the central part of the cluster, called the 1-halo
term, and the one caused by correlated large-scale structures, called
the 2-halo term.

\subsection{The 1-halo lens model}
\label{sub:1halo}
The radial density profile of the cluster main halo is modelled
considering a smoothly-truncated Navarro-Frenk-White
\citep[NFW,][]{navarro97} density profile \citep{baltz09}:
\begin{equation}
  \rho(r)=\frac{\rho_s}{(r/r_s)(1+r/r_s)^2}\Bigl(\frac{r_t^2}{r^2+r_t^2}\Bigr)^2, \label{trunc.NFW}
\end{equation}
where $\rho_s$ represents the typical matter density within the scale
radius $r_s$, and $r_t$ indicates the truncation radius, typically
expressed in terms of the halo radius $R_{200}$ - the radius enclosing
$200$ times the critical density of the Universe at the considered
redshift $\rho_c(z)$: $r_t = t\,R_{200}$ with $t$ the truncation
factor. The scale radius is commonly parametrised as
$r_s=R_{200}/c_{200}$, where $c_{200}$ represents the concentration
parameter, which is correlated with the halo mass and redshift
depending on the halo mass accretion history
\citep{maccio07,neto07,maccio08,zhao09,giocoli12b}.  The total mass
enclosed within the radius $R_{200}$, $M_{200}$, can be seen as the
normalisation of the model and a mass-proxy of the true enclosed mass
of the dark matter halo hosting the cluster \citep{giocoli12a}.
 
This truncated version of the NFW model was deeply tested in
simulations by \cite{oguri11b}, demonstrating that it describes the
cluster profiles more accurately than the original NFW profile model,
up to about $10$ times the radius enclosing $200$ times the critical
density of the Universe. In fact, one of the main advantages of the
truncation radius is that it removes the non-physical divergence of
the total mass at large radii.  Moreover, the \cite{baltz09} model
describes accurately the transition between the cluster main halo and
the 2-halo contribute \citep{cacciato09,giocoli10b,cacciato12},
providing less biased estimates of mass and concentration from shear
profiles \citep{sereno17}.  Neglecting the truncation, the mass would
be underestimated and the concentration overestimated.  As we are
considering only stacked shear profiles, the NFW model provides a
reliable description even though it is based on the assumption of
spherical symmetry. In fact, when stacking several shear profiles, the
intrinsic halo triaxiality tends to assume a spherical radial symmetry
on the stacked profile, for statistical reasons. This is the reason
why the weak lensing cross-correlation provides a direct estimate of
the mean mass for clusters in a given range of observable
properties. As reference model, we set $r_t=3$ for every amplitude and
redshift bins, as done e.g. by \citet{bellagamba19}, following the
results by \citet{oguri11b}. However, we will discuss how the mass and
concentration estimates change when assuming different values of the
truncation radius.

When analysing the weak lensing by clusters, an important source of
bias is the inaccurate identification of the lens centre. In this
study, we assume those determined by \texttt{AMICO} in the detection
procedure. However, the detection is performed through a grid which
induces an intrinsic uncertainty due to the pixel size, that is $<0.1
\text{ Mpc}/h$ \citep{bellagamba18}. Furthermore, it must be taken
into account that the galaxy distribution centre may appreciably
deviate from the mass centre of the system - or the location of the
minimum potential, especially for unrelaxed systems with ongoing
merging events. For example, in \citet{johnston07} it was found that
the Brightest Central Galaxy (BCG), defining the cluster centre, might
be misidentified and, in that case, the value of $\Delta\Sigma$ at
small scales is underestimated, biasing low the measurement of the
concentration by $\sim15\%$ and consequently underestimating the mass
by $\sim10\%$. Moreover, \citet{george12} found that the halo mass
estimates from stacked weak lensing can be biased low by 5\%-30\%, if
inaccurate centres are considered and the issue of off-centring is not
addressed.

\begin{figure}
  \includegraphics[width=\hsize]{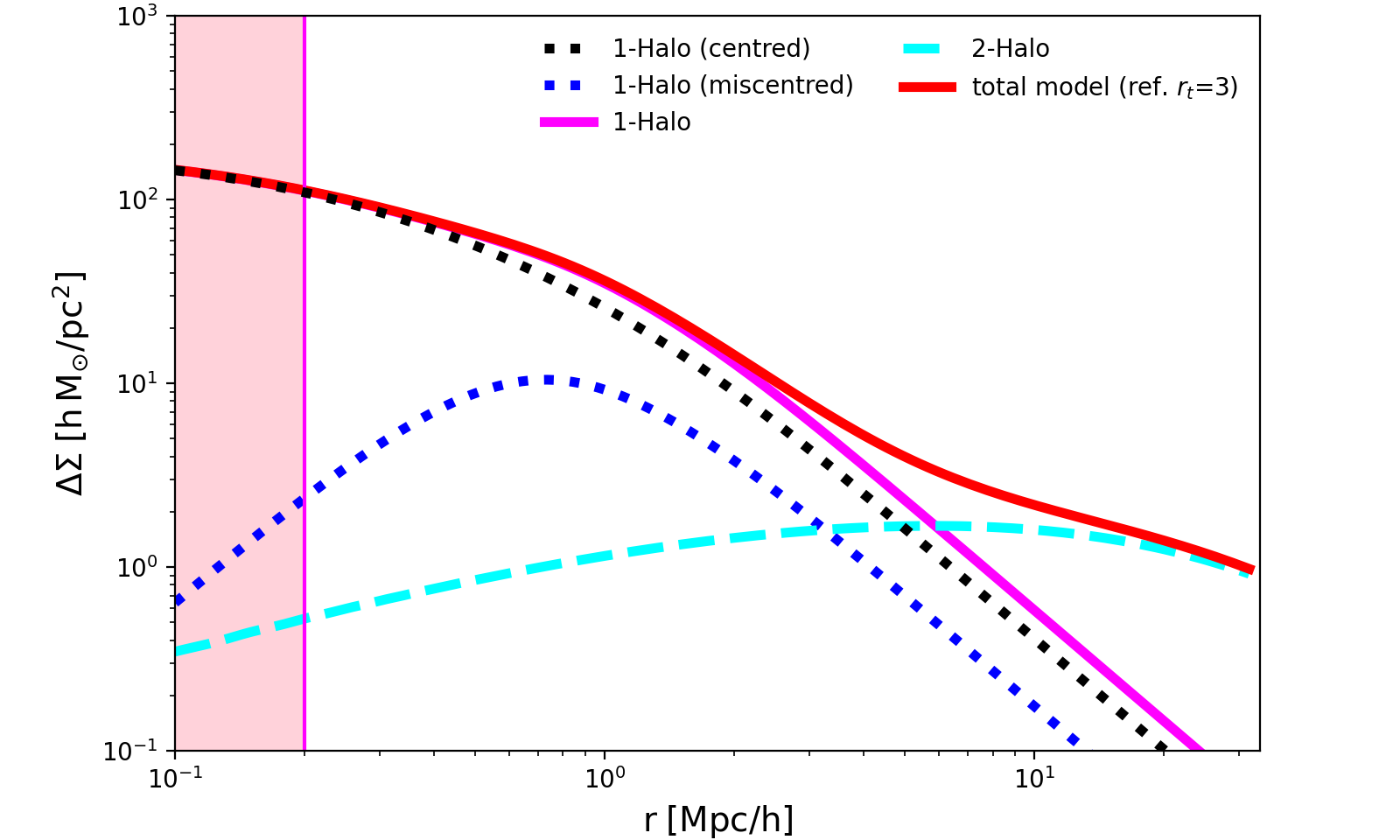}
  \includegraphics[width=\hsize]{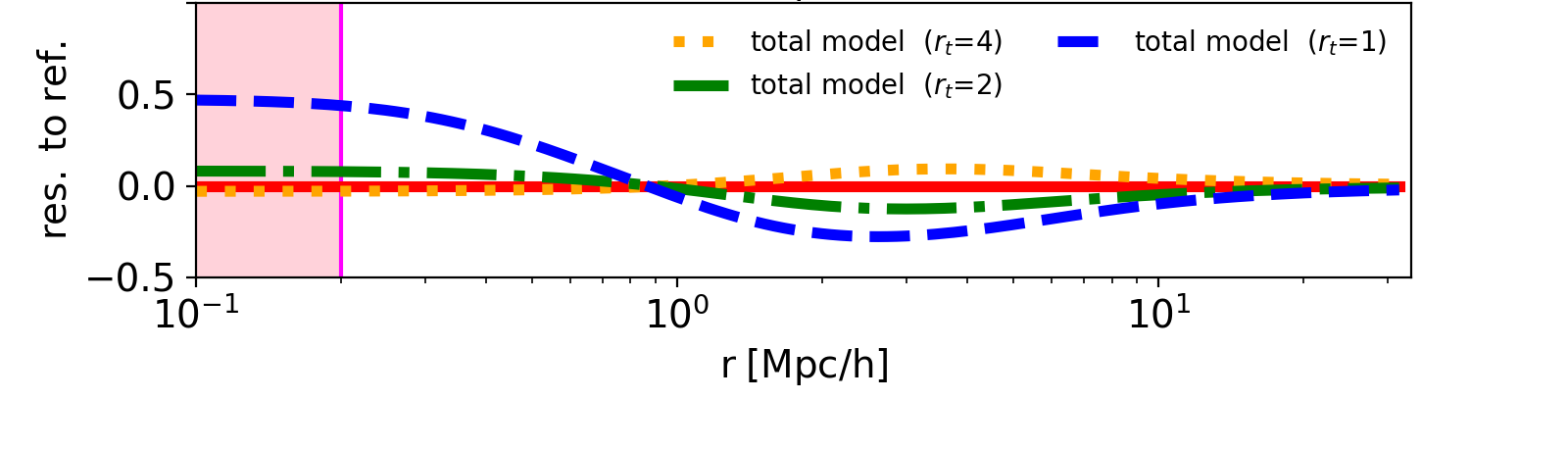}
  \includegraphics[width=\hsize]{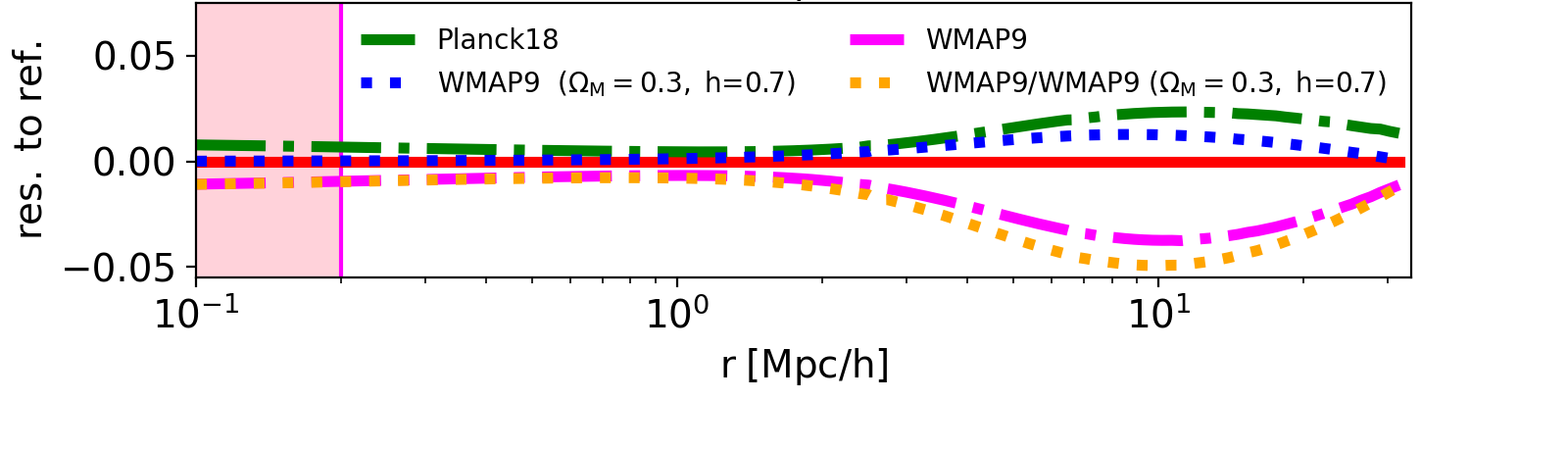}
  \includegraphics[width=\hsize]{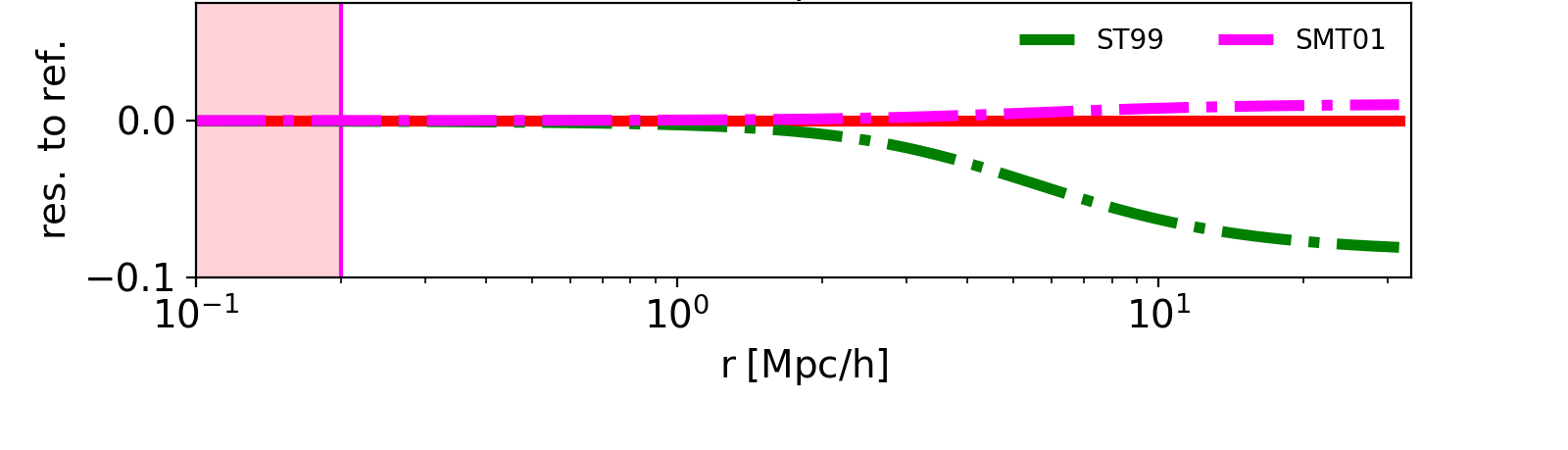}
  \caption{{\em Upper panel}: model of the projected excess of surface
    mass density, as a function of the comoving radial distance from
    the centre, for a cluster with $M_{200}=10^{14} \,M_{\odot}/h$,
    $c_{200}=3$ at redshift $z=0.3$, with $\sigma_{\rm off}=0.3$ and
    $f_{\rm off}=0.3$. The radius enclosing $200$ times the critical
    density corresponds to $R_{200}=1.47$ Mpc/$h$.  The four 1-halo
    and 2-halo terms are shown with different colours and styles, as
    reported in the labels. {\em Second panel}: the relative
    differences, with respect to the reference, of the models with
    $r_t=1,2$ and $4$. {\em Third panel}: the relative differences,
    with respect to the reference, of the models with the linear power
    spectrum estimated assuming different cosmological model
    parameters, as reported in the labels. For comparison, we display
    with yellow dotted curve the relative difference between the
    prediction adopting WMAP9 cosmology and WMAP9 but fixing
    $\Omega_M=0.3$ and $h=0.7$. {\em Bottom panel}: the relative
    differences, with respect to the reference, of two models
    constructed with either the \citet{sheth99b} or the
    \citet{sheth01b} biasing model.}
  \label{figmodelling}
\end{figure}

In our procedure we will account for this effect by considering a
second additional component in our model, which is the one produced by
haloes whose observed centres have a non-negligible deviation with
respect to the centres assumed by the stacking procedure, as provided
by the \texttt{AMICO} cluster finder. In doing so we follow the method adopted by
\citet{johnston07} and \citet{viola15}. First, we define the rms of the
misplacement of the haloes,
$\sigma_{\rm off}$, assuming an azimuthally symmetric Gaussian 
distribution. Thereby, the probability of a lens being at distance
$R_s$ from the assumed centre, or briefly the probability distribution
of the offsets, can be defined as:
 \begin{equation}
 P(R_s)=\frac{R_s}{\sigma_{\rm off}}
 \exp\Bigl[-\frac{1}{2}\Bigl(\frac{R_s}{\sigma_{\rm off}}\Bigr)^2\Bigr],
 \label{MiscentringProb}
 \end{equation}
 where $\sigma_{\rm off}$ represent the scale length, whose typical
 values found e.g. by SDSS collaboration can be of the order of
 $\sigma_{\rm off}\sim0.4$ Mpc/$h$ \citep{johnston07}. While modelling
 the halo profile, we will assume that a fraction $f_{\rm off}$ of the
 lenses is miscentred.  Now, we can introduce the azimuthally averaged
 profile of a population misplaced by a distance $R_s$ in the lens
 plane \citep{yang06}:
 \begin{equation}
 \Sigma(R|R_s)=\frac{1}{2\pi}\int_0^{2\pi}\Sigma_{\rm cen}(R)
 \Bigl(\sqrt{R^2+R_s^2+2RR_s\cos\theta}\Bigr)\,\text{d}\theta,
 \label{miscetr.averageprofile}
 \end{equation}
where $\Sigma_{\rm cen}(R)$ refers to the centred surface brightness
distribution (or briefly the centred profile) derived by integrating
$\rho(r)$ along the line-of-sight. Finally, integrating
Eq. \eqref{miscetr.averageprofile} along $R_s$ and weighing each
offset distance according to Eq. \eqref{MiscentringProb}, we can
obtain the mean surface density distribution of a miscentred halo
population:
\begin{equation}
\Sigma_{\rm off}(R)=\int P(R_s)\Sigma(R|R_s)\text{d}R_s. \label{mean sigma-off}
\end{equation}
Our final model for the 1-halo term can be written as the sum of a
centred an off-centred population:
 \begin{equation}
   \Sigma_{1h}(R)=(1-f_{\rm off})\Sigma_{\rm cen}(R)+f_{\rm off}\Sigma_{\rm off}(R).
   \label{sigma1halocen.+off}
 \end{equation} 
Generally, this 1-halo model depends on nine parameters but, when
analysing the 1-halo radial range only data-set -- where the large
scale contribution is almost negligible, we fix the cosmology to a
flat $\Lambda$-CDM with $h=0.7$, $\Omega_M=0.3$ and
$\Omega_{\Lambda}=0.7$, the truncation factor to the reference value
$r_t=3$ and the effective redshift estimated by the weight of the
stacked samples in each bin.  Therefore our analysis on the 1-halo
radial range uses a model that depends on four free parameters:
$M_{200},c_{200},\sigma_{\rm off}$ and $f_{\rm off}$. We underline
that analysing the data-set up to large scales to constrain the total
matter density parameter, we accordingly re-scale both the data and
the models to the new cosmology.
  
In our analysis we will neglect the measurements below $0.2$ Mpc$/h$
mainly for three reasons.  Firstly, from an observational point of
view, the uncertainties in the measure of photo-$z$s and shear close
to the cluster centre are large because of the contamination due to
the higher concentration of cluster galaxies. Secondly, the shear
signal analysis in close proximity to the cluster centre is sensitive
to the BCG contribution to the matter distribution, and to deviations
from the weak lensing approximation used in the profile model. Lastly,
this choice mitigates the miscentring effects. Thus, neglecting
small-scale measurements minimises the systematics possibly affecting
the estimation of the concentration $c_{200}$, which would otherwise
be overestimated - being degenerate with the $\sigma_{\rm off}$ and
$f_{\rm off}$ parameters.
 
\subsection{The 2-halo lens model}
\label{sub:2halo}
Here we introduce the 2-halo model used to describe the surface
density profiles beyond the cluster radius. The data for $r> R_{200}$
can be used to constrain the cosmological parameters. In particular,
we will focus on the total matter density, $\Omega_M$.

At scales larger than the radius $R_{200}$, the shear signal is caused
by the correlated matter distribution around the galaxy clusters and
it manifests as an increase of the mass density profiles.  In
practice, the 2-halo term characterises the cumulative effects of the
large-scale structures in which galaxy clusters are located. The
uncorrelated matter distribution along the line-of-sight does produce
only a modest contribution to the stacked shear signal.  We will model
the 2-halo term following the recipe by \citet{oguri11}.
  
It is worth noticing that the 2-halo term contribution at small radii
is expected to be negligible, becoming statistically significant only
at scales $\gtrsim 5\,\text{Mpc}/h$. The total excess surface
mass density profile will be the sum of the halo profile described by
the Eq. \eqref{sigma1halocen.+off} with the contribution due to the
matter in correlated haloes that we can write as
\citep{oguri11,oguri11b,sereno17}:
  \begin{equation}
  \Delta\Sigma_{2h}(\theta;M_{200},z) \!= \!\int
  \frac{l\text{d}l}{2\pi}J_2(l\theta)
  \frac{\bar{\rho}_{m}(z)b_h(M_{200};z)}{(1+z)^3D_l^2(z)}P_{lin,m}(k_l;z),
  \label{surf.dens.2haloterm}
  \end{equation}
where $z$ represents the cluster redshift, estimated using photometric
data, as provided by \texttt{AMICO}. The other terms in
Eq.~(\ref{surf.dens.2haloterm}) are summarised as follows:
\begin{itemize}
  \item $\theta$ is the angular scale. It is computed as the ratio
    between the projected radius, at which the shear signal is
    measured, and the corresponding lens angular diameter distance
    $D_l(z)$ $[\text{Mpc}/h]$, evaluated at the cluster redshift and
    assuming a flat $\Lambda$CDM cosmology.
  \item $J_2$ is the Bessel function of second type, which is the
    solution of the Bessel differential equation, with a singularity
    at the origin. This is a function of $l\theta$, where $l$ is the
    integration variable and the momentum of the wave vector $k_l$ for
    the linear power spectrum of matter fluctuations $P_{lin,m}$.
  \item $k_l=l/((1+z)D_l(z))$ indicates the wave vector module.
  \item $\bar{\rho}_m(z)$ represents the mean cosmic background
    density at the lens redshift, in unit of
    $M_{\odot}\text{Mpc}^{-3}h^2$.
  \item $P_{lin,m}(k_l;z)$ is the linear power spectrum of matter
    computed according to the \citet{eisenstein99} transfer
    function. For our reference case, when computing the linear matter
    power spectrum we adopt the Planck18 \citep{planck18} cosmological
    parameters though fixing $\Omega_M=0,3$, $\Omega_\Lambda=0.7$ and
    $h=0.7$. The adopted \citet{eisenstein99} transfer function model
    is accurate enough given our current measurement uncertainties. We
    tested the method using the CAMB model \citep{camb}, finding
    negligible differences.
  \item $b_h(M;z)$ indicates the halo bias, for which we adopt the
    \citet{tinker10} model, that has been successfully tested and
    calibrated using a large data-set of numerical simulations. In our
    work we will also test the robustness, and eventual biases, of our
    results using the bias models by \citet{sheth99b} and
    \citet{sheth01b}. The halo bias is expected to be constant at
    large scales and to evolve weakly with redshift.  It is expressed
    as (the square root of) the ratio between the halo and the linear
    dark matter power spectra. At fixed redshift the bias amplitude
    increases with the halo mass. However, it is typically
    parametrised in terms of the peak-high of the matter density
    field: highest peaks - which represent the seeds from which galaxy
    clusters form - tend to be more biased than the smallest ones
    \citep{mo96}.
  \end{itemize} 

\subsection{The full lens model}
The full model of the projected excess of surface mass density adopted
in this work can be read as:
\begin{equation}
  \Delta\Sigma_{model}(R)=\Delta\Sigma_{1h}(R)+\Delta\Sigma_{2h}(R),
  \label{1h+2hmodel}
\end{equation}
where $\Delta\Sigma_{1h}$ and $\Delta\Sigma_{2h}$ are computed with
Eqs. \eqref{sigma1halocen.+off} and \eqref{surf.dens.2haloterm},
respectively.

Figure \ref{figmodelling} (upper panel) shows the different components
of the adopted lens model given by Eq. \eqref{1h+2hmodel} for the
reference case of a halo with mass $M_{200} = 10^{14}M_{\odot}/h$,
$c_{200}=3$ at redshift $z_l=0.3$, with $\sigma_{\rm off}=0.3$ and
$f_{\rm off}=0.3$. The effect of the truncation radius is shown in the
second panel, where we report the relative differences, with respect
to the reference model, of the cases with $r_t=1,2$ and $4$. In
projection, we can notice that the impact of the truncation radius not
only modifies matter density distribution at scales larger than $1$
Mpc/$h$ but also reshapes the matter density distribution on small
scales to preserve the matter enclosed within $R_{200}$.  The impact
of assuming a different cosmological model when computing the linear
matter power spectrum is displayed in the third panel. Particularly, the
yellow dotted curve displays the ratio between the prediction assuming
the WMAP9 cosmological model and those adopting $\Omega_M=0.3$ and
$h=0.7$, fixing the other parameters to the flat WMAP9 case. The green
and magenta dot-dashed and the blue dotted curves display the relative
difference of the Planck18, WMAP9 and WMAP9 with $\Omega_M=0.3$ and
$h=0.7$ with respect to our reference model (we assume Planck18
cosmological parameters but setting $\Omega_M=0.3$ and $h=0.7$),
respectively.  Finally, in the bottom panel we exhibit the relative
difference between the reference model, where we assume the
\citet{tinker10} halo bias, and two alternative models constructed
assuming either the \citet{sheth99b} biasing model or the
\citet{sheth01b} one.  The halo bias model mainly affects the $\Delta
\Sigma$ profile toward large scales. We also notice that, for this
cluster mass and redshift, the \citet{sheth01b} model is slightly
above our reference one, while the \citet{sheth99b} case starts to
decrease already at $1$ Mpc/$h$ reaching almost $10\%$ difference at
scales larger than 10 Mpc/$h$. In Appendix \ref{app_sis} we test 
the impact of different bias model assumption in the estimate of 
the total matter density parameter. 
   
\section{Results}
\label{results_label}

In this section we present the main results of our analysis. Firstly,
we will model the small-scale stacked surface mass density profiles,
assessing the structural properties of the hosting halo
population. Then we will model the full lensing signal up to the
largest available scales, constraining the total matter density
parameter, $\Omega_M$.

In the first part of our analysis, when modelling the small-scale
signal coming from the central part of the clusters only, we will
limit to scales between $0.2$ and $3.16$ Mpc/$h$.  We refer to this
analysis as the $\sim 1h$-case, since most of the contribution to the
measured signal comes from the main dark matter haloes hosting the
galaxy clusters.  In fact, as evident from the upper panel of Figure
\ref{figmodelling}, the 2-halo term affects, to a small degree, only
the few last radial bins available. This makes our inference on the
cluster halo structural properties, expressed in terms of the four
1-halo (NFW) model parameters ($M_{200}$, $c_{200}$, $f_{\rm off}$ and
$\sigma_{\rm off}$), much less dependent on the uncertain details of
the halo biasing model. To test this assumption, we repeated the
following analysis including only the 1-halo contribution, finding
that the mass is slightly overestimated by few percents (see
Tab.~\ref{tab_bins} for more details). For each corresponding case,
in parenthesis we report the $\chi^2$ divided by the number of degrees of freedom. 
We recall that the highest
$S/N$ comes from small-scale lensing in this kind of measures, but it
may be affected by both theoretical and observational systematic
uncertainties \citep{mandelbaum13}.
  
Subsequently, we will model the total stacked density profiles up to
35 Mpc/$h$. In this case, we will consider Gaussian priors on the four
parameters of the 1-halo term, with the mean and the standard
deviation corresponding to the posterior distribution results.  The
main goal here is to derive constraints on $\Omega_M$ marginalised
over the 1-halo term parameters.  We will refer to this cosmological
analysis as the $1h+2h$-case.

\subsection{Constraining the cluster structural properties}   
\label{sec_haloprop}
In this Section we model the stacked shear profiles collected in the
$0.2<R<3.16$ $\text{Mpc}/h$ radial range, while in the next one we
will exploit the profiles measured up to 35 Mpc/$h$ -- where the
2-halo term contribute dominates.  In both cases we bin the data-set
adopting a logarithmic binning along the radial directions, choosing a
$d\log(r) \approx 0.1$ we have $14$ and $30$ radial bins for the two
cases, $\sim 1h$ and $1h+2h$, respectively.
In both cases the analyses are based on a set of $14$ stacked excess
surface density profiles of clusters, obtained by dividing the whole
AMICO-KiDS-DR3 cluster catalogue in redshift and amplitude, as
reported in the first four columns of Tab.~\ref{tab_bins}.

\begin{sidewaystable}
\centering	
\caption{\texttt{AMICO} cluster sample binned in z (redshift) and A
  (amplitude: as defined in Eq. \ref{AMICOAmpl}) as used in our
  analyses. The table reproduces the structure adopted by
  \citet{bellagamba19}; we report in the $\log M_{200}$ columns the
  values we obtain modelling the two cases: $\sim 1h$ (up to $3.16$
  Mpc/$h$) and $1h+2h$ (up to $35$ Mpc/$h$). The numbers, expressed in
  $\log \left( M_{200} / h^{-1} M_{\odot} \right)$, reported refer to
  the median and the $18th-82nd$ percentile widths. The values
  reported in parenthesis represent the $\chi^2$ divided by the number
  of $d.o.f.$ corresponding to each considered case. The column
  $N_{cl}$ expresses the number of clusters stacked in each bin.}
\label{tab_bins}
\begin{tabular}{lcccccccc}
\hline
$-$ & $z$ range & $z_{\text{\rm eff}}$ & $A$ range & $A_{\text{\rm eff}}$ & $\log M_{200}$: $\sim 1h$, with $1h$ mod. & $\log M_{200}$: $\sim 1h$ with $1h+2h$ mod. & $\log M_{200}$: $1h+2h$ & $N_\text{cl}$  \\
\hline
1 & $[0.10,0.30[$ & 0.190 $\pm$ 0.002 & $[0,1[$ & 0.830 $\pm$ 0.003 & $13.208_{-0.083}^{+0.078}$ (1.877) & $13.170_{-0.078}^{+0.079}$ (1.827)& $13.198_{-0.061}^{+0.059}$ (0.955)& 1066 \\
2 & $[0.10,0.30[$ & 0.207 $\pm$ 0.002 & $[1,1.55[$ & 1.213 $\pm$ 0.006 & $13.550_{-0.068}^{+0.074}$ (2.905) & $13.514_{-0.064}^{+0.071}$ (1.670)& $13.504_{-0.051}^{+0.052}$ (0.983)& 822\\
3 & $[0.10,0.30[$ & 0.212 $\pm$ 0.004 & $[1.55,2.05[$ & 1.762 $\pm$ 0.010 & $14.059_{-0.054}^{+0.058}$ (1.191) & $14.033_{-0.057}^{+0.054}$ (0.933)& $14.009_{-0.051}^{+0.048}$ (1.134)& 240\\
4 & $[0.10,0.30[$ & 0.226 $\pm$ 0.005 & $[2.05,2.75[$ & 2.350 $\pm$ 0.021 & $14.239_{-0.056}^{+0.061}$ ( 0.648) & $14.215_{-0.057}^{+0.060}$ (0.450)& $14.209_{-0.059}^{+0.053}$ (0.851)& 96\\
5 & $[0.10,0.30[$ & 0.211 $\pm$ 0.008 & $[2.75,6.5[$ & 3.259 $\pm$ 0.084 & $14.532_{-0.059}^{+0.056}$ (2.441) & $14.510_{-0.062}^{+0.056}$ (1.747)& $14.567_{-0.059}^{+0.050}$ (1.122)& 41\\
\hline
6 & $[0.30,0.45[$ & 0.378 $\pm$ 0.001 & $[0,1.15[$ & 0.954 $\pm$ 0.004  & $13.607_{-0.063}^{+0.074}$ (1.846) & $13.576_{-0.059}^{+0.062}$ (1.500)& $13.593_{-0.049}^{+0.051}$ (1.560)&  1090\\
7 & $[0.30,0.45[$ & 0.382 $\pm$ 0.002 & $[1.15,1.65[$ & 1.354 $\pm$ 0.006  & $13.870_{-0.069}^{+0.073}$ (1.322) & $13.836_{-0.068}^{+0.067}$ (0.566)& $13.860_{-0.057}^{+0.056}$ (0.596)& 762\\
8 & $[0.30,0.45[$ & 0.385 $\pm$ 0.002 & $[1.65,2.3[$ & 1.909 $\pm$ 0.012 & $14.208_{-0.055}^{+0.055}$ (1.497) & $14.175_{-0.057}^{+0.054}$ (0.953)& $14.224_{-0.058}^{+0.039}$ (0.678)& 339\\
9 & $[0.30,0.45[$ & 0.392 $\pm$ 0.004 & $[2.3,3[$ & 2.585 $\pm$ 0.022  & $14.280_{-0.078}^{+0.077}$ (1.518) & $14.256_{-0.076}^{+0.074}$ (1.023)& $14.229_{-0.067}^{+0.063}$ (1.551)& 98\\
10 & $[0.30,0.45[$ & 0.377 $\pm$ 0.007 & $[3,6.5[$ & 3.577 $\pm$ 0.071  & $14.588_{-0.073}^{+0.065}$ (2.697) & $14.560_{-0.077}^{+0.064}$ (1.818)& $14.512_{-0.055}^{+0.062}$ (0.911)& 43\\
\hline
11 & $[0.45,0.60]$ & 0.496 $\pm$ 0.001 & $[0,1.3[$ & 1.108 $\pm$ 0.004  & $13.707_{-0.082}^{+0.077}$ (1.271) & $13.669_{-0.085}^{+0.074}$ (1.008)& $13.682_{-0.068}^{+0.068}$ (1.387)& 984\\
12 & $[0.45,0.60]$ & 0.516 $\pm$ 0.002 & $[1.3,1.8[$ & 1.516 $\pm$ 0.005  & $13.877_{-0.071}^{+0.064}$ (1.586) & $13.838_{-0.064}^{+0.062}$ (1.373)& $13.848_{-0.056}^{+0.055}$ (1.181)& 889\\
13 & $[0.45,0.60]$ & 0.515 $\pm$ 0.003 & $[1.8,2.5[$ & 2.071 $\pm$ 0.011  & $14.180_{-0.075}^{+0.074}$ (3.897) & $14.145_{-0.074}^{+0.072}$ (3.496)& $14.112_{-0.069}^{+0.069}$ (1.951)& 373\\
14 & $[0.45,0.60]$ & 0.510 $\pm$ 0.004 & $[2.5,6.5[$ & 2.877 $\pm$ 0.039  & $14.377_{-0.080}^{+0.070}$ (1.914) & $14.350_{-0.079}^{+0.070}$ (1.556) & $14.314_{-0.068}^{+0.069}$ (1.400)& 119\\
\hline
\end{tabular}
\end{sidewaystable}

\begin{figure}
  \includegraphics[width=\hsize]{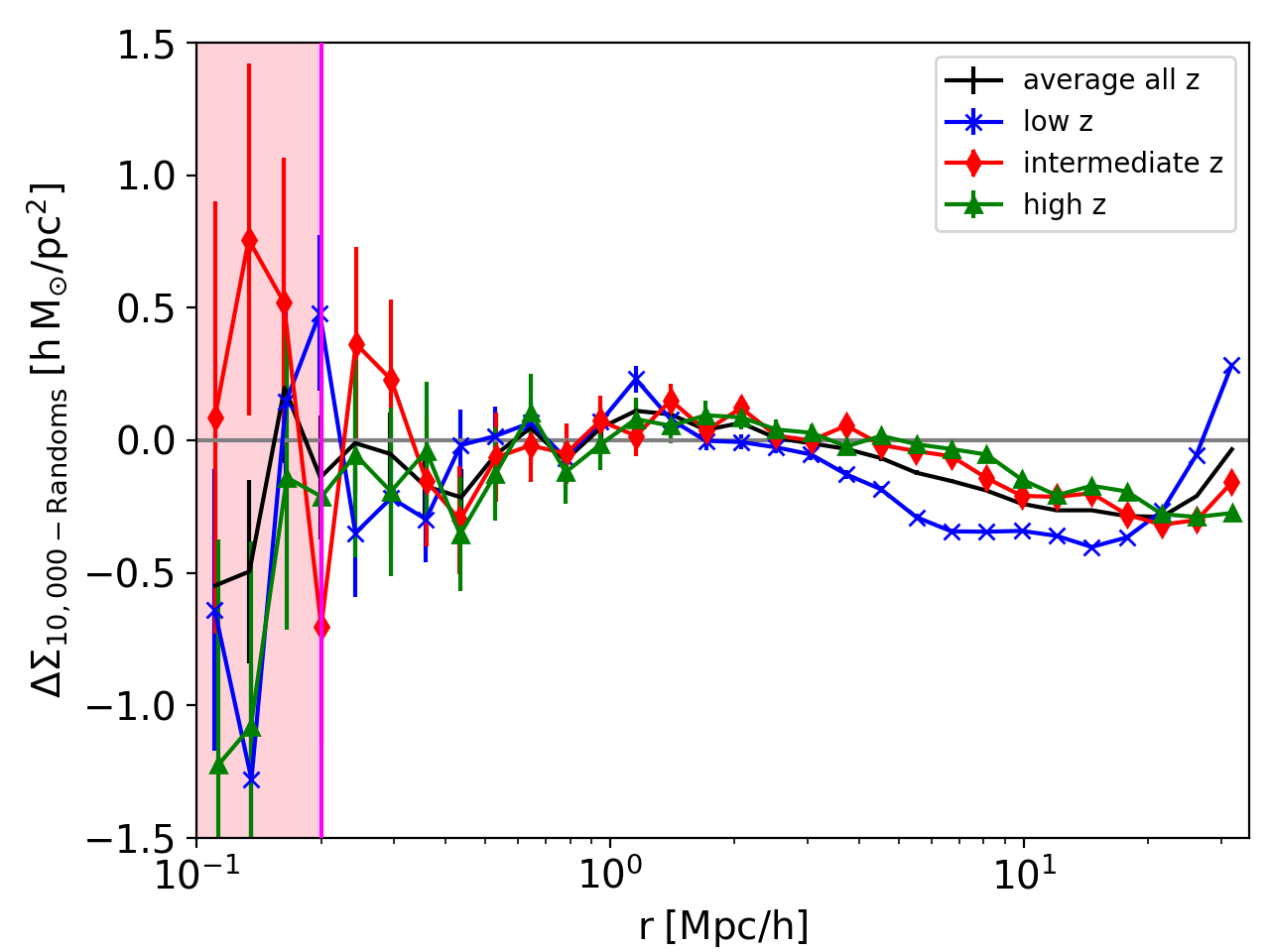}
  \caption{Excess of surface mass density profile around random
    centres for the three redshift bins in which the \texttt{AMICO}
    clusters have been divided: low ($0.1\leq z<0.3$), intermediate
    ($0.3\leq z<0.45$) and high ($0.45\leq z \leq 0.6$) redshift. We
    have created 10,000 realisations of the sample using random
    centres. The blue crosses, red diamonds and green triangles show
    the average measurements in the three redshift intervals -
    extending the profiles up to 35 Mpc/$h$ as in the
    $1h+2h$-case. The black line display the average over all
    redshifts. The error bars show square-root of the diagonal terms
    of the corresponding covariances $C_{\rm Randoms}(r,r')$. The
    vertical pink area on the left part of the figure marks the data
    points below $0.2\text{ Mpc}/h$ that we will not consider in our
    analysis.}
  \label{fig_random}
\end{figure}

\begin{figure*}
  \includegraphics[width=0.33\hsize]{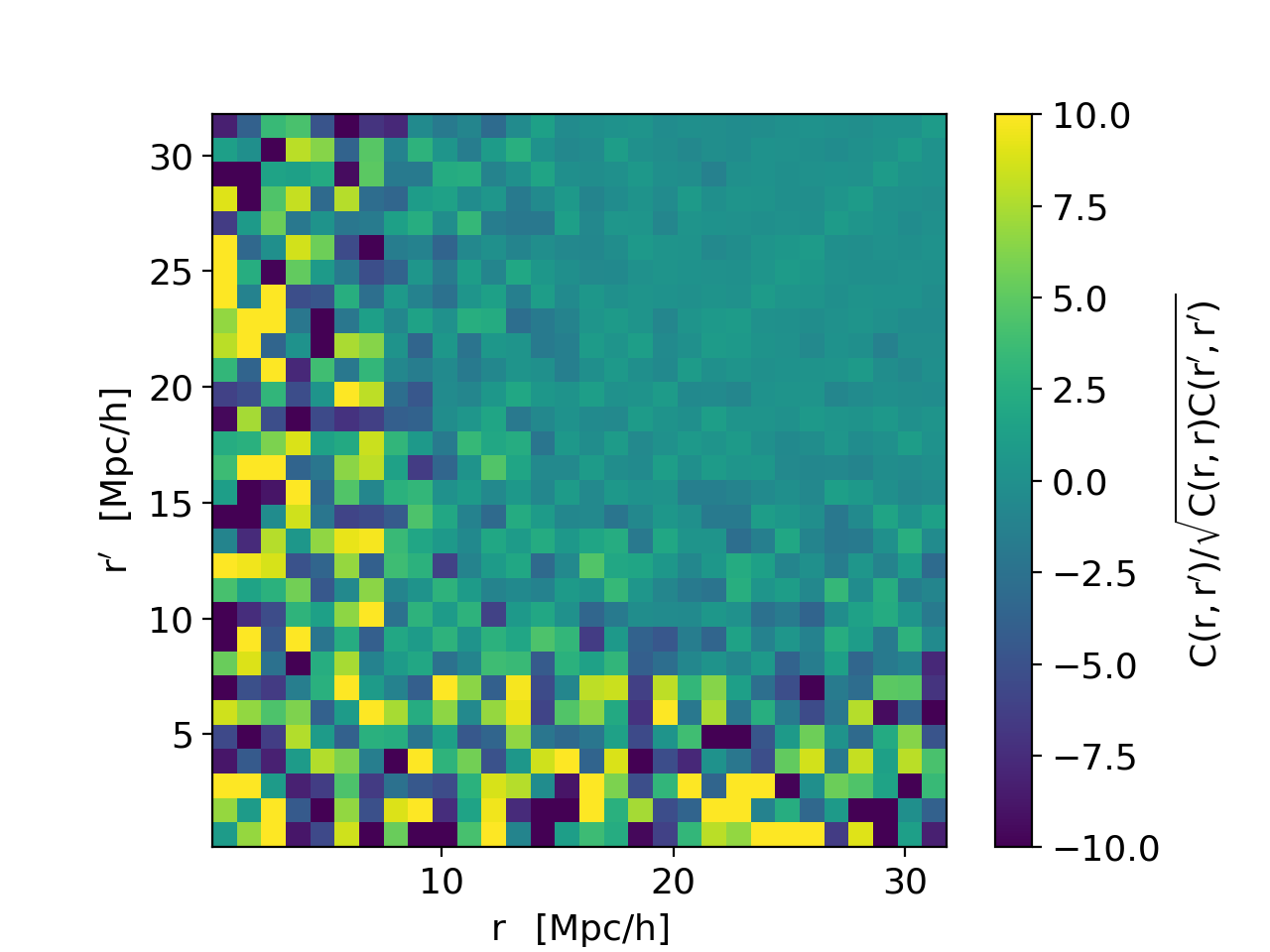}
  \includegraphics[width=0.33\hsize]{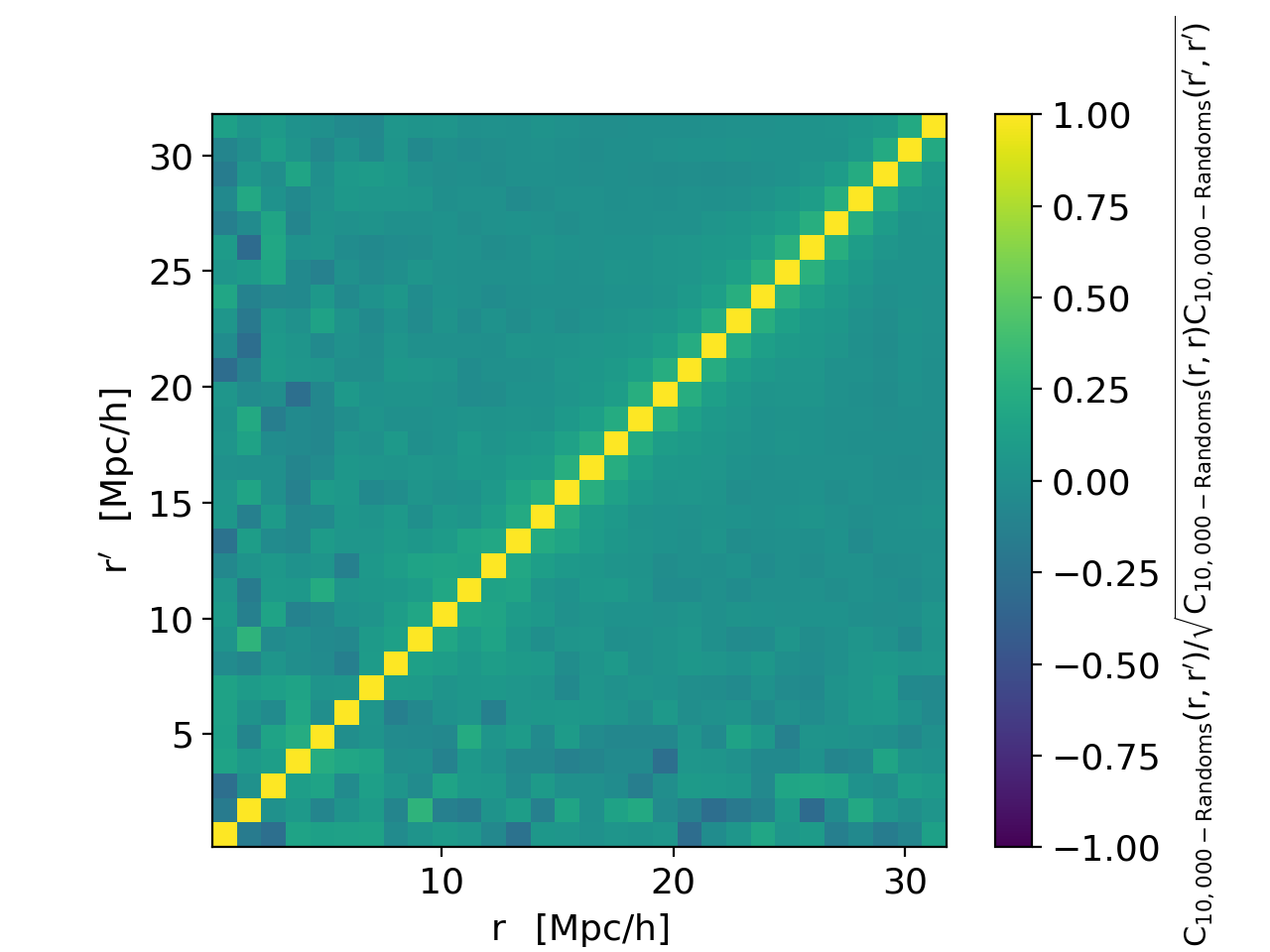}
  \includegraphics[width=0.33\hsize]{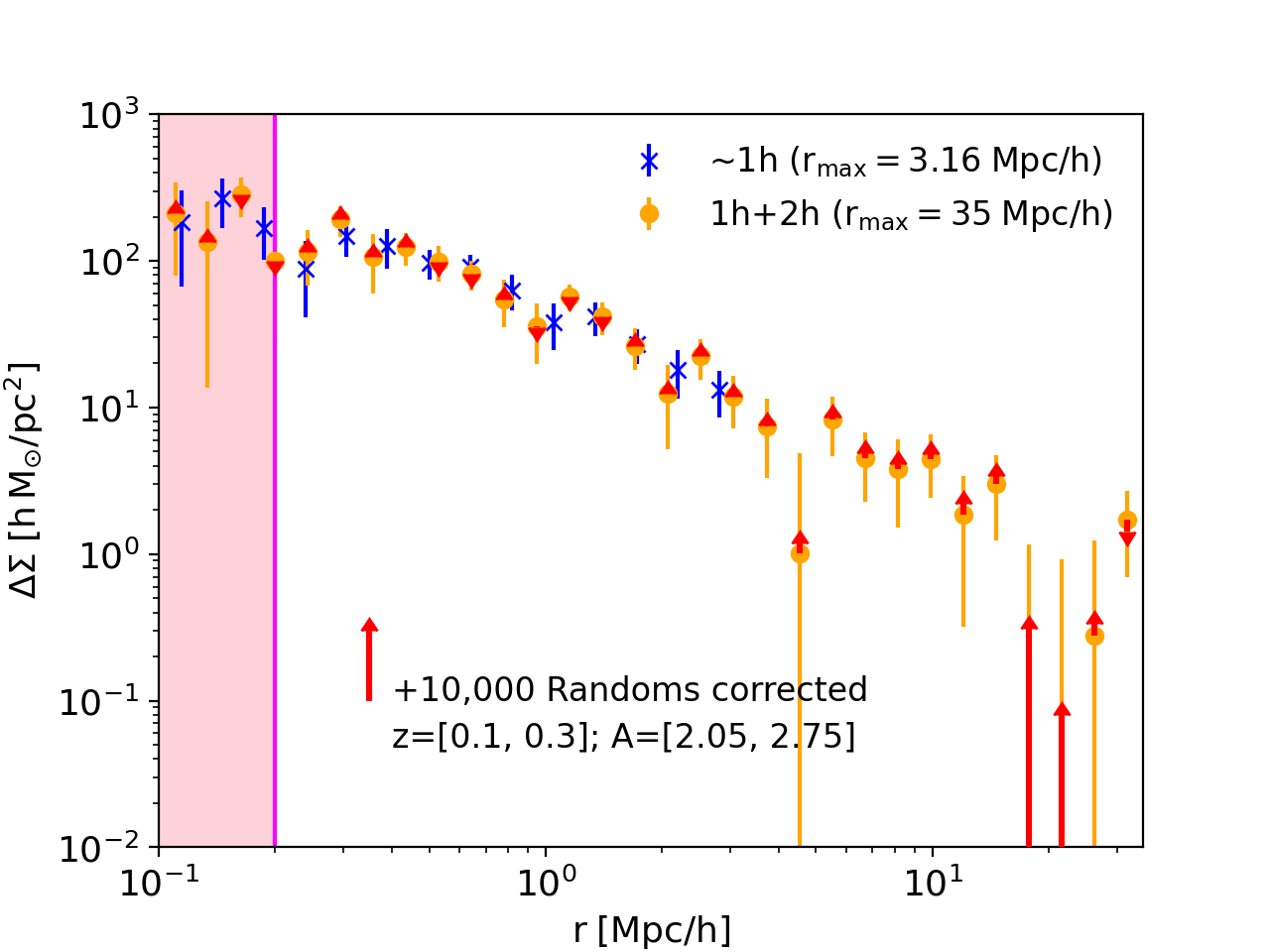}
  \caption{{\em Left panel}: the normalised covariance of the data-set
    corresponding to the bin with $0.1\leq z<0.3$ and $2.05\leq
    A<2.75$, stacking 96 detected clusters, computed with 10,000
    bootstrap realisations of the stacked profile. {\em Central
      panel}: the normalised covariance of the 10,000 realisations of
    the whole \texttt{AMICO} sample, randomly distributed in the
    survey footprints. {\em Right panel}: the stacked excess surface
    mass density profile of the $1h$-case (blue crosses) and
    $1h+2h$-case (orange circles). In both cases we subtract to the
    data points the stacked profiles around random centre. The error
    bars refer to the square-root of the diagonal elements of the
    final covariance (Eq.~\ref{eq_cov}).  The red arrows indicate the
    variation of the stacked profile when, for the corresponding
    redshift bin, the random contributions are subtracted from the
    data. We recall the reader that the blue crosses and orange filled
    circle data points refer to two different binning schemes giving
    us the possibility to constrain the cluster-halo structural
    properties and the large scale matter density distribution in
    which the clusters are located, respectively.}
  \label{fig_cov} 
\end{figure*}

In order to properly model the large-scale weak lensing signal we need
to take into account most of the systematics and uncertainties that
could emerge in the very weak regime where the excess surface mass
density becomes low.  We perform this by constructing stacked profiles
around random positions in different redshift bins, that we subtract
to the stacked cluster profiles \citep{sereno18,bellagamba19}.  We
generate 10,000 realisations of a sample of 6962 random positions
properly accounting for the KiDS-DR3 survey masks
(Fig.~\ref{figAMICOpositions}) and the cluster redshift distribution.
The number of random realisations has been selected finding a good
compromise between the computational cost needed for the analyses and
number of \texttt{AMICO} clusters used in this work.
In Fig.~\ref{fig_random} we display the excess surface mass density
profiles around the random centres for the three different redshift
range considered, as well as the average one (for simplicity we
exhibit only the data-set for the $1h+2h$-case).  The error bars are
the square-root of the diagonal terms of the corresponding covariances
$C_{\rm Randoms}(r,r')$ computed constructing the excess surface mass
density around random centres:
\begin{equation}
C_{\rm Randoms}(r,r') = \left\langle \left(\Delta \Sigma(r) - \bar{\Delta}
  \Sigma(r)\right) \left(\Delta \Sigma(r') - \bar{\Delta}
  \Sigma(r')\right) \right\rangle
\end{equation}
where $\bar{\Delta}\Sigma(r)$ indicates the average profile over the
considered random data-set.  All the average profiles are consistent
with zero, showing only a mild deviation at large distances, due to
the survey masks.

For the cluster stacked surface mass density profile, we compute the
covariance matrix $C(r,r')$ using 10,000 bootstrap realisations, for
each of the amplitude and redshift bins.  As an illustrative case, in
the left panel of Fig.~\ref{fig_cov} we display the normalised
covariance matrix for one of the bin, specifically the bin with
$0.1\le z<0.3$ and $2.05\le A<2.75$, that contains 96 clusters.  The
central panel of Fig.~\ref{fig_cov} shows the normalised $\Delta
\Sigma$ covariance computed generating 10,000 times the AMICO-KiDS-DR3
cluster sample but assigning random positions in the KiDS-DR3
field-of-view.  We notice that the off-diagonal term contributions in
this case are almost negligible, which allows us to compute the final
covariance $CC(r,r')$ for the stacked excess surface mass density
profiles, in each redshift and amplitude bins, by summing in
quadrature the diagonal terms only:
\begin{equation}
CC(r,r') = C(r,r') \sqrt{1 + \dfrac{\delta(r,r')}{n_{\rm Randoms}^2}
  \dfrac{C^2(r,r')_{\rm Randoms}}{C^2(r,r')}}
\label{eq_cov}
\end{equation}
with $n_{\rm Randoms} = 10,000$ and $\delta(r,r')$ the two-dimensional
delta Dirac function that is equal to one for $r=r'$ and zero
otherwise.  The right panel of Fig.~\ref{fig_cov} shows the stacked
$\Delta\Sigma$ profile around the $96$ clusters considered sample.
The blue crosses (the orange circles) display the $\sim 1h$-case
($1h+2h$-case) where the profiles are constructed adopting 14 (30)
radial bins up to 3.16 Mpc$/h$ (30 Mpc/$h$) minus the corresponding
profiles around 10,000 random centre.  The error bars represent the
square-root of the diagonal term of the final covariance $CC(r,r')$.
As an example, the red arrows, corresponding to each orange circle
point, indicate the variation of the $\Delta \Sigma$ profile.

In our analysis, performed adopting the \texttt{CosmoBolognaLig} \citep{marulli16}, 
we numerical invert the covariance matrix using the GSL\footnote{\href{https://www.gnu.org/software/gsl}{https://www.gnu.org/software/gs}} 
library.

Figure \ref{fig_cov} shows that the random contribution at small scale
is negligible, while it becomes relatively important at large scales
when $\Delta \Sigma$ is smaller than 10 $h M_{\odot}$/pc$^2$.  This
justifies the choice by \cite{bellagamba19} not to correct for this
effect in their small-scale analysis.

\begin{figure*}
  \includegraphics[width=0.211\hsize]{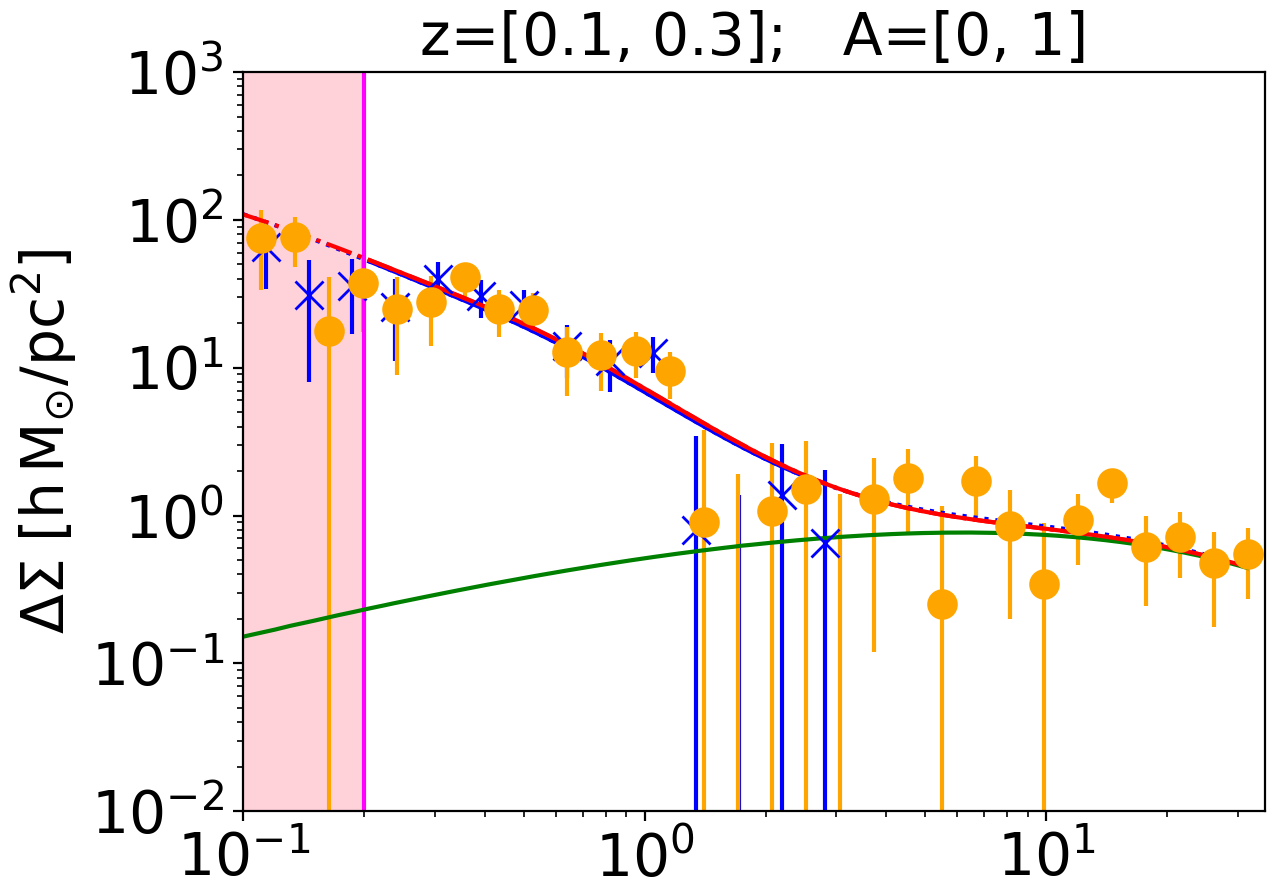}
  \includegraphics[width=0.192\hsize]{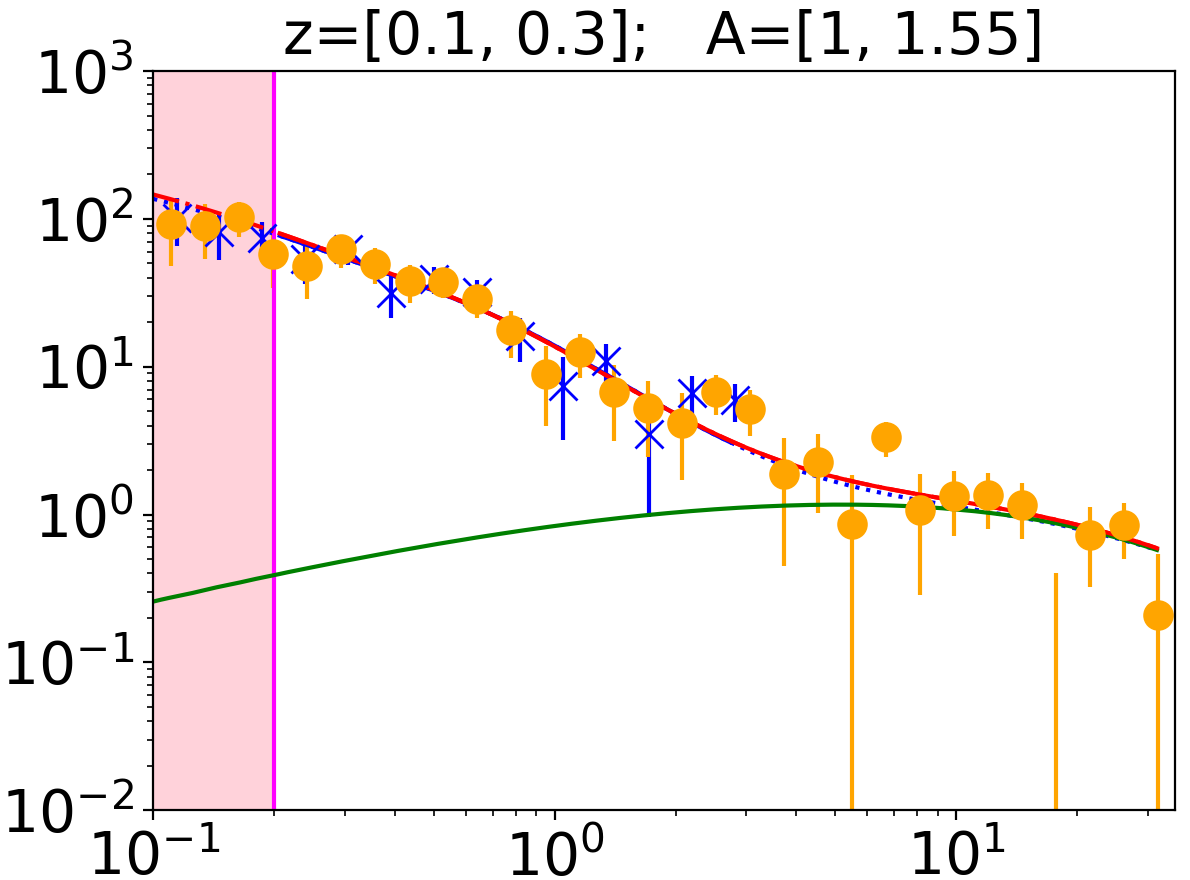}
  \includegraphics[width=0.192\hsize]{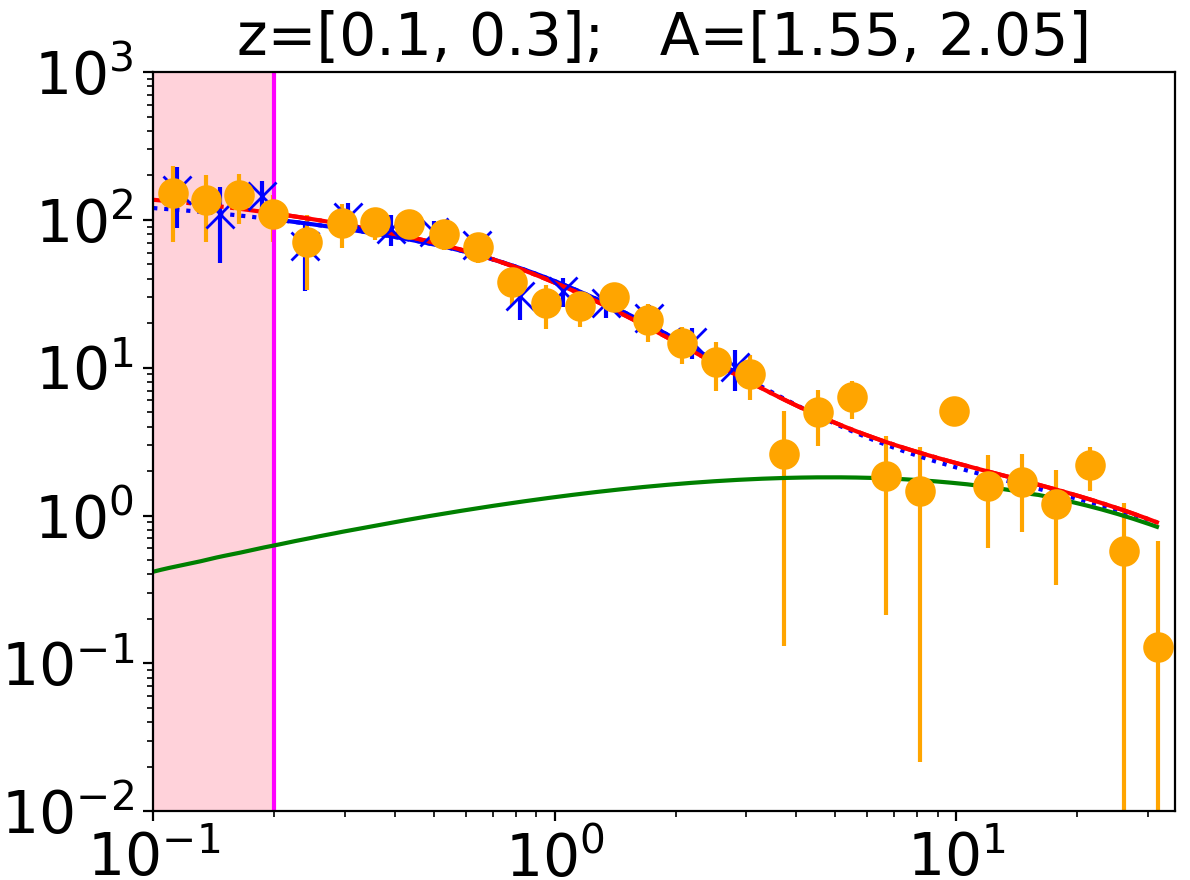}
  \includegraphics[width=0.192\hsize]{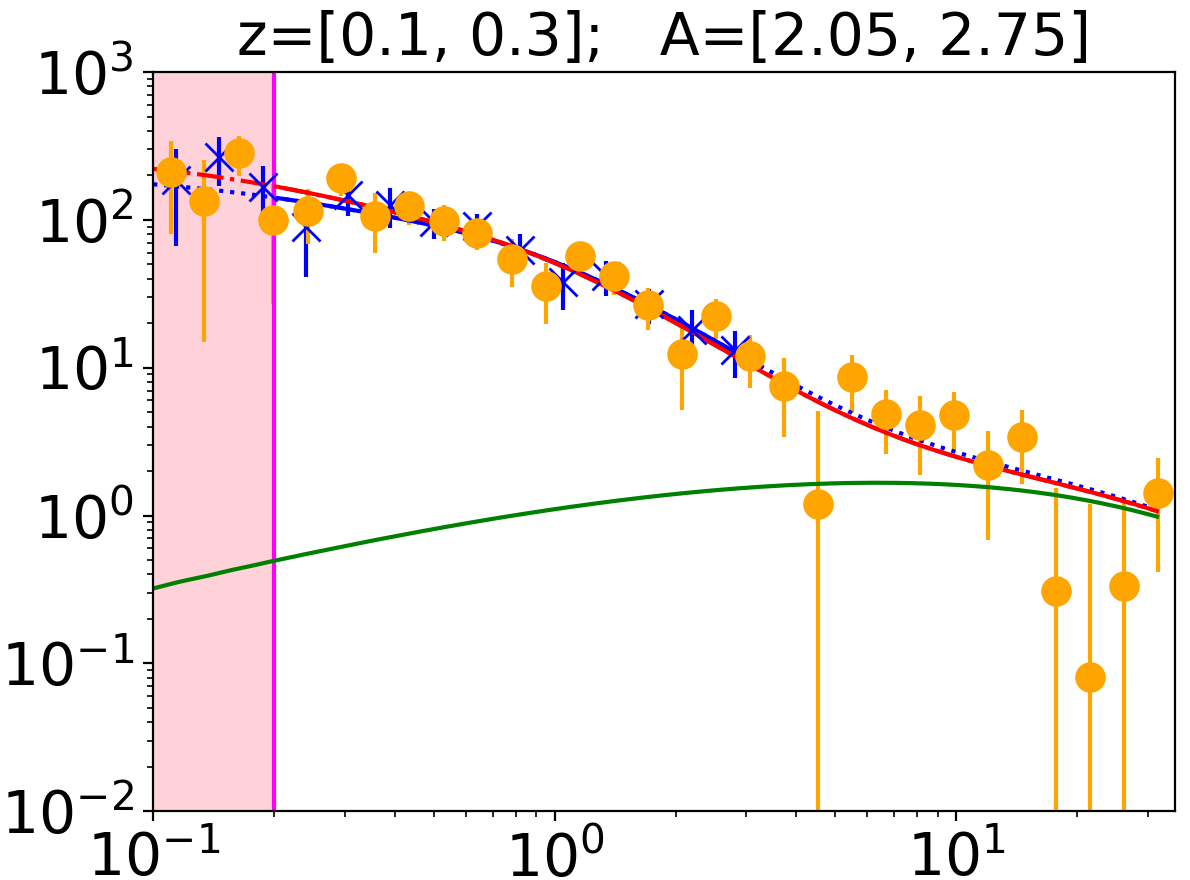}
  \includegraphics[width=0.192\hsize]{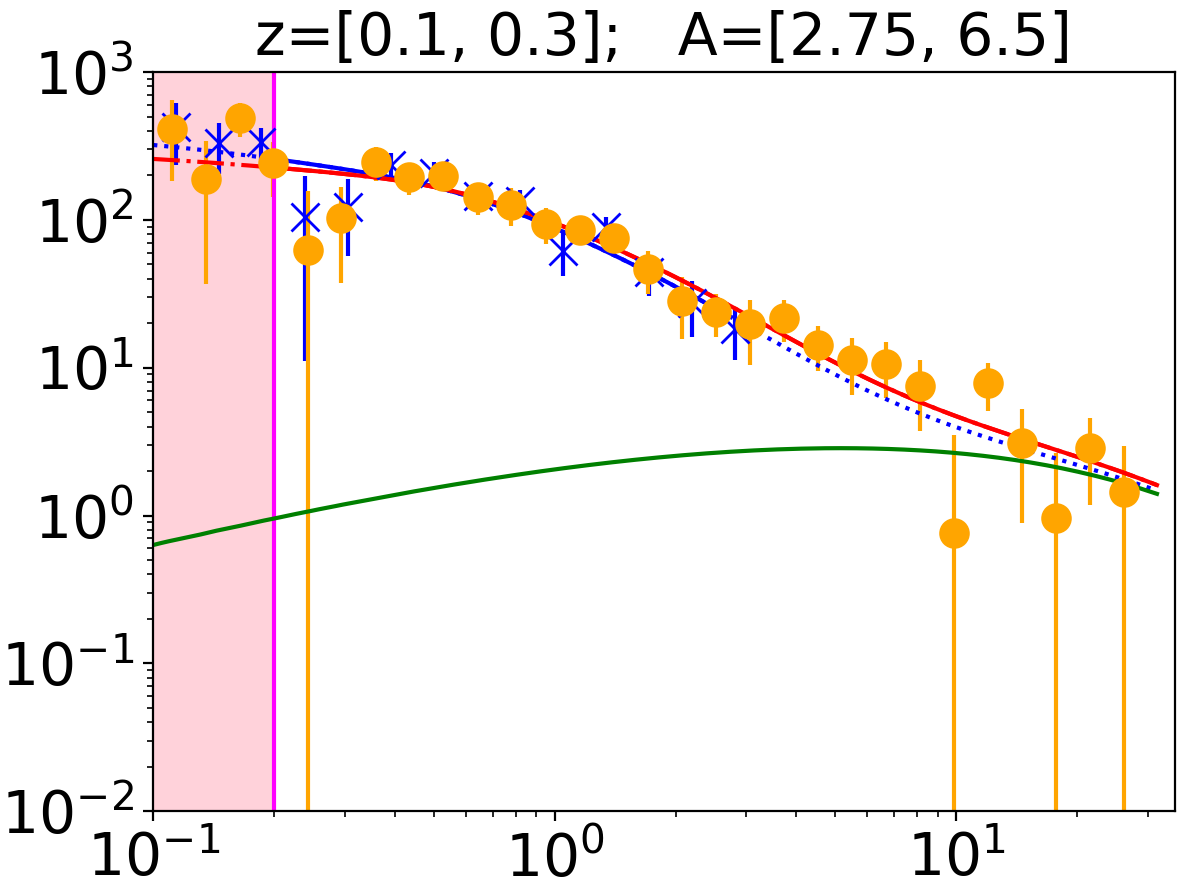}
  \includegraphics[width=0.211\hsize]{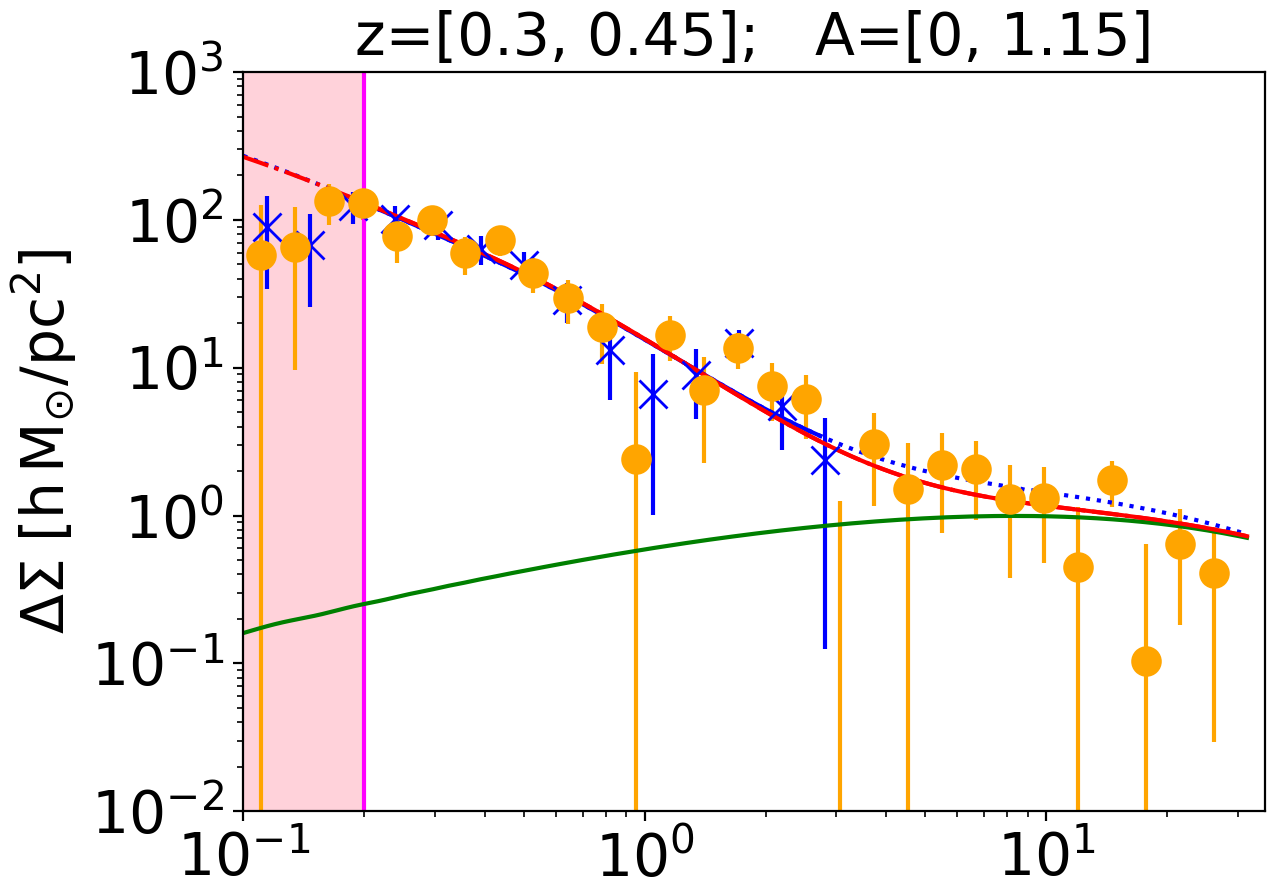}
  \includegraphics[width=0.192\hsize]{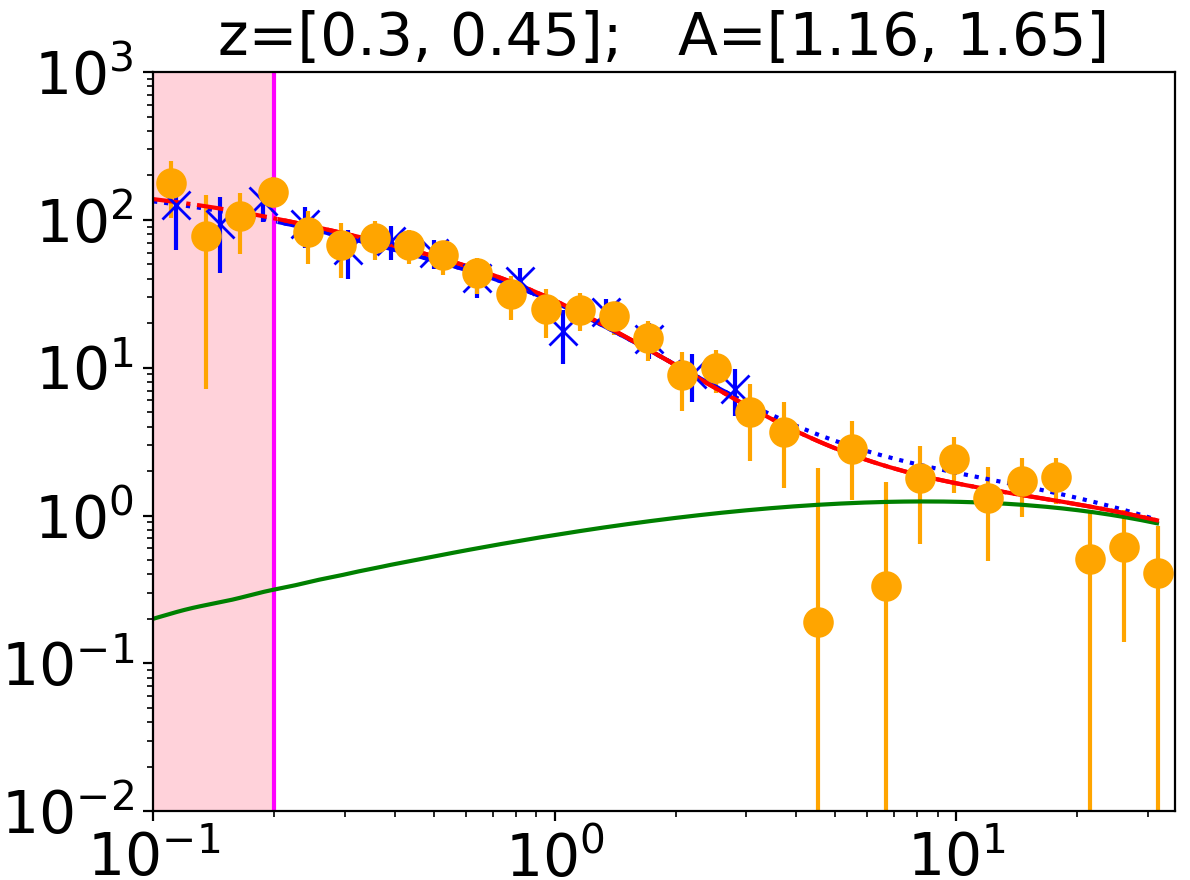}
  \includegraphics[width=0.192\hsize]{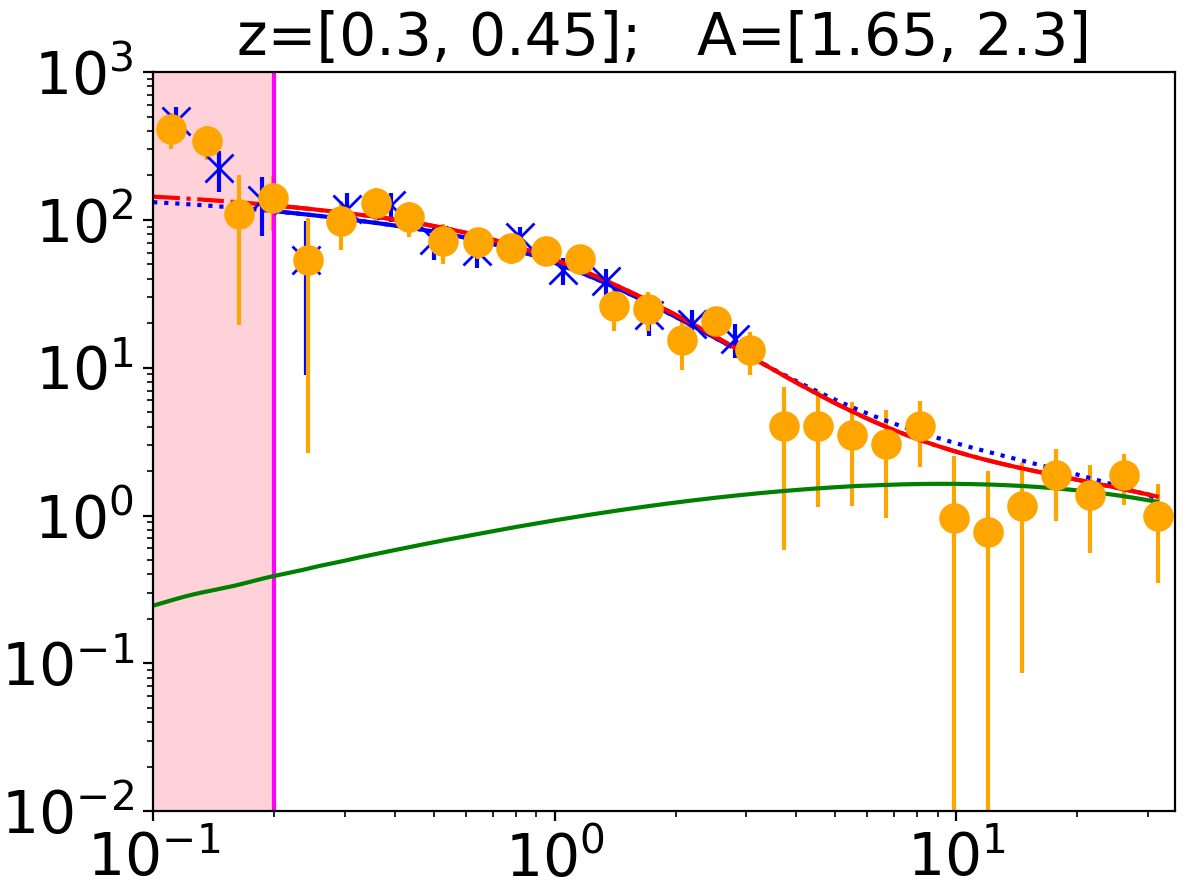}
  \includegraphics[width=0.192\hsize]{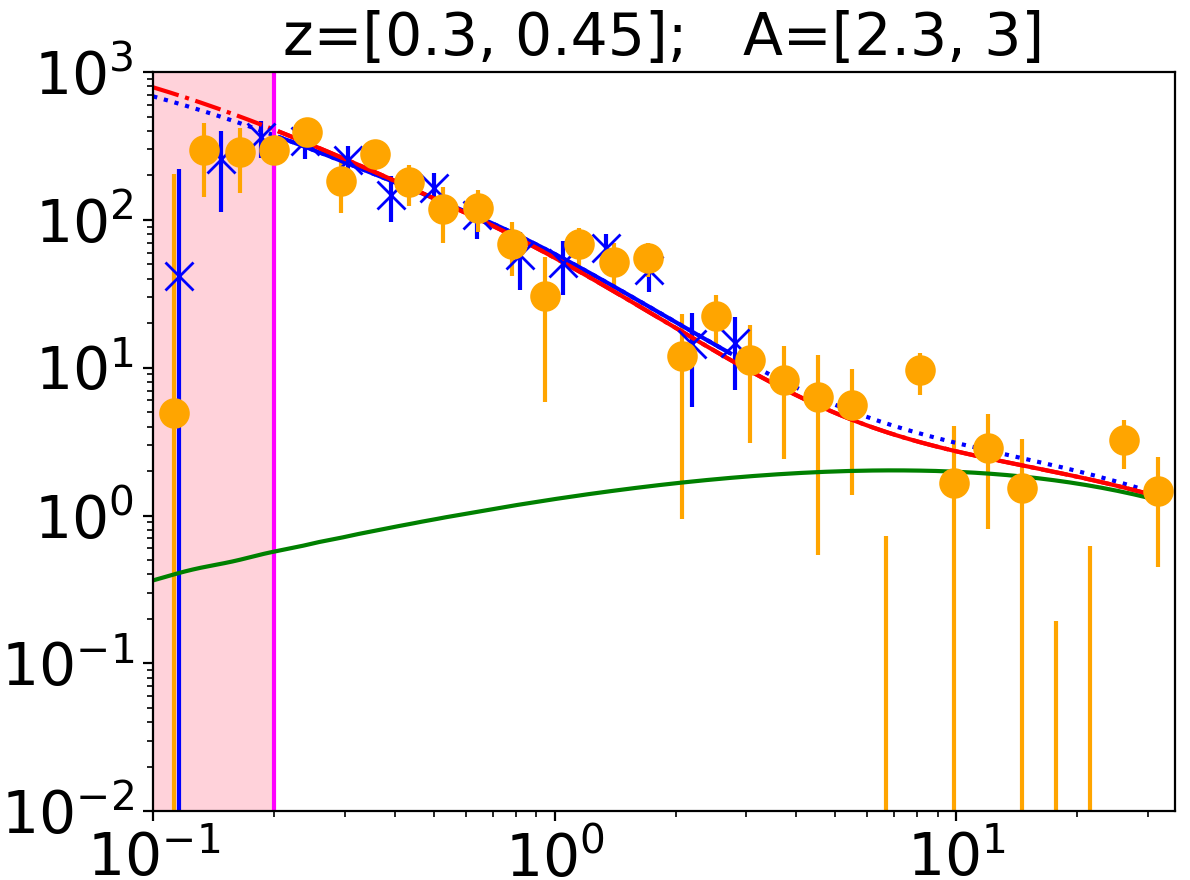}
  \includegraphics[width=0.192\hsize]{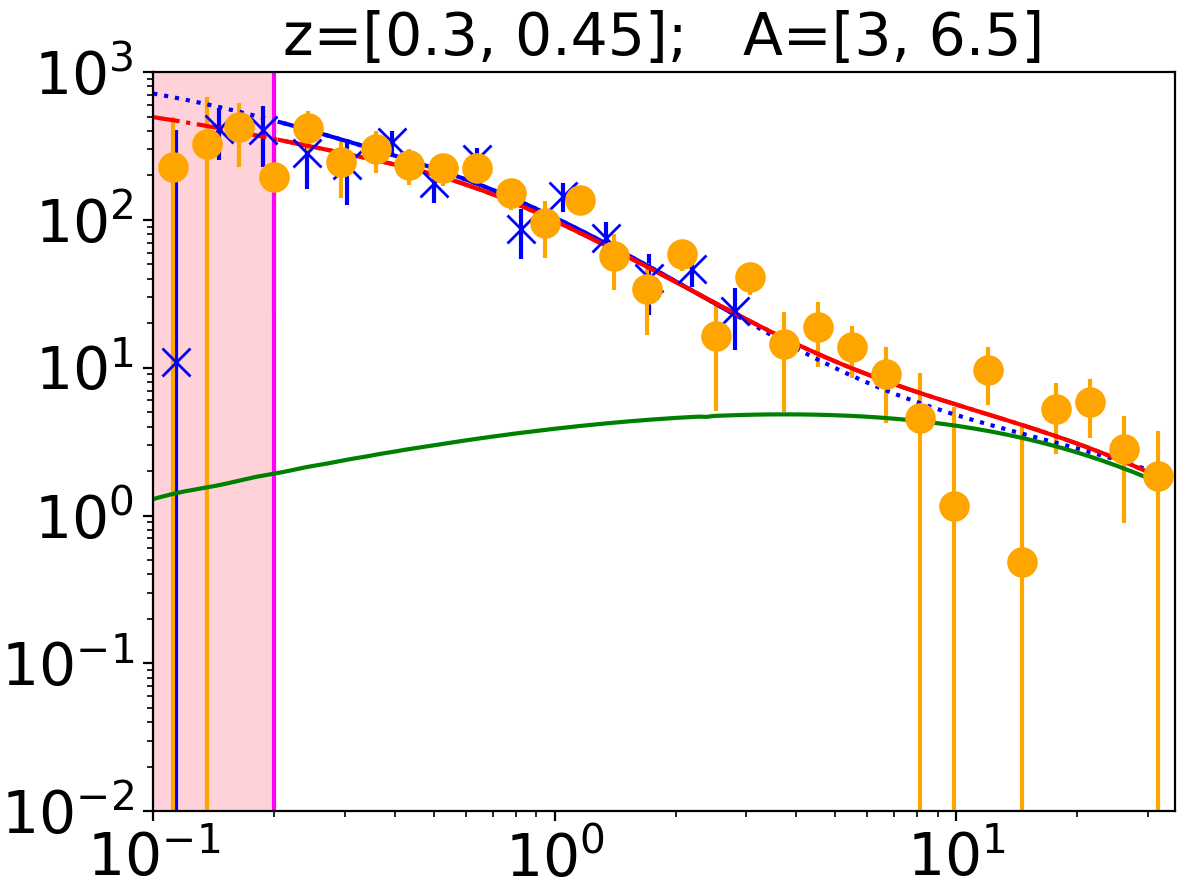}
  \includegraphics[width=0.211\hsize]{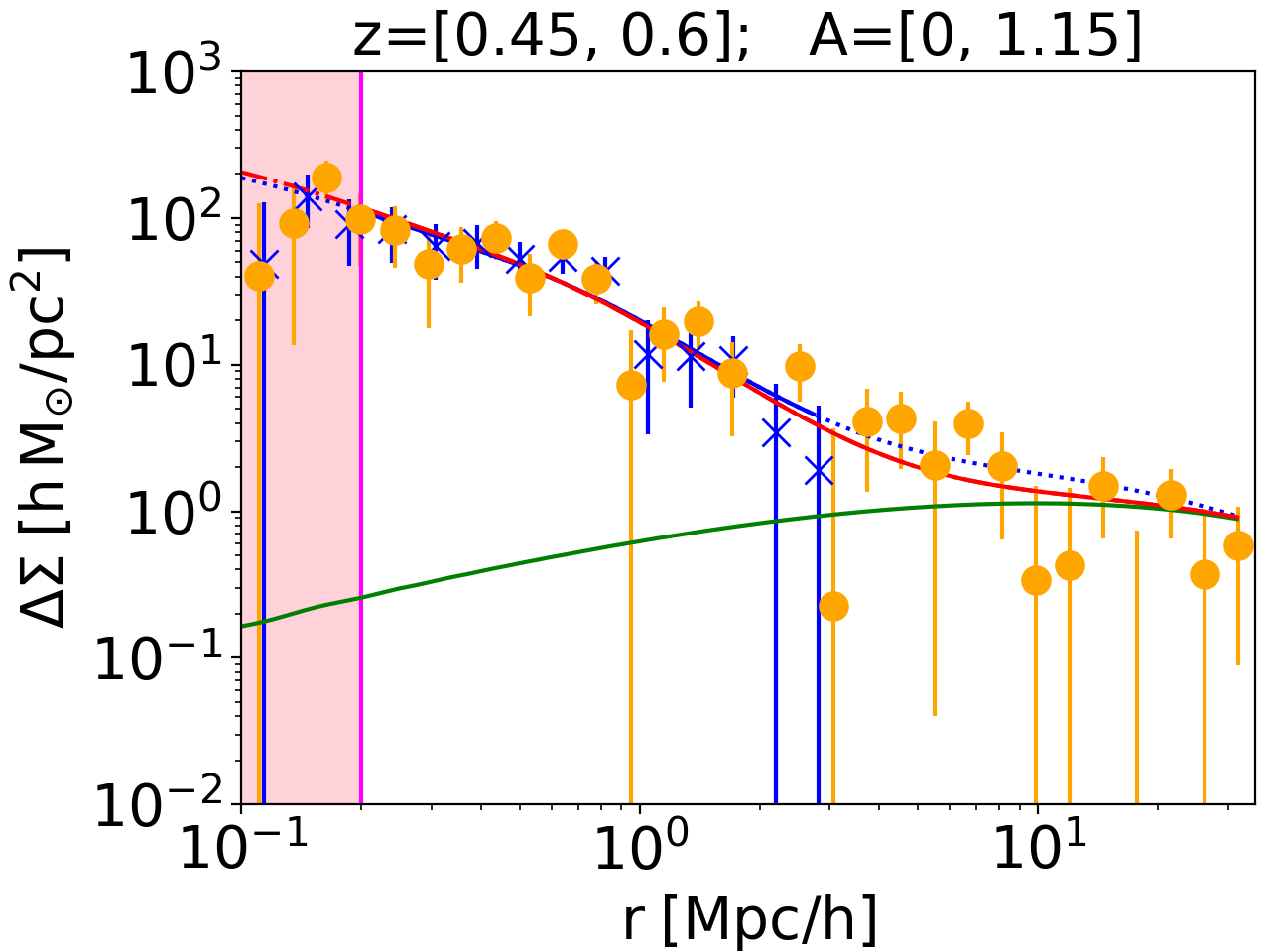}
  \includegraphics[width=0.192\hsize]{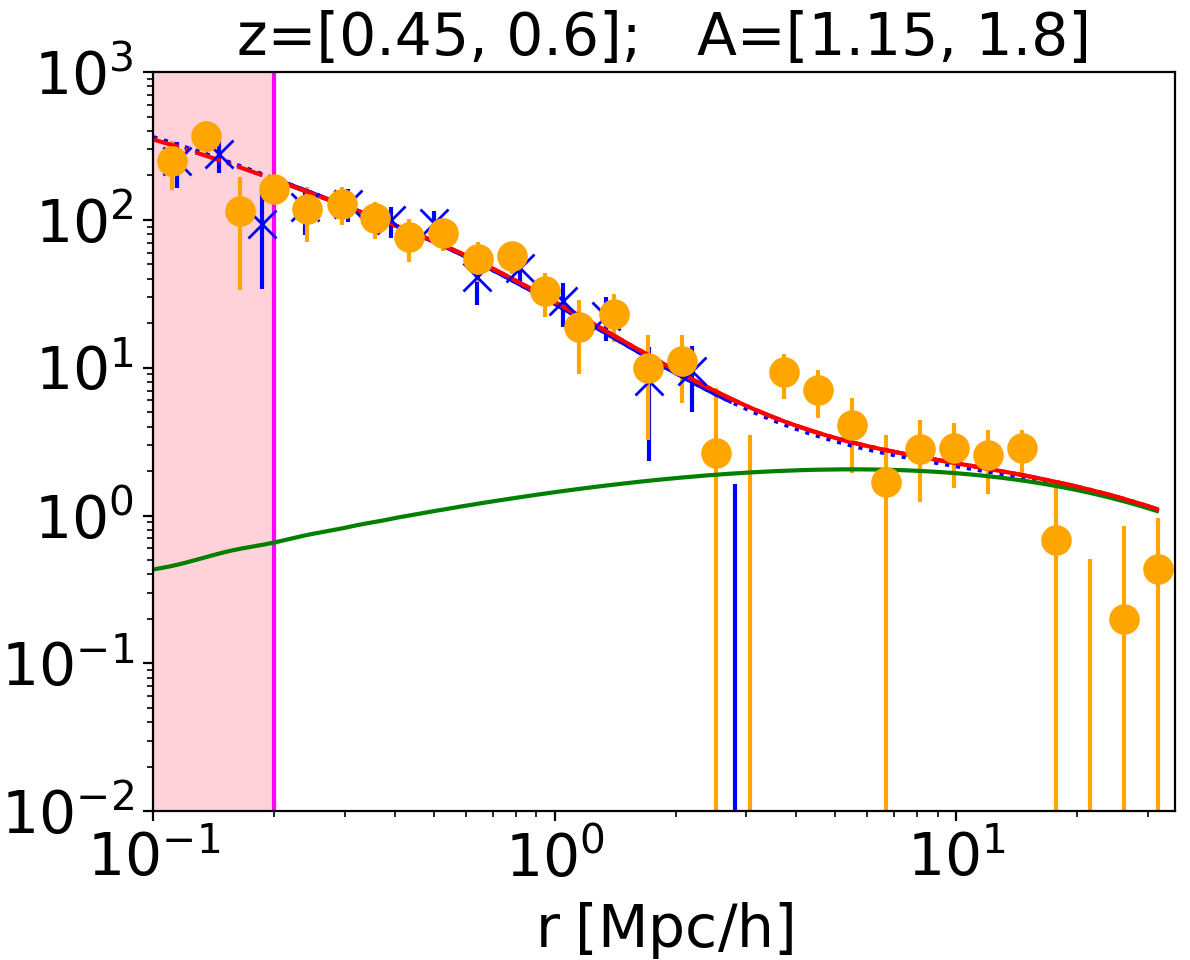}
  \includegraphics[width=0.192\hsize]{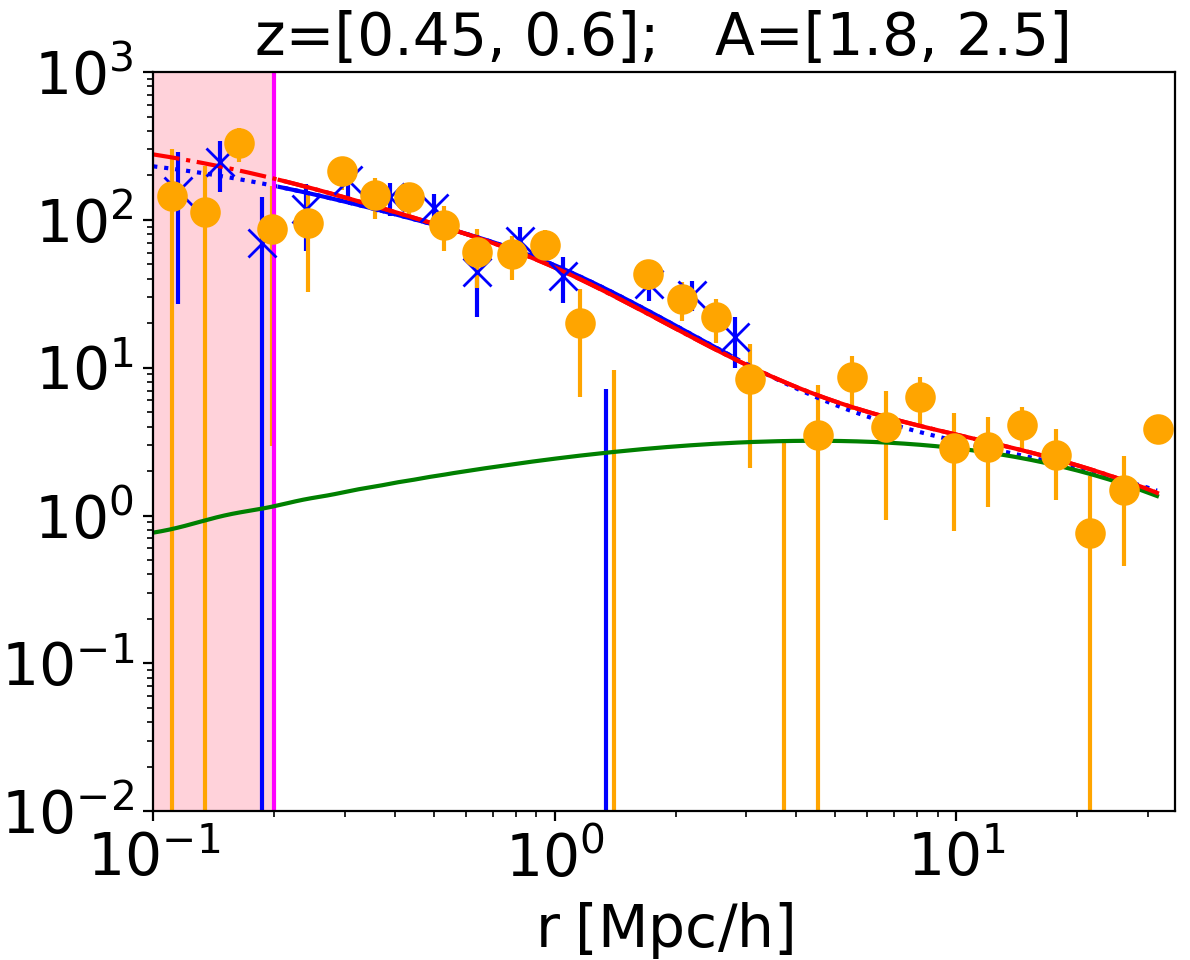}
  \includegraphics[width=0.192\hsize]{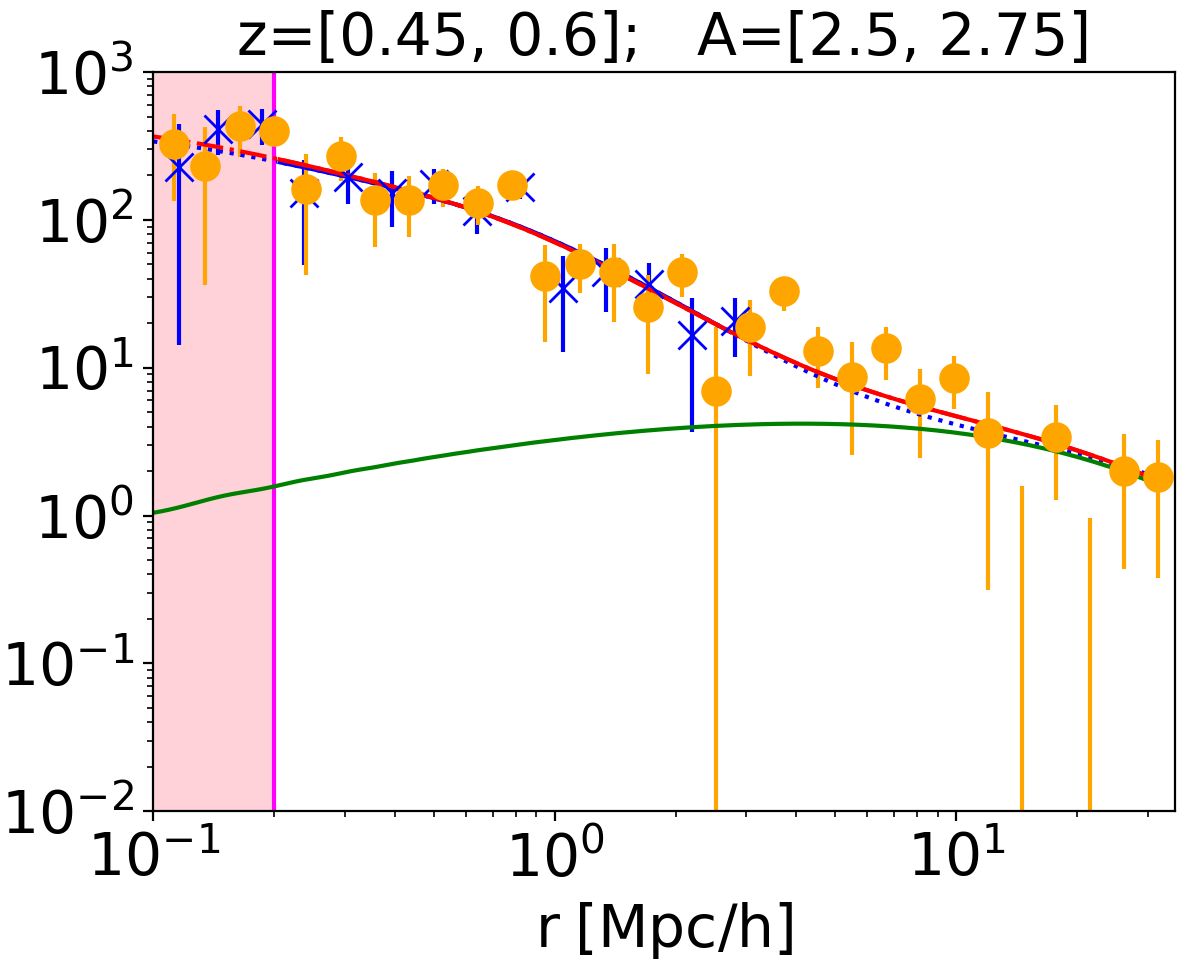}
  \includegraphics[width=0.192\hsize]{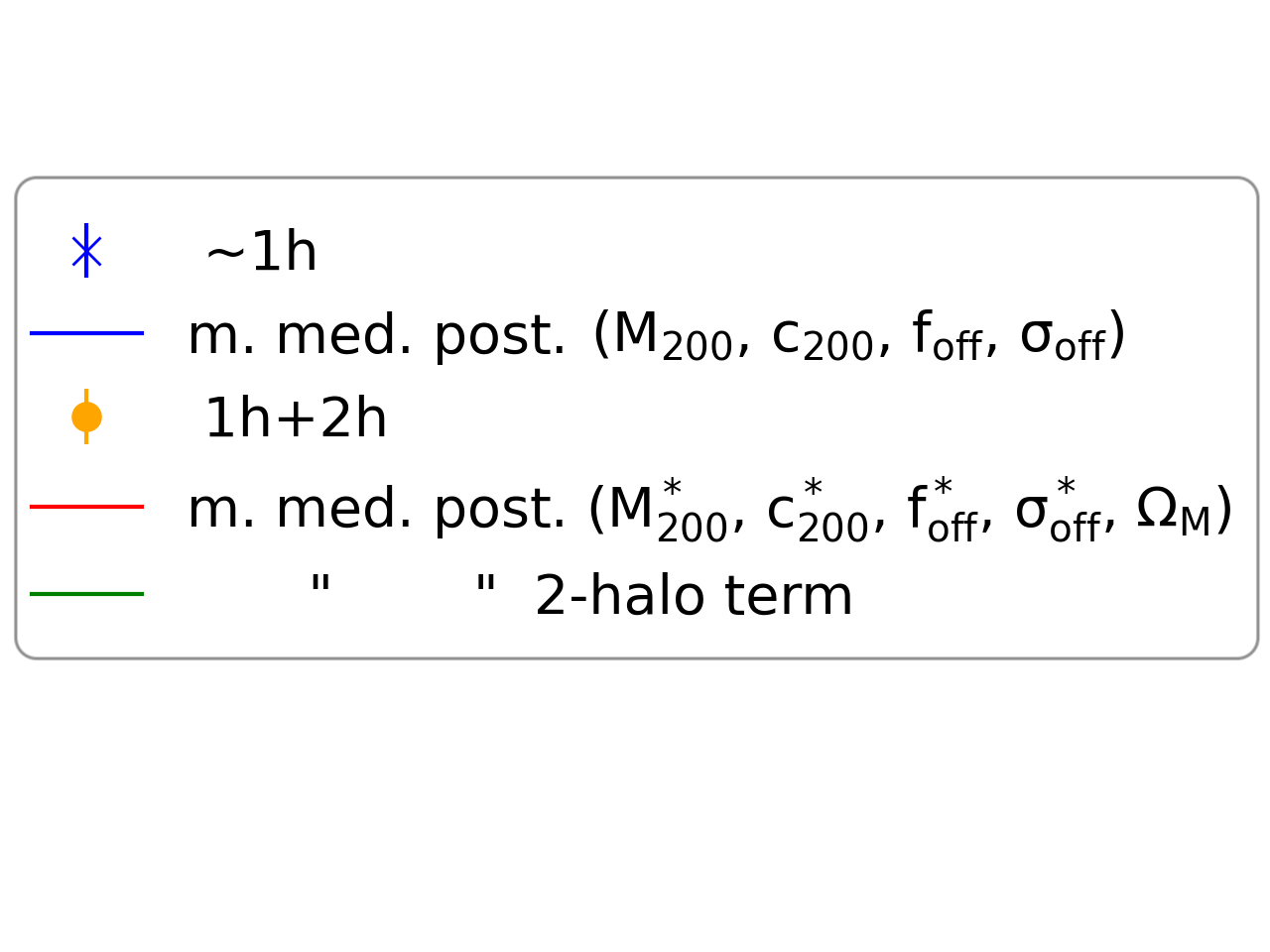}
  \caption{Stacked excess surface mass density profiles of the
    clusters divided in different amplitude (from left to right) and
    redshift (from top to bottom) bins. Blue and orange data points
    refer to the $\sim 1h$ and $1h+2h$ -cases, respectively. The
    corresponding error bars represent the square-root of the diagonal
    terms in the corresponding covariance matrix.  Blue and red solid
    curves show the best-fit models, that is the median values of the
    posterior distributions assessed with the MCMC, corresponding to
    the $\sim 1h$ and $1h+2h$ -cases, respectively. The blue and red
    curves are dotted and dashed outside the corresponding data-set,
    respectively; particularly the blue line below 0.2 and above 3.16
    Mpc/$h$ while the red only below 0.2 Mpc/$h$.  The green curves
    show the 2-halo term contribution.}
  \label{fit_prof_fit} 
\end{figure*}

Figure \ref{fit_prof_fit} shows stacked excess surface mass density
profiles of the AMICO-KiDS-DR3 clusters, divided in different
amplitude and redshift ranges.  Blue and orange data points refer to
the $\sim 1h$ and $1h+2h$ data-set -cases, respectively.  To both
cases we subtract the contributions around random centres.  The
matching error bars display the square-root of the diagonal terms in
the covariance matrix.  Blue and red solid curves show the best-fit
models, that is the median values of the posterior distributions
assessed with the MCMC, corresponding to the $\sim 1h$ and $1h+2h$
-cases, respectively.  In running our MCMC on the $\sim 1h$-case
data-set we adopt uniform priors for the $4$ structural properties
parameters, namely:
\begin{itemize}
 \item $\log \left(M_{200} / h^{-1} M_{\odot}\right)$: $[12.5-15.5]$;
 \item $c_{200}$: $[1-20]$;
 \item $f_{\rm off}$: $[0-0.5]$;
 \item $\sigma_{\rm off} / h^{-1} \mathrm{Mpc}$: $[0-0.5]$.
\end{itemize}

\subsection{Scaling Relations}

\begin{figure}
  \includegraphics[width=\hsize]{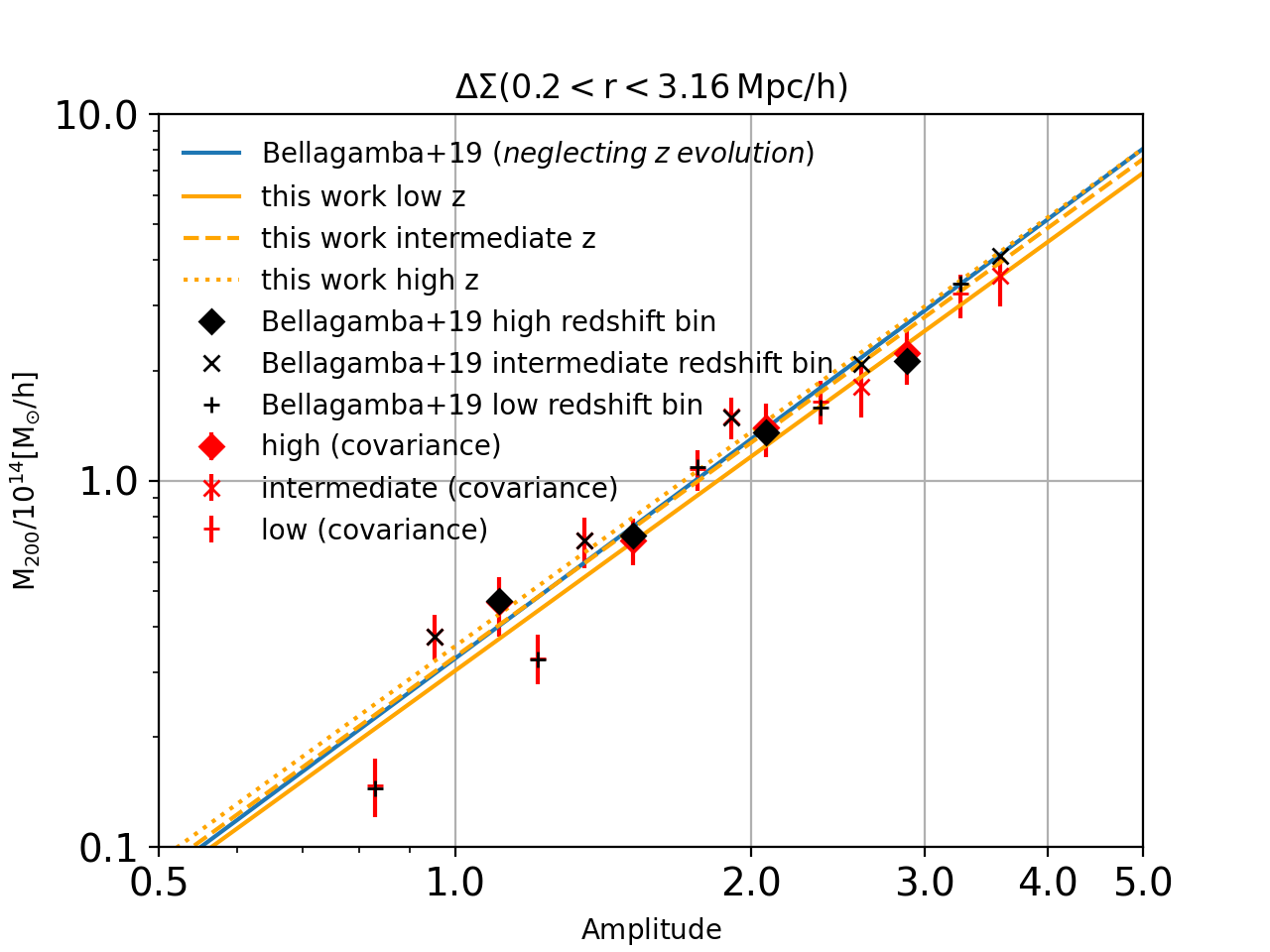}
  \caption{Mass-amplitude relation obtained in the $\sim 1h$-case,
    accounting for the full covariance in each amplitude and redshift
    interval. We compare our results (red points) with the ones by
    \citet{bellagamba19} (black data points).  The error bars show the
    $18th$ and $82nd$ percentiles of the posterior distributions.  The
    blue solid line show the scaling relation recovered by
    \citet{bellagamba19} while the orange ones exhibit our best fit
    result.
  \label{fig_scaline_relation}}
\end{figure}

In Fig.~\ref{fig_scaline_relation} we show the weak lensing
mass-amplitude relation obtained in this work from the small-scale
analysis ($\sim 1h$-case), compared to previous analysis performed on
the same data-set. Moreover, we model the trend using a linear
relation between the photometric observable and the recovered mass --
as discussed also by \citet{sereno15c} and \citet{bellagamba19}:
\begin{equation}
\log \dfrac{M_{200}}{10^{14} M_{\odot}/h} = \alpha + \beta \log
\dfrac{A}{A_{\rm piv}} + \gamma \log \dfrac{E(z)}{E(z_{\rm piv})}\,,
\label{eq_mobs}
\end{equation}
where we set $A_{\rm piv}=2$, $z_{\rm piv}=0.35$ as in
\citet{bellagamba19} and $E(z) = H(z)/H_0$. We recall the reader that,
differently from \citet{bellagamba19}, we subtract the redshift
stacked profiles around random centres and account for the full
covariance matrix in our MCMC analysis.  The red data points display
the median results of our analysis, while the error bars show the
$18th$ and $82nd$ percentiles of the posterior distributions.  The
black data points exhibit the original results by
\citet{bellagamba19}, with whom we obtain a reasonable agreement
within the error bars.  From the figure, though we notice that
accounting for the full covariance in our MCMC analysis we obtain
slightly smaller masses, and thus the slope of our linear relations
(orange lines) is slightly shallower than the one derived by
\citet{bellagamba19} (blue lines -- we display the relation neglecting
the redshift evolution in order to avoid overcrowding the figure).  We
have performed our analysis also accounting only for the $1h$ term in
the modelling function and adopting different cases for the truncation
radius.  Particularly, we discuss in Appendix \ref{app_rt} the results
adopting different truncation radii.

\begin{figure}
  \includegraphics[width=\hsize]{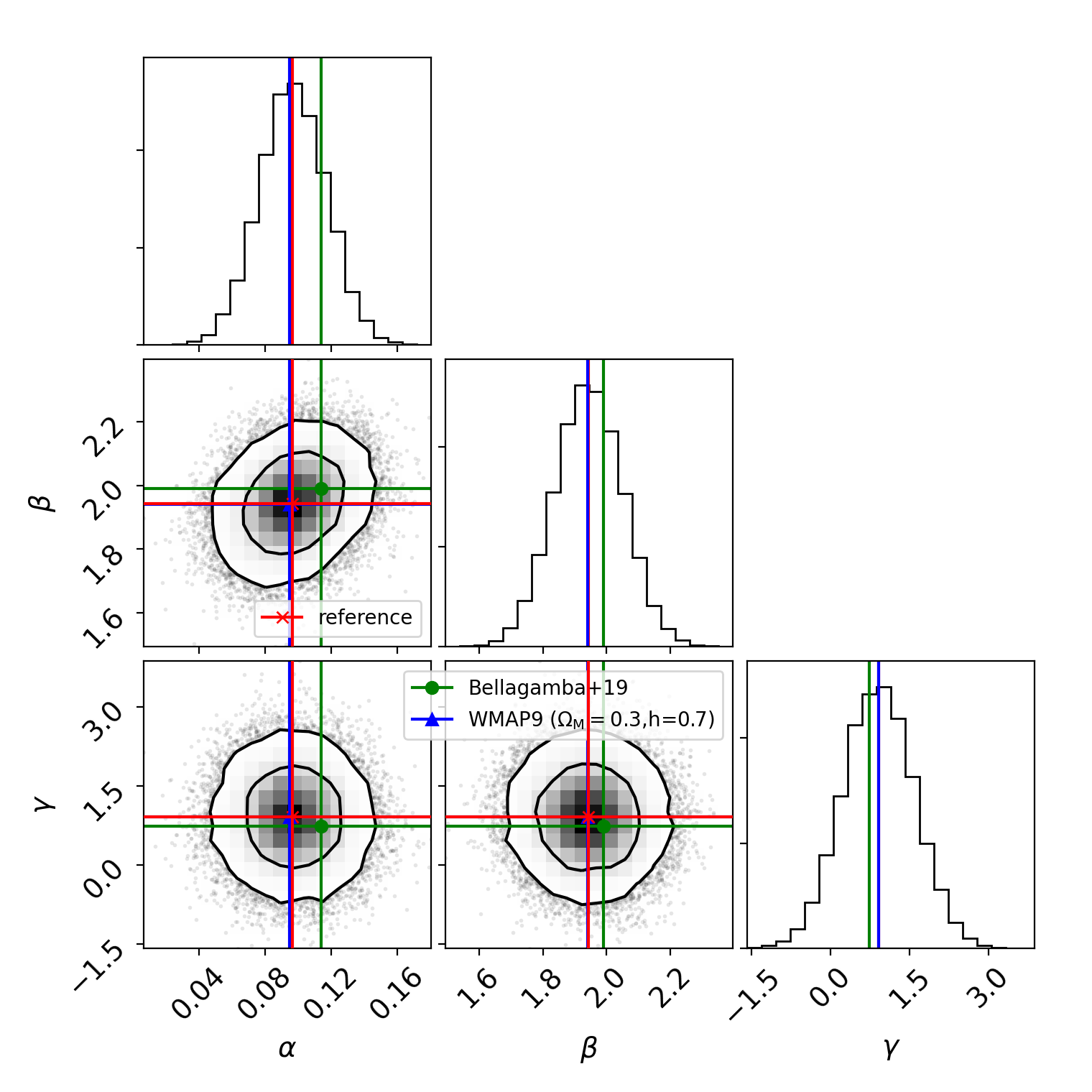}
  \caption{ Posterior distributions of the parameters of the
    mass-amplitude relation: Eq.~\ref{eq_mobs}.  The contours show the
    $68th-95th$ percentiles of the distributions.  For comparison, we
    display also the results for $\alpha$, $\beta$ and $\gamma$
    obtained by \citet{bellagamba19} (green squares) and when assuming
    a flat WMAP9 cosmology, fixing $\Omega_M=0.3$ and $h=0.7$.
      \label{fit_posterior_fit_mo_rel}}
\end{figure}

In Figure \ref{fit_posterior_fit_mo_rel} we display the posterior
distributions of the mass-amplitude model parameters
(Eq.~\ref{eq_mobs}), obtained by modelling the red data points, and
their corresponding error bars, shown in
Fig.~\ref{fig_scaline_relation}.

We follow the same fitting procedure adopted in \citet{bellagamba19},
where the amplitude is computed through a lensing-weighted average.
Moreover, we account for the systematic errors in the covariance by
summing in quadrature the uncertainties for background selection,
photo-$z$s and shear measurements \citet{bellagamba19,sereno20}.  We
assume uniform priors in the range $-1<\alpha<1$, $0.1<\beta<5$ and
$-5<\gamma<5$\footnote{We use \texttt{emcee} for this fitting
  analysis, \href{https://emcee.readthedocs.io/en/stable}{https://emcee.readthedocs.io/en/stable}.}. The obtained
median values of the posterior distributions of the mass-amplitude
relation parameters, and the $18th$ and $82nd$ percentiles around
them, are summarised in Tab.~\ref{tab_emceelinear}. In the second line
of the table we also report the scaling relation parameters for the
masses estimated on the $\sim 1h$-case data-set but adopting in the
modelling function only the $1h$ term. In this case we notice that the
slight overestimate of the masses drives the scaling relation toward a
steeper value. We remind the reader to Table~\ref{tab_bins} where we display
the median mass - expressed in term of the logarithm in base 10 -  
of the posterior distributions for the various cases considered; the values
reported in parenthesis show the $\chi^2$ divided by the number of degrees of freedom. 
\begin{table}
\centering	
\caption{ Scaling relation parameters of the mass-amplitude
  relation. The values reported are the median and $18th$-$82nd$
  percentiles of the posterior distributions.  The first line two
  lines show the results on the $\sim 1h$-case data-set adopting the
  full model (Eq. ~\ref{1h+2hmodel}) or the one that account only for
  the $1h$ term (Eq. ~\ref{sigma1halocen.+off}) for constraining the
  cluster structural properties parameters. The third line exhibit the
  results modelling the $1h+2h$-case data-set up to 35 Mpc/$h$.}
\label{tab_emceelinear}
\begin{tabular}{lccc}
\hline
case & $\alpha$ & $\beta$ & $\gamma$ \\ \hline
$\sim 1h$ &   $0.097_{-0.018}^{+0.019}$ & $1.943_{-0.097}^{+0.098}$  & $ 0.904_{-0.607}^{+0.615}$ \\
$\sim 1h$ (with $1h$ $mod$.) & $ 0.128_{-0.019}^{+0.019}$ & $1.929_{-0.099}^{+0.098}$ & $ 0.973_{-0.624}^{+0.617}$ \\
$1h+2h$ & $0.096_{-0.017}^{+0.017}$  & $1.912_{-0.085}^{+0.085}$ & $ 0.950_{-0.557}^{+0.552}$ \\
\hline
\end{tabular}
\end{table}

The contours in Fig.~\ref{fit_posterior_fit_mo_rel} display the
$68th-95th$ percentiles of the posterior distributions.  For
comparison, we show also the best-fit results obtained adopting a
different parameterisation of the matter power spectrum in the 2-halo
term (flat-WMAP9 parameters, but with $\Omega_M=0.3$ and $h=0.7$), and
the results by \citet{bellagamba19}. All results shown in
Fig.~\ref{fit_posterior_fit_mo_rel} appear statistically
consistent. We also notice and underline that the adoption of a flat-
Planck18 or WMAP9 cosmology, fixing $\Omega_M=0.3$ and $h=0.7$,
impacts mainly in the modelling of the $2h$ term that is negligible when
analysing the $\sim 1h$-case data up to $3.16$ Mpc/$h$.

\begin{figure}
  \includegraphics[width=\hsize]{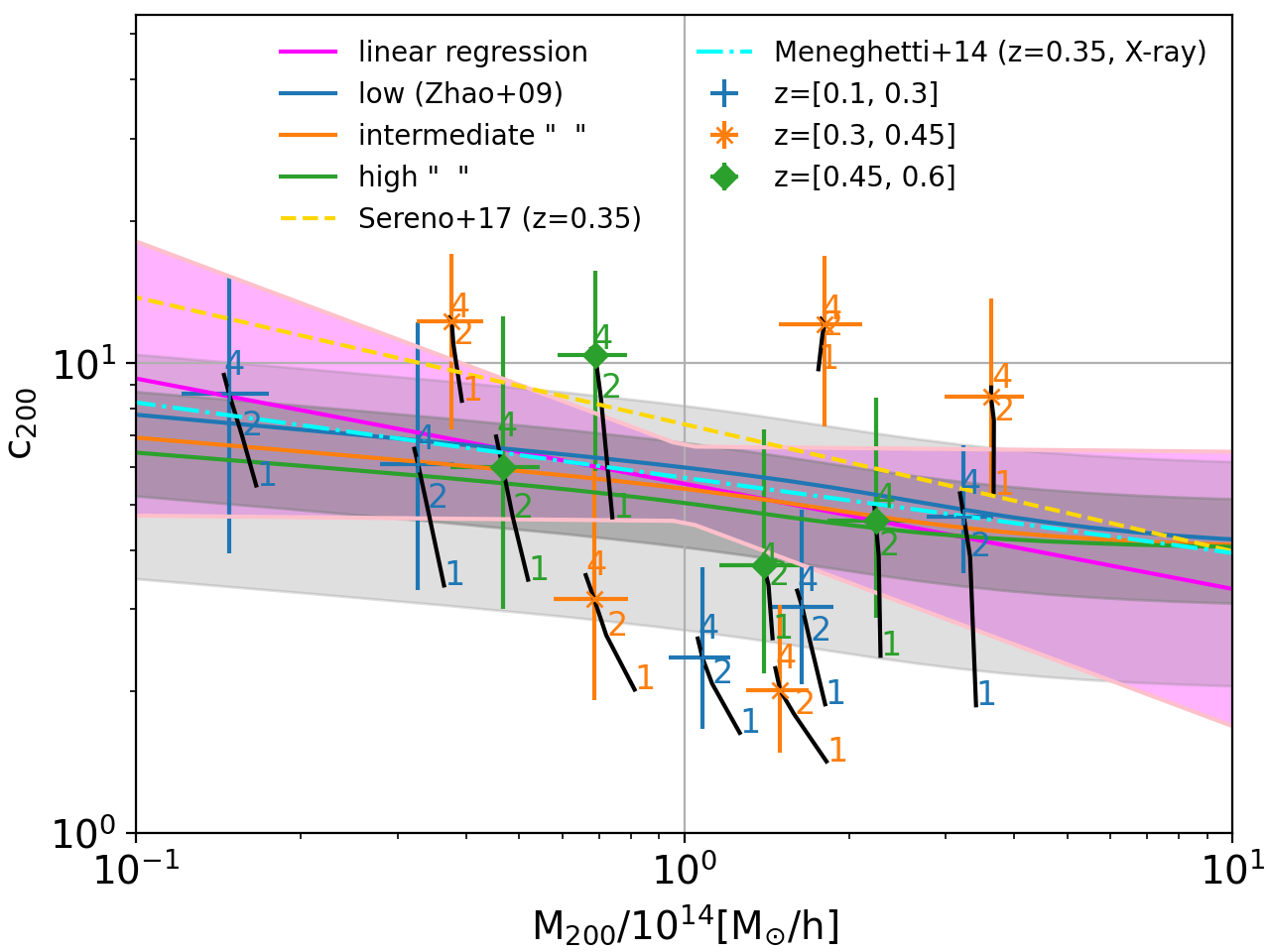}
  \caption{\label{fig_cm} Concentration-mass relation obtained from
    our stacked weak lensing analysis of the AMICO-KiDS-DR3
    clusters. Data points and error bars show the median and the
    $18th$-$82th$ percentiles of the posteriors distributions. Blue,
    orange and green colours refer to the low, intermediate and high
    redshift cluster sub-samples.  The solid curves represent the
    predictions by \citet{zhao09} in three redshift intervals. For
    each data point, the numbers show the results for the $c-M$
    relation obtained adopting the truncation radius $r_t$ equal to 1,
    2 and 4 -- in unit of $R_{200}$, respectively. The grey and
    light-grey shaded regions indicate 1 and 2-$\sigma_{\ln c }= 0.25$
    scatter as measured in numerical simulations
    \citep{dolag04,giocoli12b}. The magenta line show the linear
    regression in the $c-m$ relation computed using our reference
    results as data-set, while the shaded light-magenta region
    encloses the 1-$\sigma$ uncertainties in both the slope and the
    intercept of the relation. In this case, we use the \texttt{ODR}
    routine in \texttt{python} considering the uncertainties in both
    mass and concentration. Dashed gold and dash-dotted cyan display
    the results by \citet{sereno17} -- based on observational data --
    and \citet{meneghetti14} -- on hydrodynamical simulations,
    respectively.}
\end{figure}

As explained in Sec. \ref{sub:1halo}, the 1-halo model used in this
analysis has four free parameters: $f_{\rm off}$ and $\sigma_{\rm off}$, that
parameterise the uncertainty relative to the cluster centre
definition, and $M_{200}$ and $c_{200}$, which are related to the halo
structural properties.  Theoretical models
\citep{eke01,zhao09,ludlow12} and numerical simulations
\citep{dolag04, neto07, maccio08,duffy08,klypin14} predict that, at
fixed redshift, the halo concentration is a decreasing function of the
halo mass. Less massive haloes forming at higher redshifts tend to be
more concentrated than the more massive ones
\citep{vandenbosch02,wechsler02}; cluster size haloes are still in
their formation phase.

In Fig.~\ref{fig_cm} we display the concentration-mass relation for
our stacked cluster sample in the three redshift bins considered,
compared to the theoretical predictions for our reference cosmological
model adopting the \citet{zhao09} recipe, where the concentration is
related to the time at which haloes assemble $4\%$ of their mass. To
follow the mass accretion histories of the haloes back in time we use
the model by \citet{giocoli12b}.  We perform a log-log linear
regression modelling the 14 data points\footnote{The fitting analysis
  is done with the \texttt{Orthogonal Distance Regression} routine
  \href{https://docs.scipy.org/doc/scipy/reference/odr.html}{https://docs.scipy.org/doc/scipy/reference/odr.html},
  accounting for the uncertainties in both mass and concentration.},
though the small number of data points does not allow us to constrain
the redshift dependence of this relation.

The log-log linear relation obtained from the fitting analysis is the
following:
\begin{equation}
\log c_{200} = B \log M_{200} + C\,,
\end{equation} 
with $B=-0.22 \pm 0.21$ and $C = 0.74 \pm 0.08$, which is in good
agreement with the results from numerical simulations
\citep{duffy08,prada12,ludlow14b,meneghetti14} and from observational
data \citet{mandelbaum08,merten14,sereno17} in the corresponding mass
ranges. The shaded light-magenta region encloses the 1-$\sigma$
uncertainties in both the slope and the intercept of the relation.
For comparison and for each data point, the numbers exhibit the
results for the $c_{200}-M_{200}$ relation obtained adopting the
truncation radius $r_t$ equal to 1, 2 and 4 -- in unit of $R_{200}$,
respectively.  In the figure, we also display with a dashed gold line
the finding by \citet{sereno17} modelling the relation of the PSZ2LenS
clusters and with a dot-dashed cyan one the results by
\citet{meneghetti14} fitting the results of hydrodynamical simulations
for the 2D fit of the X-ray selected cluster sample. Both relations
refer to an intermediate redshift of our cluster sample corresponding
to $z=0.35$.

\subsection{Constraining the total matter density parameter}

\begin{table*}
\centering
\caption{Priors considered in the various redshift and amplitude bins. The different lines 
are ordered as in Tab.~\ref{tab_bins}. For the halo structural properties (first four columns) 
we consider Gaussian priors with mean and standard deviations derived from the posteriors 
of the $\sim 1h$-case analysis, while for the total matter density parameter (last column) 
we assume a uniform prior. \label{tab_priors}}
\begin{tabular}{lccccc}
\hline
$-$ & $\log M_{200}$ & $c_{200}$ & $f_{\rm off}$ & $\sigma_{\rm off}$ &  $\Omega_{M}$\\ \hline
1 & $13.170 \pm 0.088$ & $9.3 \pm 5.2 $ & $0.25 \pm 0.14$ & $0.22 \pm 0.13$ & $[0.06 - 0.6]$ \\ 
2 & $13.517 \pm 0.074$ & $7.4 \pm 4.6 $ & $0.28 \pm 0.14$ & $0.26 \pm 0.13$ & "   " \\
3 & $14.034 \pm 0.059$ & $2.7 \pm 1.3 $ & $0.27 \pm 0.14$ & $0.24 \pm 0.14$ & "   " \\
4 & $14.214 \pm 0.067$ & $3.6 \pm 2.0 $ & $0.28 \pm 0.14$ & $0.26 \pm 0.14$ & "   " \\ 
5 & $14.507 \pm 0.065$ & $5.2 \pm 2.1 $ & $0.31 \pm 0.14$ & $0.20 \pm 0.10$ & "   " \\ \hline

6 & $13.579 \pm 0.069$ & $12.1 \pm 4.5 $ & $0.22 \pm 0.14$ & $0.23 \pm 0.15$ & "   " \\
7 & $13.835 \pm 0.073$ & $4.2 \pm 3.1 $ & $0.29 \pm 0.14$ & $0.29 \pm 0.13$ & "   " \\
8 & $14.176 \pm 0.059$ & $2.3 \pm 1.1 $ & $0.28 \pm 0.14$ & $0.26 \pm 0.13$ & "   " \\
9 & $14.255 \pm 0.082$ & $12.0 \pm 4.5 $ & $0.24 \pm 0.14$ & $0.26 \pm 0.15$ & "   " \\
10 & $14.557 \pm 0.074$ & $9.3 \pm 4.3 $ & $0.26 \pm 0.14$ & $0.24 \pm 0.14$ & "   " \\ \hline

11 & $13.664 \pm 0.088$ & $7.5 \pm 4.8 $ & $0.29 \pm 0.14$ & $0.24 \pm 0.12$ & "   " \\
12 & $13.837 \pm 0.069$ & $10.7 \pm 4.6 $ & $0.25 \pm 0.14$ & $0.22 \pm 0.13$ & "   " \\ 
13 & $14.143 \pm 0.082$ & $4.9 \pm 3.4 $ & $0.29 \pm 0.14$ & $0.29 \pm 0.14$ & "   " \\
14 & $14.345 \pm 0.082$ & $5.7 \pm 3.6 $ & $0.29 \pm 0.14$ & $0.25 \pm 0.13$ & "   " \\ 
\hline
\end{tabular}
\end{table*}

In this section we will present the results when modelling the
large-scale $\Delta \Sigma$ profiles up to 35 Mpc/$h$, $1h+2h$-case.
The large-scale signal is sensitive to the linear power spectrum
normalisation and to the host halo bias
\citep{sereno15b,sereno17,sereno18}.  In this work we follow the
approach of fixing the halo bias-cluster mass relation -- as well as
its redshift dependence -- to the results obtained from the analyses
of various numerical simulations \citep{sheth99b,sheth01b,tinker10}
while leaving the matter density parameter $\Omega_M$ as a free
quantity to vary in our analysis. The $\sigma_8$ parameter, measuring
the power spectrum fluctuations smoothed with a top-hat window
function of aperture 8 Mpc/$h$, will be obtained as a derived quantity
from our MCMC and the corresponding cosmological model.

We employ the same Bayesian approach and the $1h+2h$ modelling
function used in the previous section, introducing an extra degree of
freedom, that is the total matter density parameter $\Omega_M$ --
assuming a flat Universe, i.e. $\Omega_{\Lambda} = 1 - \Omega_M$.  In
doing so, we consistently account for the cosmology dependence of the
conversion from angular to comoving and physical coordinate in our
data-set -- i.e. the radial profiles of $\Delta \Sigma$. The stacked
weak lensing profiles shown before have been computed assuming a
reference flat cosmology with $\Omega_M=0.3$ and $h=0.7$. In the
Bayesian analysis, when varying the matter density parameter, we take
into account the dependence on $\Omega_M$ not only on $r$ but also on
$\Delta \Sigma$ (see Eq.~\ref{eq_scrit}).

The red curves shown in Fig.~\ref{fit_prof_fit} display the best-fit
models computed from the MCMC Bayesian analysis, corresponding to the
median values of the posterior distributions. As described in
Sec. \ref{sub:2halo} and above, the considered model has five free
parameters, that is $M_{200}$, $c_{200}$, $f_{\rm off}$, $\sigma_{\rm
  off}$ and $\Omega_M$.  In modelling the $1h+2h$-case data-set we
assume Gaussian priors for the four 1-halo model parameters with mean
and standard deviation corresponding to the $1h$-case modelling
posteriors, while we let $\Omega_M$ vary between $0.06$ and $0.6$,
uniformly (see Tab.~\ref{tab_priors}). The green curves show the
corresponding 2-halo term contribution.

In our analyses we decided to perform this twofold approach -- first
performing the fit of the halo structural property parameters and
then, while limiting them to a smaller Gaussian interval, constraint
the total matter density parameter -- for two reasons. The first is
related to the wide range of scales probed by our stacked weak lensing
profiles and the second is connected to the number of parameters that
we vary -- four for the 1-halo term and one for the 2-halo -- our
procedure allows separating in a self-consistent way the weights of
the small and large scale contributions.

\begin{figure}
  \includegraphics[width=\hsize]{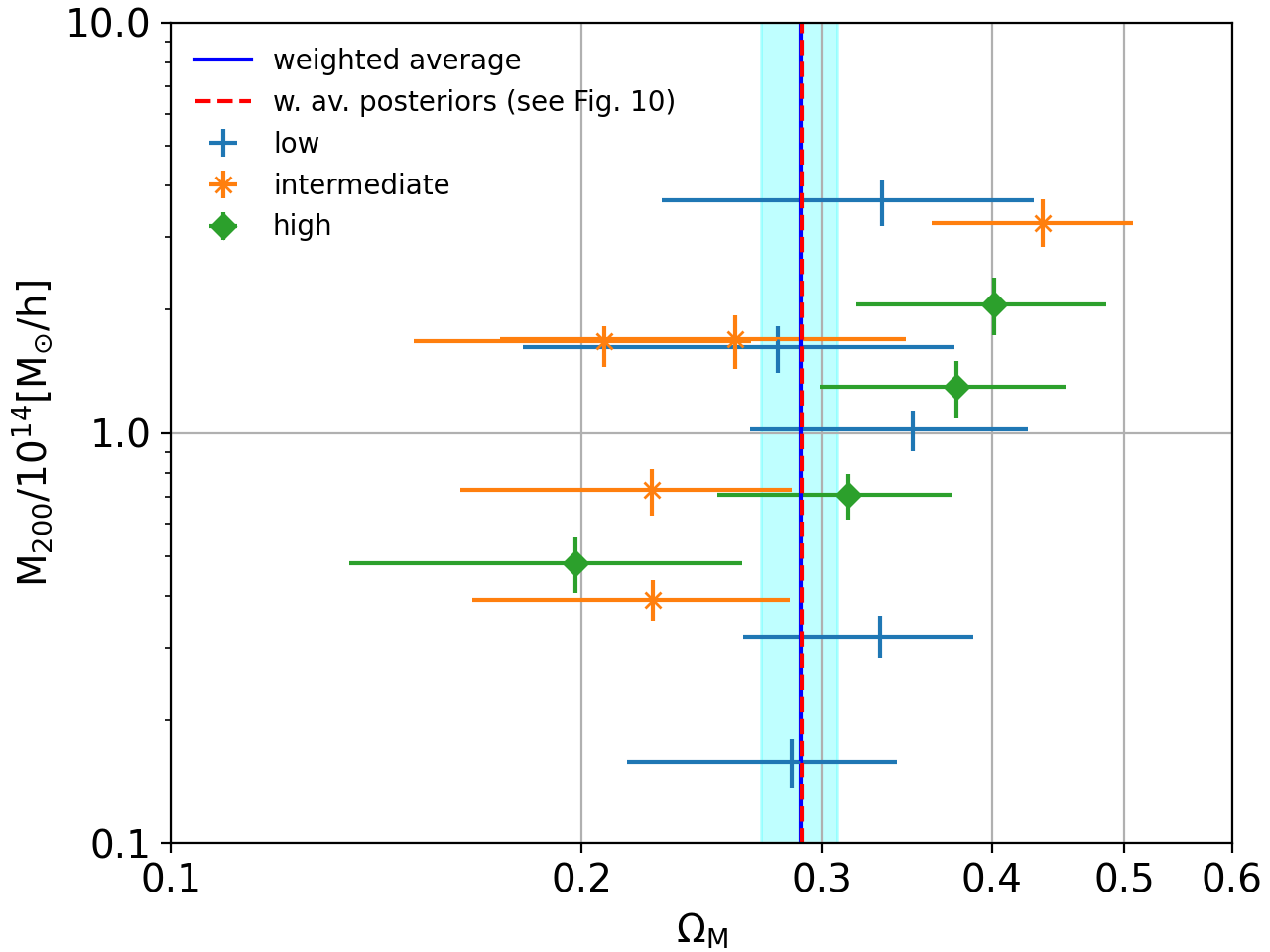}
  \caption{Correlation between the recovered cluster mass and total
    matter density parameter from our stacked weak lensing
    analysis. The different coloured data points refer to three
    redshift ranges considered, as indicated in the legend.
    The blue solid line 
    displays the weighted average of the various estimated $\Omega_M$ the 
    the different amplitude and redshift bins; the cyan shaded region 
    encloses the one $\sigma$ uncertainty. The dashed red line exhibits 
    the weighted average of the posterior distributions as in Fig.~\ref{fig_planck_wmap}.
  \label{fig_m_omega}}
\end{figure}

Fig.~\ref{fig_m_omega} displays the relation between the recovered
weak lensing mass and total matter density parameter $\Omega_M$.  From
the figure, we note that there is no particular correlation.  The data
points display the medians and the error bars the $18th-82nd$
percentiles of the posterior distributions.  The different coloured
data points refer to the three redshift ranges considered. 
The blue solid vertical line and the cyan shaded region display the weighted
average of the various $\Omega_M$ estimates and its associated uncertainty.  

The last
line of Tab.~\ref{tab_emceelinear} displays the mass-amplitude
relation parameters (see Eq.~\ref{eq_mobs}) obtained modelling the
$1h+2h$-case data-set.

\begin{figure}
  \includegraphics[width=\hsize]{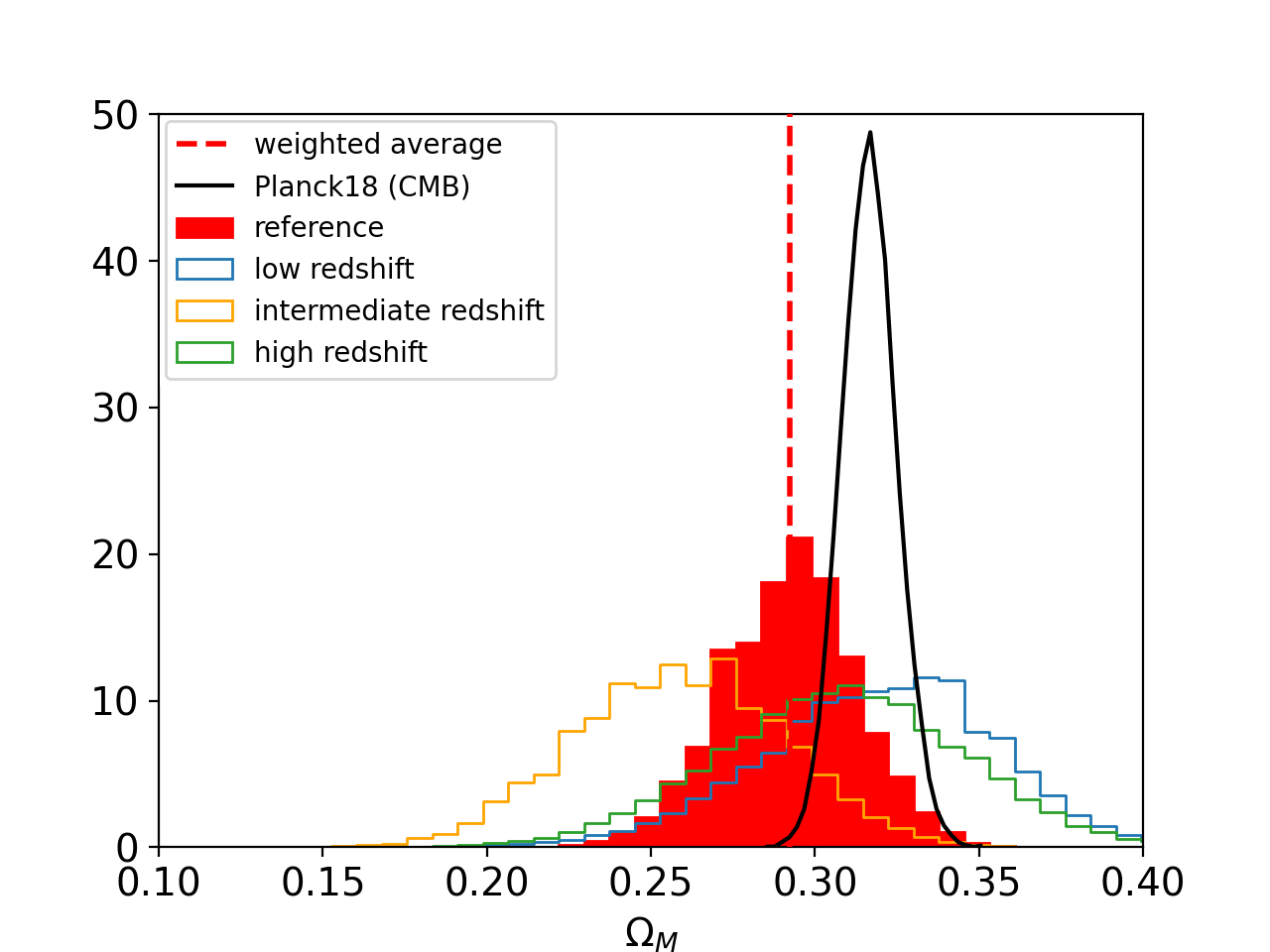}
  \caption{Posterior distributions of the total matter density
    parameter $\Omega_M$. Blue, orange and green histograms show the
    results for the low, intermediate and high redshift cluster
    samples, respectively. The red-filled histogram exhibit the
    posterior distribution combining all redshifts together, the
    dashed red line indicates the median of the distribution. The
    black curve displays the posterior distribution obtained from the
    analysis of the CMB data by \citet{planck18}.
  \label{fig_planck_wmap}}
\end{figure}

As discussed in Sec.~\ref{models_label}, we assumed Planck18 as the
reference cosmological model, but fixing $\Omega_M=0.3$ and $h=0.7$,
since the distances and the stacked weak lensing data have been
computed assuming those parameters \citep{bellagamba19}.  In
Fig.~\ref{fig_planck_wmap} we show the posterior distributions for the
measured $\Omega_M$ parameter in the three redshift bins considered,
as well as the result obtained by combining all redshifts together.
In this case, we have fixed all the cosmological parameters to the
flat Planck18 ones, varying only $\Omega_M$ uniformly between
$[0.06-0.6]$.  The one halo term parameters are varied in the same
range as done for the $\sim 1h$-case data-set, but assuming a Gaussian
distribution for the priors with mean and standard deviation as from
the $\sim 1h$-case posteriors.  The solid black line exhibits the
posterior distribution of $\Omega_M$ from the analysis of the CMB data
performed by the \citet{planck18}. In the figure, we notice a small
shift toward lower values of $\Omega_M$ for the intermediate redshift
bin, however consistent with the other redshift intervals within the
measured uncertainties. As underlined by \citet{maturi19}, this
interval may suffer from slightly larger photometric errors that could
impact the modelling of the estimated tangential shear profile, we
expect to further investigate this in the next AMICO-KiDS data
release.

Table ~\ref{tab_om} reports the $\Omega_M$ constraints obtained from
the stacked weak lensing analyses at the three redshift ranges
considered, and the combined one.

In our cosmological analysis, we are in good agreement with
\citet{lesci20} and \citet{nanni21} who analysed the same AMICO-KiDS-DR3 data-set.
This is remarkable since our analysis is independent from
cluster-clustering and cluster-counts, but we still find a consistent
value of $\Omega_M$.

We discuss the model systematics of our results in more details the
two appendices. In Appendix \ref{app_rt} we examine the impact of the
truncation radius in modelling the $\sim 1h$-case; while in Appendix
\ref{app_sis} we discuss their impact in recovering the total matter
density parameter -- modelling the $1h+2h$-case data-set, as well as
the use of a different reference linear power spectra and various
halo-bias models.

\begin{table}
\centering
\caption{Constraints on $\Omega_M$ obtained from the stacked weak
  lensing analyses at the three redshift ranges considered, and the
  combined one. \label{tab_om}}
\begin{tabular}{|l|c|}
\hline
 & $ \Omega_M $  \\ \hline
\underline{all redshifts} & $\mathbf{0.29 \pm 0.02}$ \\
low & $0.32 \pm 0.04 $\\ 
intermediate & $ 0.26 \pm 0.03$ \\ 
high &  $ 0.31 \pm 0.04 $ \\ \hline
\end{tabular}
\end{table}

\section{Summary \& Conclusions}
\label{summary_label}

In this work, we presented new cosmological results from the analysis
of weak lensing signal produced by a galaxy cluster sample detected
from the KiDS-DR3 survey \citep{dejong17, hildebrandt17}.
Specifically, we exploited photometric redshifts with relative
probability distribution functions \citep{kuijken15, dejong17}, a
global improved photometric calibration, weak lensing shear catalogues
\citep{kuijken15, hildebrandt17} and lensing-optimised image data.
Our cluster sample, used to build the stacked shear profiles for the
cosmological analysis, was obtained using the \texttt{AMICO} algorithm
\citep{bellagamba18} in \citet{maturi19}. The redshift-amplitude
binning of the cluster sample has been constructed as in
\citet{bellagamba19}.

Our studies support the use of the stacking method in weak lensing to
calibrate galaxy cluster scaling relations, in agreement with previous
results.  The robustness of the AMICO-KiDS-DR3 data-set enabled us to
infer cosmological parameters from the stacked shear
measurements. Modelling the shear profiles of these clusters up to
large scales, we inferred robust cosmological constraints on
$\Omega_M$, assuming a reference recipe for the halo-bias parameter
\citep{tinker10}.

We constructed two data-sets from the same cluster and galaxy shear
catalogues. The first data-set has been used for the $\sim 1h$-case
analysis, in which we binned the stacked radial excess surface mass
density profiles in $14$ intervals from 0.1 to 3.16 Mpc/$h$ as in
\citet{bellagamba19}, while the second data-set has been considered
for the $1h+2h$-case, where the analyses has been performed in 30 bins
up to 35 Mpc/$h$. In order to keep under control all the observational
systematic uncertainties of the survey, we subtracted to the
measurements the signal around random centres, randomising 10,000
times the positions of the $6962$ \texttt{AMICO} clusters. We also accounted
for the full covariances on the final $\Delta \Sigma$ profiles. We
split our data-sets in $14$ stacked measures in redshift and
amplitude.  We used a halo model approach developed in our MCMC
routines in the \texttt{CosmoBolognaLib} \citep{marulli16}.  Modelling
the $\sim 1h$-case data-set, we recovered with good accuracy the
cluster halo structural model parameters ($M_{200}$, $c_{200}$,
$f_{\rm off}$, $\sigma_{\rm off}$) and the mass-amplitude relation
parameters. We also studied the impact of adopting only the $1h$ term
in our modelling function, finding a small positive bias in mass of
the order of few percents.  In addition, we modelled the recovered
concentration and mass relation of the $14$ samples finding very good
agreements with other observational studies and numerical simulation
results: more massive haloes tend to be less concentrated than the
smaller ones. However, the small data-set and the mild expected
redshift dependence of the relation does not allow us to model the
redshift evolution of the $c_{200}-M_{200}$ relation.

We employed the $1h+2h$-case data-set to simultaneously constrain the
cluster structural properties and the $\Omega_M$ parameter.  Our
results are fully consistent with the theoretical predictions of the
$\mathrm{\Lambda}$CDM model.  Marginalising over the cluster model
parameters, we obtained $\Omega_M=0.29 \pm 0.02$, from the combined
posterior distribution, assuming a flat geometry. This value is fully
consistent with the independent results on the same cluster sample
performed by \citet{lesci20} and \citet{nanni21}, and also in agreement
with the findings from recent CMB experiments
\citep{wmap9,planck18}. 
Higher than the low value of the total matter density parameter estimated from counts and weak lensing signal of redMaPPer clusters from the Dark Energy Survey (DES) Year 1 \citep{abbott20} and slightly lower then the one computed by \citet{planck18} analysing CMB data.

We expect to reduce the uncertainties on our finding by enhancing the
number of clusters to stack, looking forward to the new cluster
catalogues from the future KiDS data releases, and afterwards from the
data coming from future wide field surveys like the ESA-Euclid mission
\citep{euclidredbook}.  In fact, upcoming deeper galaxy surveys will
allow us to detect and stack lensing clusters in small redshift bins
and to extend the analysis up to higher redshifts, enabling us to
break parameter degeneracies and limit systematic uncertainties
\citep{oguri11}.
   
The accuracy in the estimate of cosmological parameters will
significantly benefit from the advancements planned for KiDS when it
will reach its final coverage of $1350\,\text{deg}^2$ of sky, with a
median redshift of $\sim0.8$.
 
The lensing signal can be incremented by considering a larger number
of clusters (which can be attained with either deeper or wider
surveys) and background sources (deeper surveys), and a wider survey
area, in order to collect the 2-halo term lensing signal up to
$50\text{ Mpc}/h$.  These aspects will be reached by ESA-Euclid
mission, which will provide a significantly improved data-set compared
to the one used in this work. The Euclid wide survey is expected to
cover an area of $\sim15000$ deg$^2$ and it will detect approximately
$n_g\sim 30$ galaxies per square arc-minute, with a median redshift
larger than $0.9$.

We conclude underling that our methodology of finding clusters, and
our cosmological statistical analyses in modelling the weak lensing
shear profiles around them, will represent a reference starting point
and a milestone for cluster cosmology in near future experiments.
 
\section*{Acknowledgments}
Based on data products from observations made with ESO Telescopes at
the La Silla Paranal Observatory under programme IDs 177.A-3016,
177.A-3017 and 177.A-3018, and on data products produced by
Target/OmegaCEN, INAF-OACN, INAF-OAPD and the KiDS production team, on
behalf of the KiDS consortium. OmegaCEN and the KiDS production team
acknowledge support by NOVA and NWO-M grants. Members of INAF-OAPD and
INAF-OACN also acknowledge the support from the Department of Physics
\& Astronomy of the University of Padova, and of the Department of
Physics of Univ. Federico II (Naples).  We acknowledge the KiDS
collaboration for the public data realises and the various scientists
working within it for the fruitful and helpful discussions.  We
acknowledge the grants ASI n.I/023/12/0, ASI-INAF n.  2018-23-HH.0 and
PRIN MIUR 2015 Cosmology and Fundamental Physics: illuminating the
Dark Universe with Euclid".  CG and LM are also supported by PRIN-MIUR
2017 WSCC32 ``Zooming into dark matter and proto-galaxies with massive
lensing clusters''.  CG acknowledges support from the Italian Ministry
of Foreign Affairs and International Cooperation, Directorate General
for Country Promotion.  MS acknowledges financial contribution from
contract ASI-INAF n.2017-14-H.0 and contract INAF mainstream project
1.05.01.86.10.  JHD acknowledges support from an STFC Ernest
Rutherford Fellowship (project reference ST/S004858/1).  We thank the
reviewer for her/his useful comments that helped us to improve the
presentation of our results.  \\ ~\\

All authors have contributed to the scientific preparation of this
work.\\ {\em Particularly}: CG has lead the paper preparing the
data-set, performing the computational analyses and organising the
manuscript; FM, LM, MS, AV and LG have helped with the development of
some codes and the scientific interpretation of the results; MM, MR,
FB, MR have implemented the code for cluster detection and lead the
first lensing analyses of the same data-set; SB, SC, GC, JHD, LI, GL,
LN and EP participated to the discussion and interpretation of the
results.

\appendix

\section{Model systematics: 1h term}
\label{app_rt}

In this section we discuss the impact of different assumptions for the
truncation radius in the $1h$-term modelling function of the projected
density profile (see Eq.~\ref{trunc.NFW}).  As a first point, we
examine the results modelling the $\sim 1h$-case data-set up to $3.16$
Mpc/$h$ and then discuss how the different assumptions impact on the
total matter density parameter.

\begin{figure}
  \includegraphics[width=\hsize]{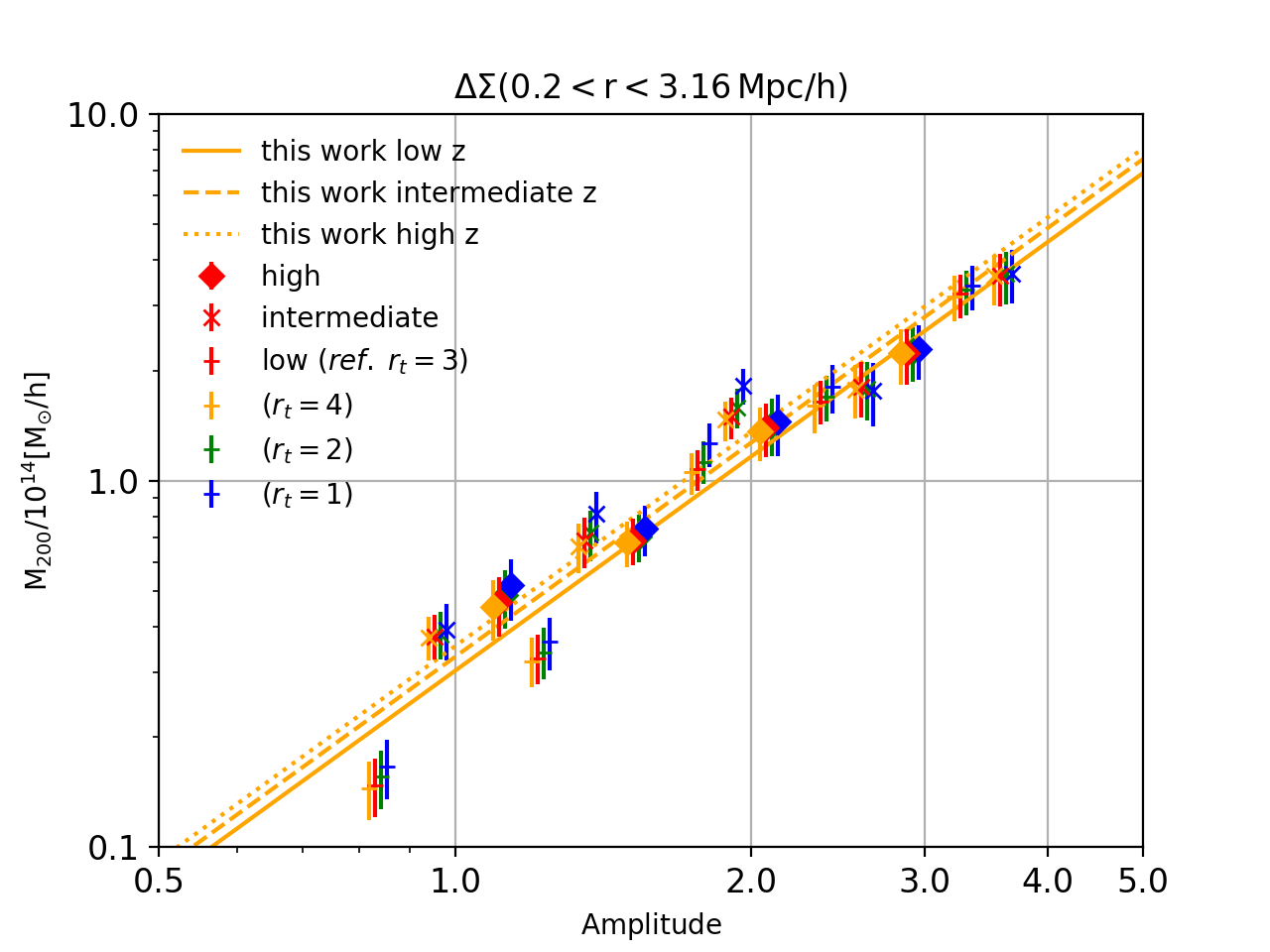}
  \caption{Mass-amplitude relation obtained using different modelling
    of the truncation radius; orange, green and blue refer to the
    cases with $r_t=4$, $r_t=2$ and $r_t=1$, respectively. The red
    data points show the results for our reference model with $r_t=3$.
    We slightly displace the data points along the x-direction,
    computed with different truncation radius definitions, to make the
    results more visible.  The orange lines exhibit the best fit
    results for our reference case for the $\sim 1h$-case data-set as
    in the first line of Tab. ~\ref{tab_emceelinear}.
  \label{fig_scaline_relation2}}
\end{figure}

In Fig. ~\ref{fig_scaline_relation2} we exhibit the mass-amplitude
relation obtained using different values of the truncation radius:
$r_t$=1, 2 and 4, with respect to our reference case $r_t=3$ (red data
points). The orange lines display the best fit results for our
reference case as in Tab. ~\ref{tab_emceelinear}. In our modelling
function we include also the effect of the 2h term for our reference
cosmology (Planck18 with $\Omega_M=0.3$ and $h=0.7$) and assuming the
\citet{tinker10} halo-bias relation. At fixed redshift and amplitude
bin, the data points and the error bars show the median and the
$18th-82nd$ percentiles of the corresponding posterior $M_{200}$
distributions.  From the figure we notice a small trend -- totally
consistent within the uncertainties, and not always in the same
direction -- with the truncation radius: on average lower truncation
radius slightly overestimate the mass.  In Tab.~\ref{tab_rt_emcee} we
report the results of the mass-amplitude relations parameters
(Eq.~\ref{eq_mobs}) for the different assumption on the $r_t$
parameter. We notice that a lower truncation radius in the $1h$ term
model tends to produce a marginally steeper mass-amplitude relation
When the truncation radius varies from $r_t=1$, to $r_t=4$ $\alpha$ --
in Eq.~\ref{eq_mobs} -- it changes from $0.138$ to $0.090$.

\begin{table}
\centering	
\caption{ Scaling relation parameters of the mass-amplitude
  relation. The values reported are the median and $18th$-$82nd$
  percentiles of the posterior distributions. The different lines show
  the results on the $\sim 1h$-case data-set adopting different values
  of the truncation radius $r_t$.}
\label{tab_rt_emcee}
\begin{tabular}{lccc}
\hline
case & $\alpha$ & $\beta$ & $\gamma$ \\ \hline
$r_t=1$ & $0.137_{-0.019}^{+0.019}$ & $1.900_{-0.104}^{+0.102}$ & $0.684_{-0.646}^{+0.635}$ \\
$r_t=2$ & $0.111_{-0.019}^{+0.019}$ & $ 1.934_{-0.098}^{+0.097}$ & $ 0.831_{-0.617}^{+0.626}$ \\
$r_t=3$ & $0.097_{-0.018}^{+0.019}$ & $1.943_{-0.097}^{+0.098}$ & $ 0.904_{-0.607}^{+0.615}$ \\
$r_t=4$ & $0.090_{-0.019}^{+0.019}$  & $1.949_{-0.097}^{+0.097}$ & $ 0.906_{-0.621}^{+0.618}$ \\
\hline
\end{tabular}
\end{table}

\section{Model systematics: 2h term}
\label{app_sis}

\begin{figure}
\includegraphics[width=\hsize]{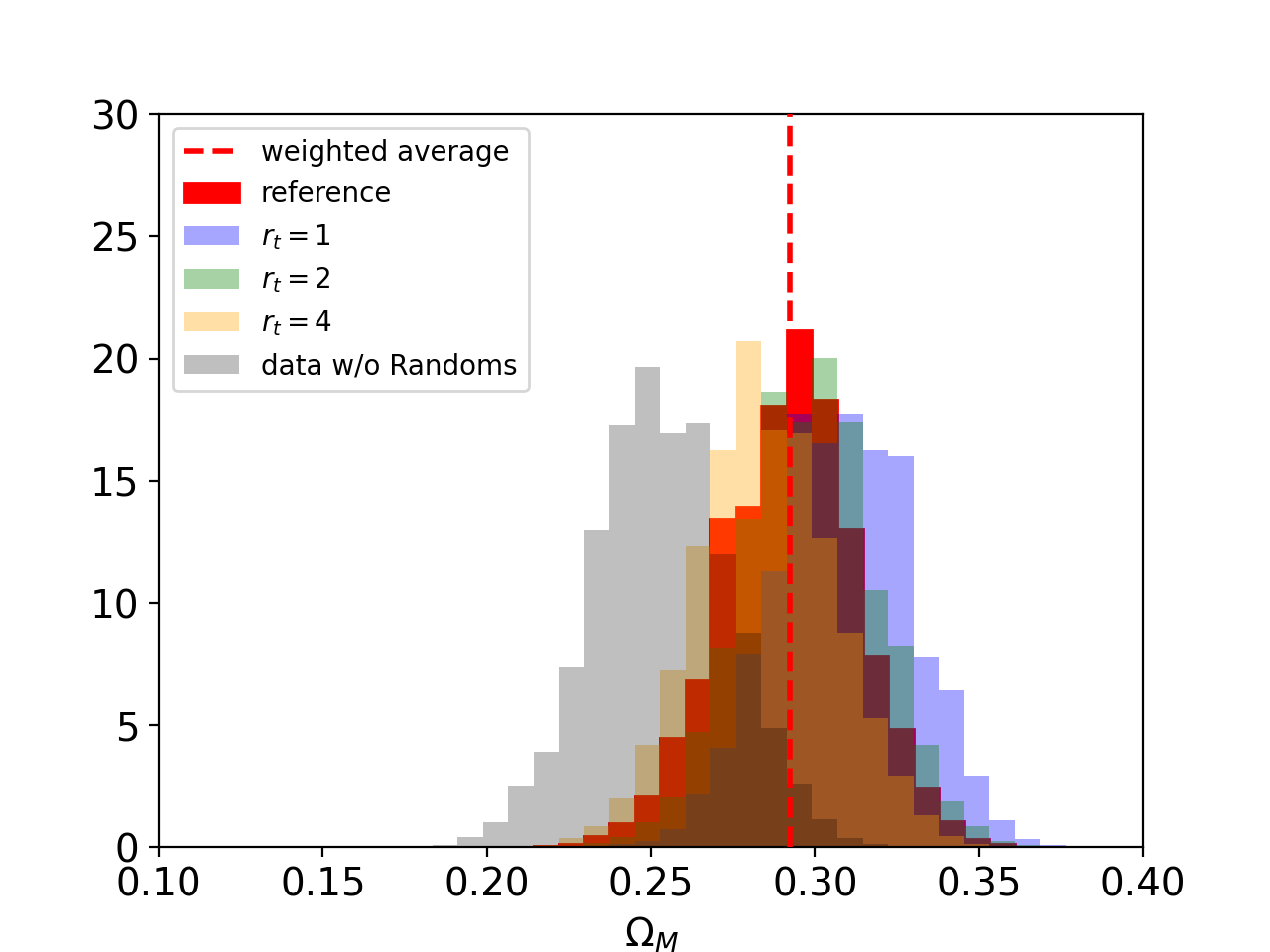}
\caption{\label{fig_om_rt} Combined posteriors, in redshift, for the recovered total
  matter density parameter assuming different values of the truncation
  radius. The red-filled histogram shows the posterior of our reference model
  with $r_t=3$. Gold, green and blue histograms show the
  distributions of $\Omega_M$ assuming $r_t=4$, $2$ and $1$,
  respectively. The grey histogram shows the result obtained from the
  stacked profiles estimated without accounting for the signal around
  random centres. The vertical red dashed line marks the weighted average of the
  reference posterior distribution.}
\end{figure}

In this section, we investigate how different model assumptions might
affect also the recovered total matter density parameter. In
Fig.~\ref{fig_om_rt} we exhibit the combined posterior distributions,
in redshifts, of $\Omega_M$ obtained assuming different values of the
truncation radius. We find a degeneracy between the effects of
$\Omega_M$ and the truncation radius, with a small value of the
truncation radius being compensated by an increase of the total matter
density parameter. Specifically, we find a systematic deviation of the
total matter density parameter of the order of $5.6\%$, $2.1\%$ and
$-2.5\%$ for $r_t=1$, $2$ and $4$ with respect to our reference case
$r_t=3$, respectively.  In the same Figure we show also the result
obtained from the stacked profiles estimated without accounting for
the signal around random centres. In this case we get an underestimate
$\Omega_M$ of approximately $13\%$.

\begin{figure}
  \includegraphics[width=\hsize]{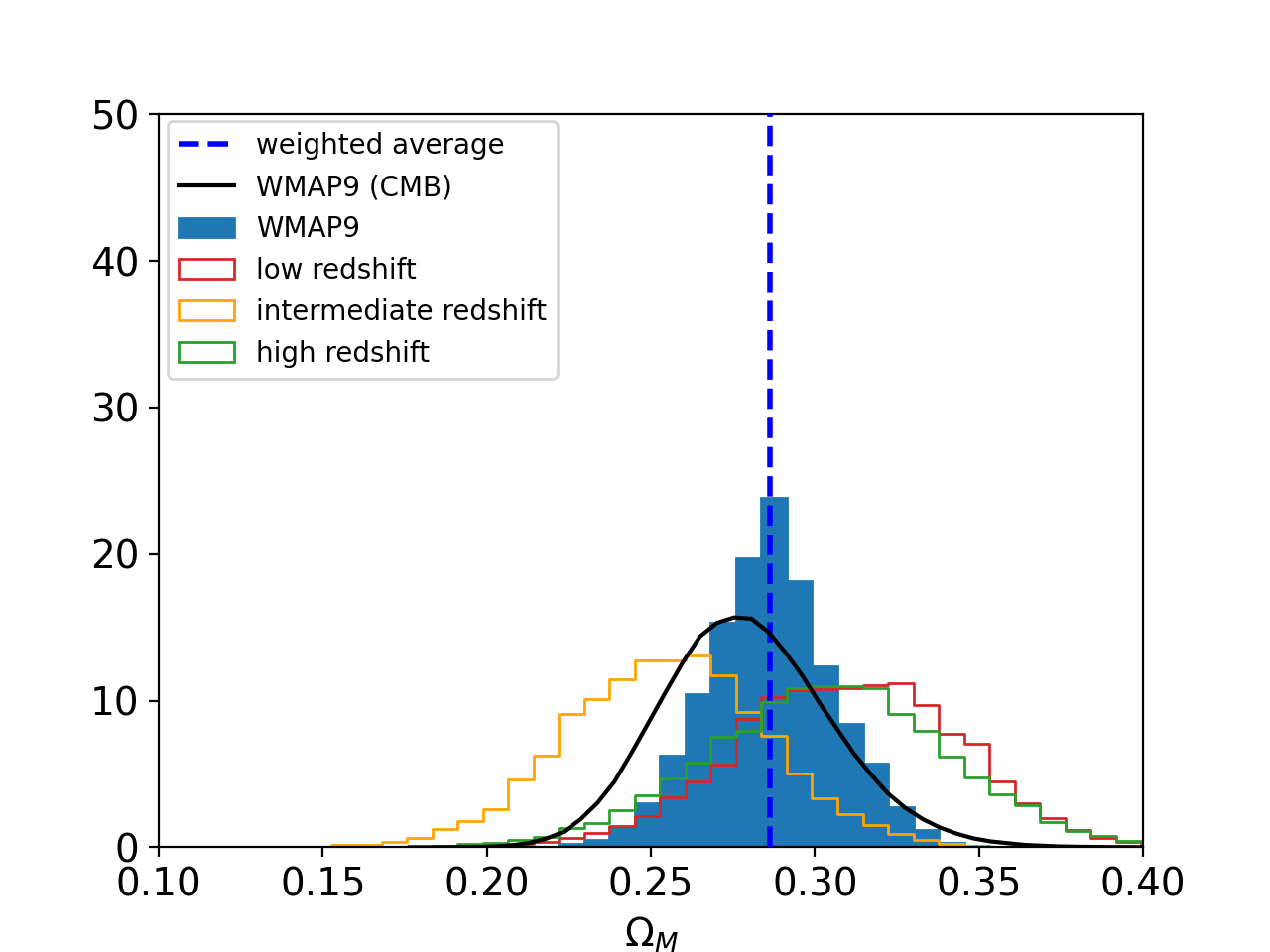}
  \caption{Posterior distributions of the total matter density
    parameter $\Omega_M$. Blue, orange and green histograms show the
    results for the low, intermediate and high redshift cluster
    samples, respectively. The red-filled histogram exhibit the
    posterior distribution combining all redshifts together, the
    dashed red line indicates the median of the distribution. The
    black curve displays the posterior distribution obtained from the
    analysis of the CMB data by \citet{wmap9}.}
  \label{fig_wmap}
\end{figure}

In Fig.~\ref{fig_wmap} we show the posterior distribution of the total
matter density parameter assuming a linear power spectrum for the $2h$
term adopting WMAP9 \citep{wmap9} cosmological parameters with
$h=0.7$.  The blue-filled histogram show the results combing all
redshifts together, while red, orange and green histograms display the
results for the low, intermediate and high redshift samples,
respectively. We notice full consistence in our posterior
distributions, particularly, when adopting a Planck18 or a WMAP9
cosmological parameters -- and fixing $h=0.7$ -- for the linear power
spectrum when recovering $\Omega_M$. Using WMAP9 we slightly
underestimate by $2\%$ the total matter density parameter with respect
to our reference case based on Planck18.

\begin{figure}
\includegraphics[width=\hsize]{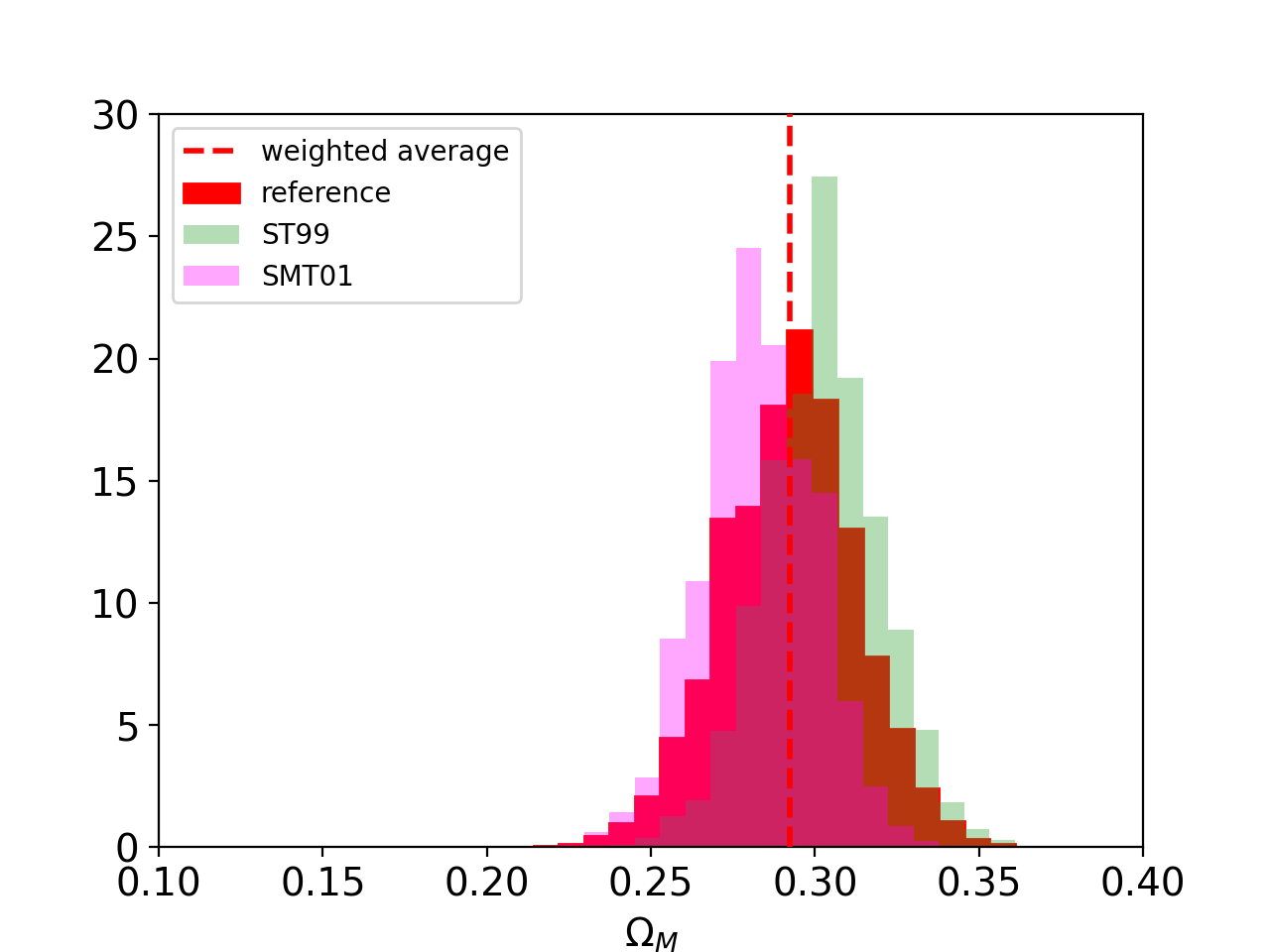}
\caption{\label{fig_om_bias} Combined posterior distributions obtained
  for different halo bias model. The red, gold and blue histograms
  show the cases obtained with \citet{tinker10} - our reference case,
  \citet{sheth99b} and \citet{sheth01b} halo biases, respectively.}
\end{figure}

Considering that the normalisation of the 2-halo model depends on the
halo bias parameter, we do expect our results to be dependent also on
the assumed bias model. Figure \ref{fig_om_bias} shows the impact of
different bias model assumptions on the $\Omega_M$ posterior
distributions. Assuming the \citet{sheth99b} or \citet{sheth01b}
biasing models we obtain a weighted average of the combined posteriors
for $\Omega_M$ that is $3.5\%$ and $-3.2\%$ different than the one
obtained with our reference model, based on the \citet{tinker10}
halo-bias recipe.

The deviations observed when changing the cosmology, the bias model
and the truncation radius are all comparable but smaller than the
statistical error of our measurement.  Those place some perspective on
the systematics requirements for future data.

\label{appendix_label}
\bibliographystyle{aa}
\bibliography{paper}
\label{lastpage}
\end{document}